\documentclass[]{jfm}

\usepackage{graphicx}
\usepackage{newtxtext}
\usepackage{newtxmath}
\usepackage{natbib}
\usepackage{hyperref}

\usepackage{overpic}
\usepackage{multirow}
\usepackage{derivative}
\usepackage{amsmath}

\usepackage[svgnames]{xcolor}

\hypersetup{
    colorlinks = true,
    urlcolor   = blue,
    citecolor  = black,
}

\newcommand{\RomanNumeralCaps}[1]

\title{Dissecting coherent motions in extreme wall shear stress events within adverse pressure gradient turbulent boundary layers}

\author{Leandro J. O. Silva\aff{1},
  William R. Wolf\aff{1} \corresp{\email{wolf@fem.unicamp.br}}
}

\affiliation{\aff{1}Universidade Estadual de Campinas, Campinas, SP, 13083-860, Brazil}

\begin{document}
\maketitle

\begin{abstract}
Coherent motions associated with extreme wall shear stress events are investigated for adverse pressure gradient turbulent boundary layers (APG-TBLs). The analyses are performed using wall-resolved large eddy simulations of a NACA0012 airfoil at angles of attack of $9$ and $12$ deg. and Reynolds number $4\times10^5$. The suction side exhibits attached TBLs which develop under progressively stronger APGs. A quadrant decomposition of Reynolds shear stress shows that sweeps and ejections dominate the momentum exchange between the mean and fluctuating fields, with the intensity of sweeps near the wall growing more rapidly with APG strength. Probability density functions of wall shear stress reveal a higher frequency of backflow events and an increased distribution symmetry with stronger APGs. Extreme positive and backflow events are examined using space–time correlations and conditional statistics. Their temporal scales expand with APG strength, consistent with a reduced convective velocity. Conditional averages show that backflow events originate from inner-layer sweep motions bringing high-momentum fluid toward the wall, followed by ejections that drive local deceleration. In such cases, the intensity of ejections is modulated by the APG strength. The dynamics of coherent turbulent structures and their interactions are examined using conditional flow field analyses. For extreme positive events, stronger APGs lead to shorter high-speed streaks, while the associated sweep motions generate spanwise velocities that increasingly influence the near-wall dynamics. In the case of backflows, stronger APGs shorten low-speed streaks and amplify high-speed structures associated with sweep motions, promoting spanwise alignment of vortical structures. Overall, APGs modify the structure and dynamics of extreme near-wall events by reshaping the balance and spatial organization of sweep- and ejection-dominated motions.

\end{abstract}



\section{Introduction}
\label{sec:headings}

Turbulent boundary layers (TBLs) play a central role in the field of fluid mechanics. Their dynamics strongly affect the performance of aerodynamic and marine devices, pipelines, ducts, and turbomachinery, besides also exerting a major influence on environmental flows in the ocean and atmosphere. 
Despite the inherent chaotic nature of turbulence, coherent motions are commonly observed in TBLs \citep{robinson1991}, where they play a key role in the transfer of turbulent kinetic energy (TKE) between the inner and outer layers. These motions often exhibit intermittent features, commonly associated with bursting events recognized as important contributors to TKE production in the near-wall region. Such events also impact the transition to turbulence and momentum transfer, leaving footprints on the wall shear stress, and hence, impacting drag production.

The dynamics of TBLs were investigated in an earlier study by \citet{kline1967}, who conducted experiments in an open water channel with particular emphasis on the near-wall region. By combining flow visualization and measurements, the authors analyzed the coherent motions in TBLs developing under various pressure gradients. 
They concluded that the dynamics of the streak breakup, beginning with its formation, lift-up, oscillation, bursting, and ejection, play an important role in the entire TBL, being a dominant process in the energy transfer between the inner and outer regions. The authors also observed that, in the presence of adverse pressure gradients (APGs), bursting events exhibited greater intensity, whereas under favorable pressure gradients (FPGs), this mechanism was attenuated. \citet{corino1969} employed flow visualization techniques to characterize fluid motions in the vicinity of a pipe wall. Their work highlighted the importance of ejections, particularly how their interaction with the mean shear contributes to the maintenance of turbulence. These previous studies drew attention to the significance of coherent motions, contributing to subsequent analyses of the mechanisms governing TKE production and the wall shear stress.

The role of coherent motions in the Reynolds shear stress was investigated by \citet{wallace1972} through experiments in turbulent channel flows. The authors analyzed the individual contributions of streamwise and wall-normal velocity fluctuations, concluding that ejection events had a dominant effect on the Reynolds stress and TKE distributions. 
In the same year, \citet{willmarth1972} employed conditional sampling analysis to identify intense bursting events that exhibited large contributions to Reynolds stresses and TKE. Such extreme events corresponded to ejection motions. The studies by \citet{wallace1972} and \citet{willmarth1972} were the precursors of quadrant analysis, a technique that classifies streamwise and wall-normal velocity fluctuations within a Cartesian plane, where the second and fourth quadrants correspond to ejections and sweeps, respectively, and the first and third quadrants represent outward and inward interactions.
Although the previous references have emphasized the role of ejections, \citet{offen1974} investigated the relationship between bursts and sweeps in a TBL using dye injection and bubble wire visualization. Their observations revealed that the bursting phenomenon follows a cycle, beginning with a sweep originating in the logarithmic region, which generates disturbances that ultimately trigger a burst. This finding led to the model of the aforementioned bursting `quasi-cycle' process \citep{offen1975}. 

Beyond the visualization techniques used to understand the dynamics of sweeps and ejections in TBLs, conditional analysis from hot-wire probe measurements has also been employed \citep{wallace1972, willmarth1972}. \citet{morrison1992} also applied conditional sampling techniques to isolate the contributions of sweeps and ejections in zero-pressure-gradient (ZPG) boundary layers developing over smooth and rough-to-smooth walls. A complete description of the behavior of sweeps and ejections in both inner and outer layers revealed that both types of motion exhibited high spatial coherence, efficiently transporting momentum, and supporting the hypothesis that they form part of a large-scale inertial structure. 

Previous investigations have primarily focused on elucidating the dynamics of coherent motions and their role in momentum transport and TKE production. It was shown that these motions are crucial for the near-wall dynamics \citep{jimenez1999}, directly impacting the wall shear stress. This aspect was studied by \citet{orlandi1994}, who performed simulations of a simplified two-dimensional model of a plane transversal to the flow, combined with the minimal flow unit methodology \citep{jimenez1991}. The numerical model consisted of a pair of counter-rotating vortices embedded in a uniform streamwise shear field.
The results showed that the longitudinal velocity profile was lifted at the center of the vortex pair, producing a low speed streak, while the outer regions were forced toward the wall, generating regions of high wall shear. The latter mechanism was associated with a sweep motion that transported high-speed fluid from the outer layer to the near-wall region. 

Understanding the physical process related to wall shear stress provides the foundation for developing flow control strategies. Moreover, such analysis offers insights of drag generation mechanisms in cases of positive stress, and reverse flow dynamics in cases of negative stress, the latter being related to local boundary layer separation \citep{simpson1989}. In addition, patterns of instantaneous wall shear stress reveal the footprints of near-wall streaks, whose modulations reflect the influence of large-scale outer-layer structures \citep{abe2004}. Particular attention should be warranted to the study of extreme wall shear stress events, as they significantly contribute to skin-friction drag and local separation. Understanding their dynamics is essential for advancing turbulence modeling and developing strategies for drag reduction.

When considering the generation of extreme positive wall shear stress events, a mechanism similar to that described by \citet{orlandi1994} is observed. \citet{sheng2009} performed channel flow experiments at $Re_\tau = 1470$ using digital holographic microscopy to simultaneously measure both wall shear stress components and the three-dimensional velocity field. Their analysis showed that regions of large wall shear stress magnitude are typically associated with single or paired quasi-streamwise vortices that induce a sweep motion and spanwise momentum transport. The authors also found that regions of low skin friction were related to paired streamwise vortices that promoted ejection motions. 
Subsequently, \citet{hutchins2011} conducted experiments in a high Reynolds number TBL using hot-film shear-stress sensors together with a traversing hot wire to conditionally sample high and low skin-friction events. They reported that high-speed structures associated with high positive skin friction exhibit an important role on the modulation of small-scale activity near the wall, whereas this activity diminishes away from the wall. The opposite trend was observed for low-speed structures, which are associated with low skin-friction events, besides being more predominant on the small scales away from the wall. 

The relative contributions of small- and large-scale structures to Reynolds shear stress were investigated by \citet{gomit2018} through particle image velocimetry together with wall shear stress sensors in a high Reynolds number TBL experiment. They decomposed the friction velocity fluctuations into four quartiles, each one containing $25\%$ of the events. Their findings revealed that extreme events in the first and last quartiles are predominantly associated with large and small scales, respectively, both contributing substantially to the total Reynolds shear stress.
The topic of extreme wall shear stress events was further examined by \citet{pan2018}, who analyzed snapshots from a direct numerical simulation (DNS) dataset of a TBL developing over a smooth wall. They assessed the role of large-scale structures in both extreme positive and negative wall shear events. It was found that extreme positive events are linked to sweep motions that transport high-momentum fluid from the outer layer towards the wall, in a mechanism similar to a splatting process.
More recently, \citet{guerrero2020} performed DNS of turbulent pipe flows across a range of Reynolds numbers to investigate the mechanisms underlying extreme positive and negative wall shear stress events. Their conditional analysis demonstrated that extreme positive events are related to strong quasi-streamwise vortices in the buffer region, which transport high streamwise momentum from the overlap and outer layers towards the wall.

Negative wall shear stress events can also occur in wall-bounded turbulent flows, despite early claims that such events would not exist \citep{eckelmann1974}. 
Subsequent experimental and numerical investigations have demonstrated that these events, although rare, are indeed present in turbulent flows, being associated with the occurrence of backflows. 
\citet{lenaers2012} performed DNS of turbulent channel flows at various Reynolds numbers to quantify the characteristics of backflow events. They observed a Reynolds-number dependence where backflow events become more frequent and occur farther away from the wall, as the friction Reynolds number increases. Their results also demonstrated that backflow regions exhibit an average diameter of 20 viscous units, and are caused by an oblique vortex located outside the viscous sublayer. Using images of flexible micropillars, \citet{brucker2015} demonstrated the presence of backflow in a ZPG-TBL over a flat plate. This previous work showed that backflow regions coincide with areas of large spanwise gradients of wall shear stress due to splatting.
More recently, \citet{cardesa2019} analyzed DNS datasets of turbulent channel flow at multiple Reynolds numbers by characterizing regions with negative streamwise velocity to investigate the dynamics of backflow events. They found that backflow structures result from complex interactions between regions of high and low spanwise vorticity away from the wall. It was also shown that these structures become more elongated in the spanwise direction at higher wall-normal positions. 

\citet{guerrero2020} examined extreme negative wall shear stress events associated with backflows and reported that their formation can be associated either with a single strong oblique vortex, or with a pair of counter-rotating vortices that transport momentum from the wall toward the outer layer. In a follow-up study, \citet{guerrero2022} observed that backflow events are often preceded by the asymmetric collision of two large-scale structures of high and low speed. The result of this interaction is a vortex that generates the negative wall shear stress and triggers the breakdown of the low-speed large-scale structure. In such cases, high momentum from the high-speed structure is transported from the outer region towards the wall, intercepting the low-speed structure located near the wall, engulfing a small portion of the latter. The authors associated this phenomenon with the near-wall self sustaining process of turbulent boundary layers.

The influence of APGs on backflow events was examined by \citet{vinuesa2017} through analysis of a DNS dataset of a NACA4412 airfoil. The authors focused on the TBL developing on the suction side, which exhibits an increasing APG in the streamwise direction. They reported backflow levels of $30\%$ near the trailing edge, where a very strong APG condition is reached. Conditional analysis based on extreme wall shear stress events indicated that, under strong APG conditions, the wall shear stress orientation has a higher probability to align either with or against the freestream direction. 
More recently, \citet{bross2019} investigated the interaction of coherent flow structures in an APG-TBL using particle tracking velocimetry. Their findings indicate that backflow events are triggered when a low-momentum large-scale structure interacts with and decelerates an existing low-speed streak.
The mechanism described by the authors differs from those reported by \cite{guerrero2020} and \citet{guerrero2022}, 
highlighting the uncertainty regarding the influence of APGs on the generation of both positive and negative extreme wall shear stress events.

When considering APG-TBLs, several studies have examined how adverse pressure gradients affect turbulence statistics, for instance, the energization of outer-layer structures and the amplification of large-scale motions \citep{monty2011, harun2013}. Although the fundamental energy transfer mechanisms remain broadly similar to those in ZPG flows \citep{gungor2022, deshpande2024}, their intensity is substantially modified by the presence of an APG. Recent work has also demonstrated that an FPG upstream of an APG can strongly influence the inner-layer dynamics, while exerting a weaker effect on the outer region \citep{parthasarathy2023}. Moreover, studies of airfoils at high angles of attack revealed that the resulting strong APGs affect not only the Reynolds stresses but also the TKE production mechanisms and the anisotropy along the turbulent boundary layer \citep{silva2024}. While these studies focused on fundamental turbulence statistics of APG-TBLs, they did not address the occurrence of extreme events or their associated coherent motions.

To advance the understanding of how APGs influence near-wall coherent structures responsible for momentum transport and wall shear stress generation, the present work examines the effect of increasing APG intensities on turbulent structures responsible for extreme positive and backflow events. Particular attention is given to the spatial support and dynamics of sweep and ejection motions, and their interactions with near-wall streaks and vortical  structures.
To this end, wall-resolved large eddy simulations (LES) are performed for a NACA0012 airfoil at angles of attack of $9$ deg. and $12$ deg., focusing on the development of the TBL along the suction side. The simulations are conducted at a chord-based Reynolds number of $Re = 4 \times 10^5$ and a freestream Mach number of $M = 0.2$. Despite the relatively high angles of attack, no mean flow separation is observed.
Further details of the flow configurations analyzed, as well as the theoretical formulation and numerical methods employed, are presented in \S \ref{sec:methodology}. The characterization of the mean flow, including the pressure-gradient parameter, a study of momentum balances, and the decomposition of the Reynolds shear stress, is discussed in \S \ref{sec:mean}. Section \S \ref{sec:extreme} investigates the occurrence of extreme wall shear stress events under mild and strong adverse pressure gradients. This analysis is supported by a statistical characterization of events using probability density functions and space-time correlations (\S \ref{sec:characterization}); conditional statistics of extreme events, focusing on the evolution of mean flow profiles and the contribution of individual Reynolds shear stress quadrants (\S \ref{sec:conditional}); and the examination of the coherent structures responsible for backflow and extreme positive event dynamics (\S \ref{sec:coherent_structures}). Finally, the main conclusions regarding the role of APGs on extreme shear stress events and their associated coherent structures are summarized in \S \ref{sec:conclusions}.

%

\section{Theoretical and numerical methodology}\label{sec:methodology}

Wall-resolved large eddy simulations are conducted to solve the compressible mass, momentum, and energy balance equations written in general curvilinear coordinates $(\xi^i \mbox{, } i=1,2,3)$, in which the velocity vector is represented in the contravariant $(U^i)$ form. The balance equations are expressed as 
\begin{equation}\label{eq:continuity_equation}
\pdv{\rho}{t} +(\rho U^j)_{,j} = 0 \mbox{ ,}
\end{equation}
\begin{equation}\label{eq:momentum_equation}
\pdv{\rho U^i}{t} +\left(\rho U^iU^j\right)_{,j} =  - \left(g^{ij}p\right)_{,j} + \tau^{ij}_{,j} \mbox{ ,}
\end{equation}
and,
\begin{equation}\label{eq:energy_equation}
\pdv{E}{t} +\left[(E+p)U^j \right]_{,j} = -q^j_{j} + \left(\tau^{ij}g_{ik}U^k\right)_{,j} \mbox{ ,}
\end{equation}
where time, density, and pressure are denoted by $t, \rho,$ and $p$, respectively. The contravariant and covariant metric tensors are represented by $g^{ij}$ and $g_{ij}$, respectively, and subscripts preceded by comma $(.)_{,j}$ indicate covariant derivatives (see Appendix \ref{appA} and \citet{aris1989} for more details). The Einstein summation convention applies for repeated indices.

The equations are solved in a non-dimensional form using the freestream properties and the airfoil chord $L$. 
Further details regarding the non-dimensionalization procedure are provided by \citet{lui2022}. Additionally, considering a Newtonian fluid that obeys Fourier's law, the viscous stress tensor $\tau^{ij}$, heat flux $q^j$, and total energy $E$ are described, respectively, by
\begin{equation}\label{eq:viscous_stress_tensor}
\tau^{ij} = \mu\dfrac{M}{\Rey}\left(g^{jk}U^i_{,k} 
          + g^{ik}U^j_{,k} -\dfrac{2}{3}g^{ij}U^k_{,k}\right) \mbox{ ,}
\end{equation}
\begin{equation}\label{eq:heat_flux}
q^j = -\dfrac{M}{\Rey\Pran}\mu g^{ij}T_{,i} \mbox{ ,}
\end{equation}
and,
\begin{equation}\label{eq:total_energy}
E = \dfrac{p}{\gamma-1}+\dfrac{1}{2}\rho g_{ij}U^iU^j \mbox{ .}
\end{equation}
Here, $\Rey$ is the Reynolds number based on the airfoil chord and freestream velocity, $M$ is the freestream Mach number, $\Pran$ is the Prandtl number, and $T$ is the temperature. Assuming a calorically perfect gas, the system of equations is closed by the equation of state
\begin{equation}\label{eq:equation_of_state}
p = \dfrac{(\gamma-1)}{\gamma}\rho T\,\mbox{,}
\end{equation}
where $\gamma$ is the ratio of specific heats. The dynamic viscosity $\mu$ is computed by Sutherland's law, expressed by
\begin{equation}\label{eq:sutherlandslaw}
\mu = \left[(\gamma-1)T\right]^{\frac{3}{2}}\dfrac{1+S_\mu}{T(\gamma-1)+S_\mu}\,\mbox{,}
\end{equation}
where $S_\mu$ is the non-dimensional Sutherland constant.

The computational domain is constructed using an overset grid approach, in which an O-type grid, conforming to the airfoil surface, is embedded within a background Cartesian H-type grid. The present O-grid is built using a hyperbolic grid generator that enforces orthogonality near the airfoil surface. Hence, throughout this work, the instantaneous velocity components in contravariant form are denoted by $U^1 = U_t$, $U^2 = U_n$, and $U^3 = W$, representing the tangential, wall-normal, and spanwise components, respectively, with respect to the curvilinear coordinates $\xi^1$, $\xi^2$, and $\xi^3$. Subscripts $t$ and $n$ are used to indicate properties evaluated in the tangential and normal directions with respect to the airfoil surface. When needed, instantaneous velocities in the Cartesian coordinate system are written as $U, V$ and $W$ with respect to the $x$, $y$ and $z$ directions, respectively. It is important to highlight that, for the present simulations with spanwise homogeneous flows, $\xi^3 = z$.

The spatial derivatives of the governing equations are discretized using a sixth-order compact finite-difference scheme implemented on a staggered grid \citep{nagarajan2003}. The communication between the overlapping zones of the O- and H-grids is ensured using a fourth-order compact interpolation scheme \citep{bhaskaran2010}. To suppress numerical instabilities caused by grid stretching and interpolation across grid interfaces, a sixth-order compact filter \citep{lele1992} is applied away from the airfoil wall, as shown by \citet{silva2024}.
The temporal integration of the governing equations is performed using a hybrid implicit-explicit approach. An implicit second-order scheme is employed on the O-grid, whereas an explicit third-order Runge-Kutta scheme is applied on the Cartesian mesh. Characteristic boundary conditions based on Riemann invariants are applied in the farfield, together with a sponge layer that prevents reflections of acoustic waves. At the airfoil surface, no-slip adiabatic conditions are imposed and in the spanwise direction periodic boundary conditions are applied. The present numerical framework has been validated in previous studies of compressible flows over airfoils for different conditions \citep{wolf2012,wolf2012DU,ricciardi2022,lui2022,miotto2022}. Further details about the numerical methodology can be found in \citet{nagarajan2003,bhaskaran2010,wolf2011}.

The simulations are performed for a NACA0012 airfoil at $9$ and $12$ deg. angle of attack, with freestream Mach number $M = 0.2$ and chord-based Reynolds number $Re = 4 \times 10^5$. The airfoil trailing edge is rounded at $99\%$ of the chord with a curvature radius of $0.0015L$ to obtain smooth metric tensors.
Table \ref{tab:grid} presents the number of grid points $\left(N_{\xi^1} \times N_{\xi^2} \times N_{\xi^3}\right)$ for the individual grids, as well as the respective spatial resolution in terms of wall units for both simulations. The present simulations meet the requirements for wall-resolved LES as recommended by \citet{georgiadis2010}. The airfoil spanwise length $L_z$ is also presented in table \ref{tab:grid}. The present values of this parameter are chosen to resolve at least 5 times the length of the boundary-layer displacement thickness at the trailing edge in order to minimize the effects of spanwise periodic boundary conditions. As shown by \citet{silva2024}, the spanwise two-point correlations of velocity fluctuations decay significantly along the span for various positions along the entire airfoil chord.
\begin{table}
  \begin{center}
\def~{\hphantom{0}}
\begin{tabular}{ccccccc}
AoA (deg.) & Grid type & Grid size ($N_{\xi^1} \times N_{\xi^2} \times N_{\xi^3}$) & $L_z/L$ & $\Delta \xi^{1+}_{max}$ & $\Delta \xi^{2+}_{max}$ & $\Delta \xi^{3+}_{max}$        \\ [3pt]
\multirow{2}{*}{9}  & O-grid    & $1200 \times 170 \times 144$ & \multirow{2}{*}{$0.12$} & \multirow{2}{*}{40} & \multirow{2}{*}{0.50} & \multirow{2}{*}{18} \\
                    & H-grid    & $960 \times 599 \times 72$   &                &           &                              &            \\
\multirow{2}{*}{12} & O-grid    & $1200 \times 170 \times 288$ & \multirow{2}{*}{$0.24$} & \multirow{2}{*}{36} & \multirow{2}{*}{0.45} & \multirow{2}{*}{15} \\
                    & H-grid    & $960 \times 599 \times 144$  &                 &           &                              &                 
\end{tabular}
  \caption{Grid details and near-wall resolution in wall units for $9$ and $12$ deg. angle of attack.}
  \label{tab:grid}
  \end{center}
\end{table}

\begin{figure}
 \centering
 \begin{overpic}[width=1\textwidth]{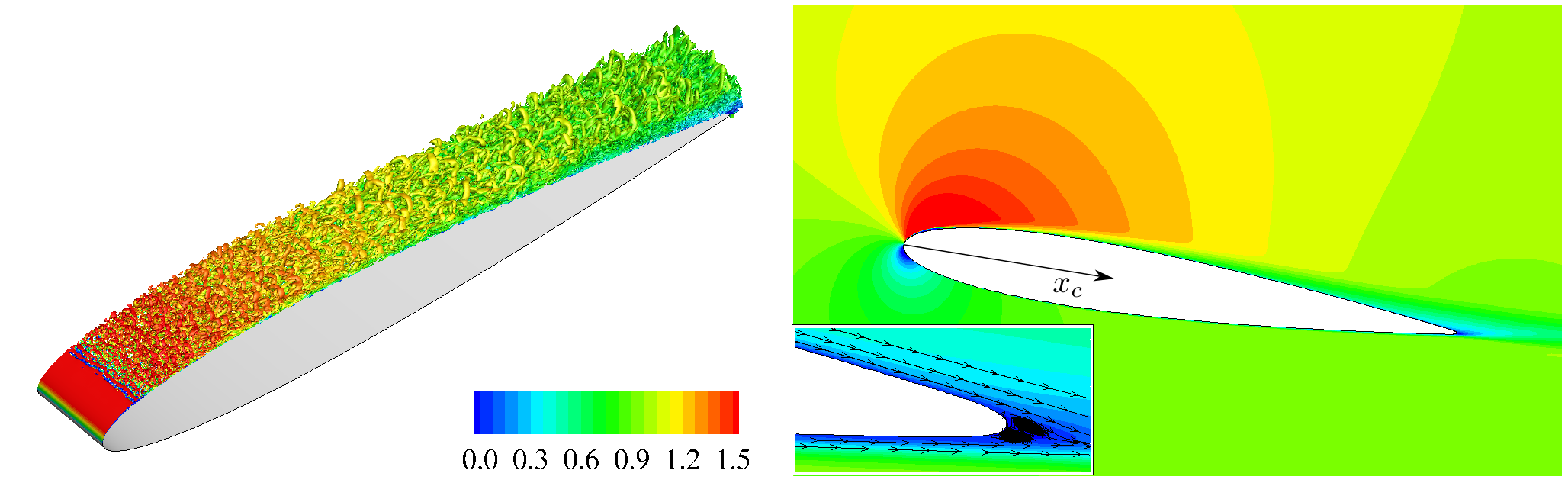}
	\put(1,28){(\textit{a})}
        \put(51.5,28){(\textit{b})}
        \put(36.25,7){$U/U_{\infty}$}
  \end{overpic}
  \begin{overpic}[width=1\textwidth]{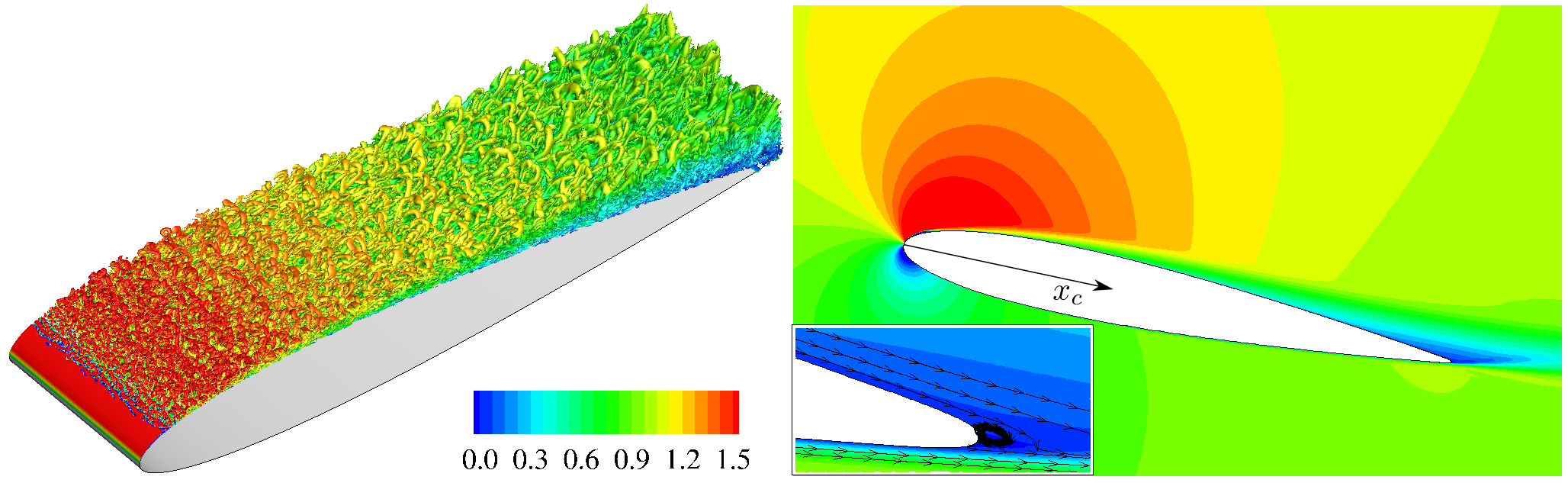}
	\put(1,28){(\textit{c})}
        \put(51.5,28){(\textit{d})}
        \put(36.25,7){$U/U_{\infty}$}
  \end{overpic}
\caption{Simulation results presented for $9$ and $12$ deg. angle of attack in the top and bottom rows, respectively. Iso-surfaces of $Q$-criterion colored by the streamwise velocity $U$ normalized by the freestream velocity $U_\infty$ are presented in (\textit{a}) and (\textit{c}). Mean flow fields are depicted in (\textit{b}) and (\textit{d}) using the same contour levels of the left column plots.}
\label{fig:snapshots}
\end{figure}
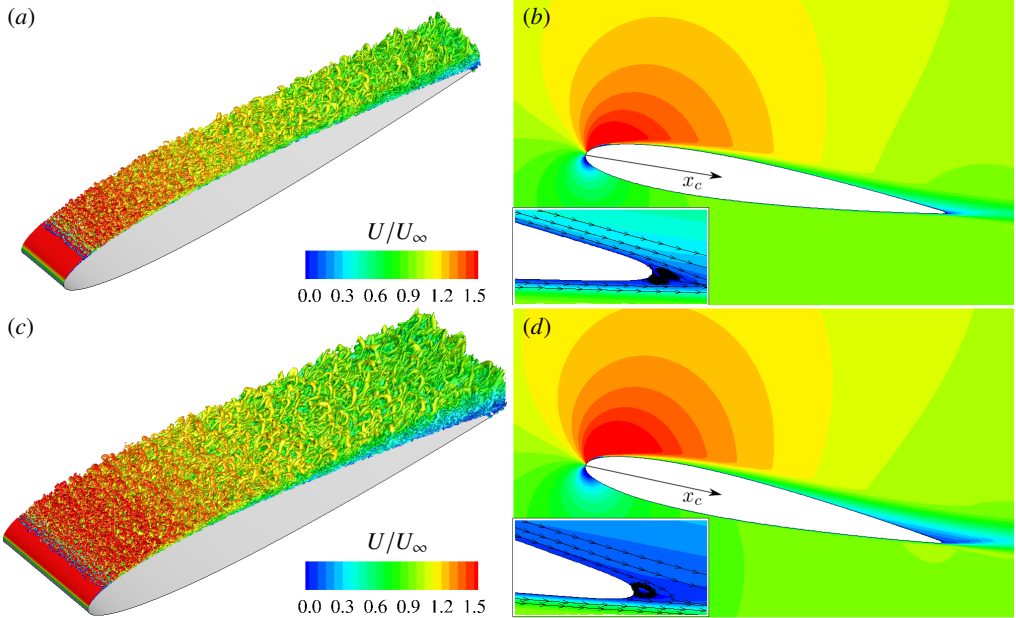

A numerical tripping is applied on the suction side of the airfoils in the region $0.04 \leq x_c \leq 0.09$ to promote bypass transition and prevent the formation of Tollmien–Schlichting waves, where $x_c$ is the position measured along the airfoil chord, as shown in figure \ref{fig:snapshots}.
The tripping consists of a random spanwise and streamwise disturbance, introduced with minimal amplitude to trigger transition without introducing excessive perturbations to the flow. As a consequence of tripping and the APG, a turbulent boundary layer develops along the suction side of the airfoils, whereas the pressure side maintains a laminar boundary layer. Figures \ref{fig:snapshots}(\textit{a}) and (\textit{c}) present snapshots for the $9$ and $12$ deg. angle of attack simulations, respectively, illustrating the presence of fine-scale turbulent structures. Figures \ref{fig:snapshots}(\textit{b}) and (\textit{d}) present the mean flow fields, where the insets provide an enlarged view of the trailing edge region demonstrating that, despite the high angle of attack, the mean flow remains attached upstream of the trailing edge.
More details about the flow conditions, tripping setup, and grid resolution can be found in \citet{silva2024}.

\section{Mean flow characterization}
\label{sec:mean}

This section presents the mean flow characterization of the investigated TBLs. The results are expressed using a Reynolds decomposition, where the velocity components are written as $U^i = \left<U^i\right> + u^i$. Here, the angle brackets denote the combined temporal and spanwise average, while the lowercase letters represent the corresponding fluctuations. Similarly, scalar variables are expressed as $f = \left<f\right> + f'$, with the primed quantities denoting fluctuations.

\begin{figure}
 \centering
 \begin{overpic}[width=1\textwidth]{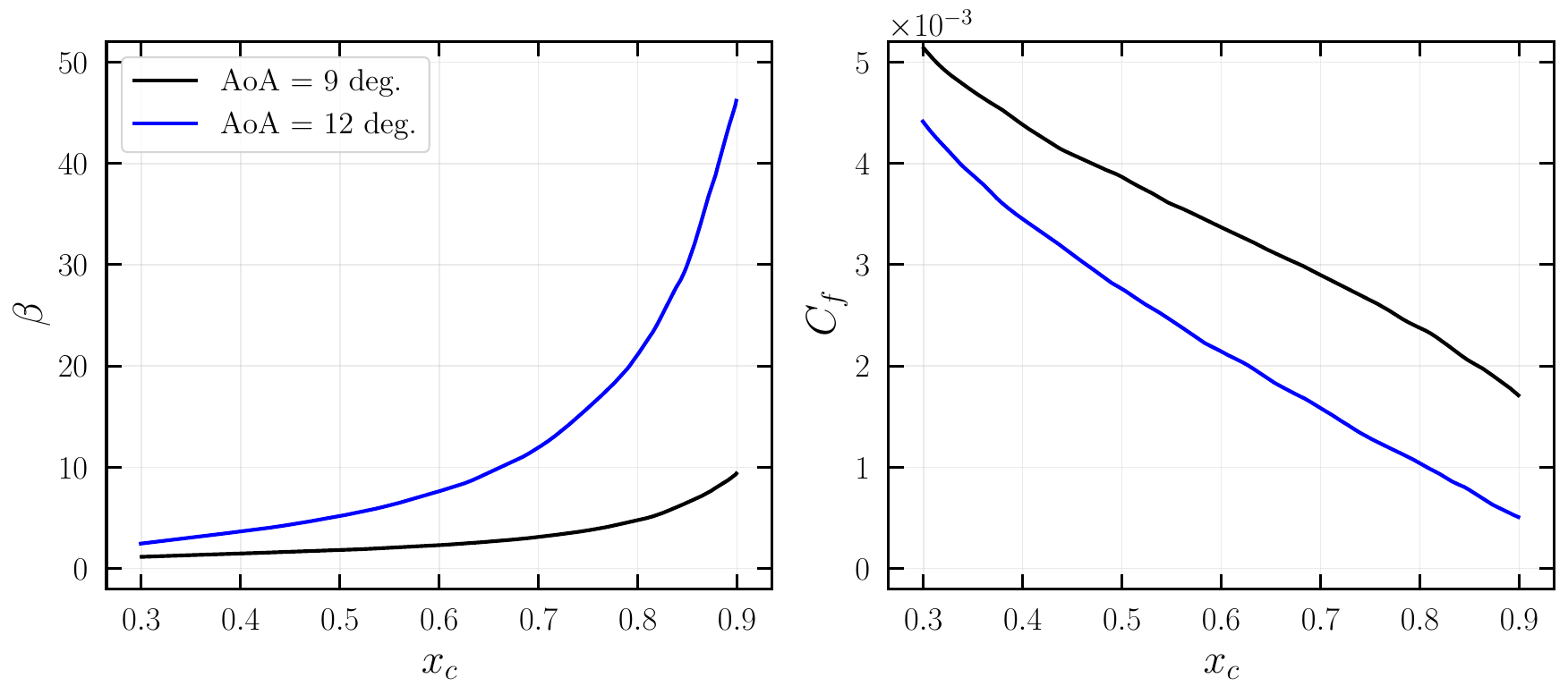}
	\put(3,42.5){(\textit{a})}
        \put(53,42.5){(\textit{b})}
  \end{overpic}        
\caption{Evolution of the (\textit{a}) pressure-gradient parameter $\beta$ and (\textit{b}) skin-friction coefficient $C_f$ along the airfoil chord.}
\label{fig:beta}
\end{figure}

Due to the high angles of attack, the TBLs develop over increasing APGs that can be locally quantified by the pressure gradient parameter $\beta = (\delta^*/\left<\tau_w\right>)\mathrm{d}\left<p\right>/\mathrm{d}x_t$. Here, $\delta^*$ is the displacement thickness, $\tau_w = \mu\left(\left<U^1\right>\right)_{,2}$ is the wall shear stress and $\mathrm{d}\left<p\right>/\mathrm{d}x_t=\left(\left<p\right>\right)_{,1}$ is the wall pressure gradient in the direction tangential to the airfoil surface. Figure \ref{fig:beta} presents the chordwise distribution of $\beta$ and the friction coefficient $C_f = 2\left<\tau_w\right>/\rho_\infty U_\infty^2$ for the $9$ and $12$ deg. angle of attack. 
In these figures, we present the results starting from the position $x_c = 0.30$ to avoid the influence of the tripping region. It can be observed from figure \ref{fig:beta}(\textit{a}) that the difference in angle of attack leads to a significant difference between the pressure gradients, particularly in the trailing edge region, where the $12$ deg. case exhibits a pronounced increase in the pressure gradient. In figure \ref{fig:beta}(\textit{b}) the chordwise evolution of the friction coefficient $C_f$ reflects the APG effect, where an increasing APG is associated with a reduction in $C_f$.
In order to further investigate the effects of mild, moderate and strong APG conditions, the subsequent analyses will place particular emphasis on the airfoil chord positions $x_c = 0.5, 0.7$ and $0.9$, where the pressure gradient parameters are given, respectively, by $\beta = 1.8$, $3.1$, and $9.4$ for 9 deg, and  $\beta = 5.2$, $12.0$, and $46.2$ for $12$ deg.


An analysis of the mean momentum balance equations is conducted to elucidate the influence of the APG on the individual terms comprising the balance. As discussed in section \ref{sec:methodology}, the Navier-Stokes equations are solved in general curvilinear coordinates using contravariant vector components, which require the use of covariant derivatives. This approach enables a direct evaluation of the flow properties along the tangential and wall-normal directions. Within this framework, the mean momentum equations can be expressed in the tangential and wall-normal directions, respectively, as:
\begin{equation}
  \left<U^j\right>\left(\left< U^1\right>\right)_{,j} + \frac{1}{\left<\rho\right>}(g^{1j}\left<p\right>)_{,j} + \left<u^1 u^j\right>_{,j} - \frac{1}{\left<\rho\right>}\left<\tau^{1j} \right>_{,j} = 0 \mbox{ ,}
  \label{eq:momentum_xt}
\end{equation}
and
\begin{equation}
  \left<U^j\right>\left(\left< U^2\right>\right)_{,j} + \frac{1}{\left<\rho\right>}(g^{2j}\left<p\right>)_{,j} + \left<u^2 u^j\right>_{,j} - \frac{1}{\left<\rho\right>}\left<\tau^{2j} \right>_{,j} = 0\mbox{ .}
  \label{eq:momentum_yn}
\end{equation}
%

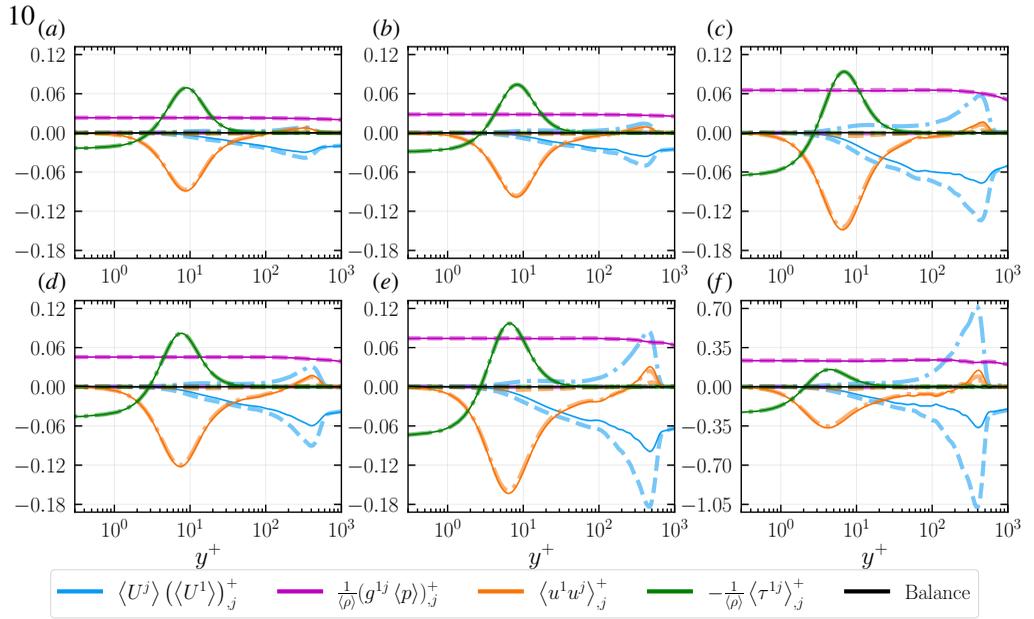
\begin{figure}
 \centering
 \begin{overpic}[width=1\textwidth]{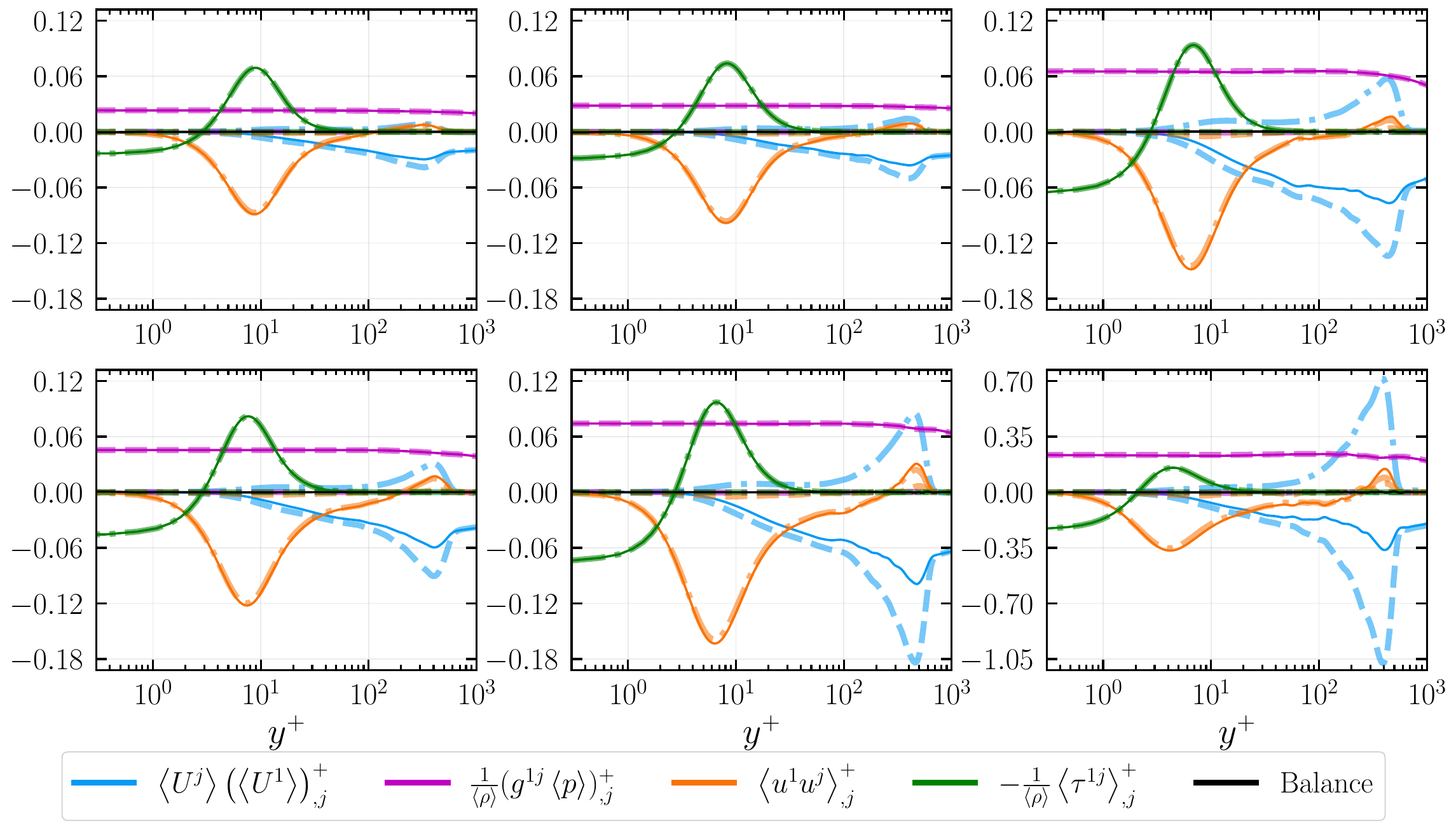}
	\put(3,57.5){(\textit{a})}
        \put(35.8,57.5){(\textit{b})}
        \put(68.55,57.5){(\textit{c})}
        \put(3,32.75){(\textit{d})}
        \put(35.8,32.75){(\textit{e})}
        \put(68.55,32.75){(\textit{f})}
  \end{overpic}        
\caption{Individual terms of the wall-tangential mean momentum equation at chordwise positions (\textit{a,d}) $x_c = 0.5$, (\textit{b,e}) $0.7$, and (\textit{c,f}) $0.9$. Solid lines denote the contracted terms in equation \ref{eq:momentum_xt}, while dashed and dot-dashed lines correspond to contributions from components $j=1$ and $j=2$, respectively. Top and bottom rows show results for angles of attack $9$ and $12$ deg, respectively. All terms are non-dimensionalized by $u_\tau^3/\nu$.}
\label{fig:momentum_t}
\end{figure}

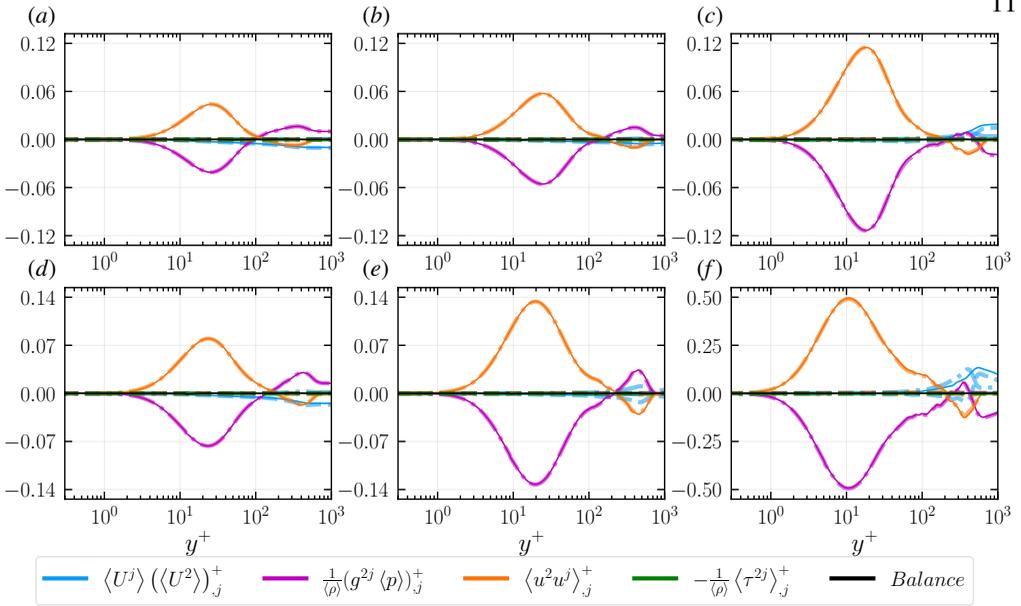
\begin{figure}
 \centering
 \begin{overpic}[width=1\textwidth]{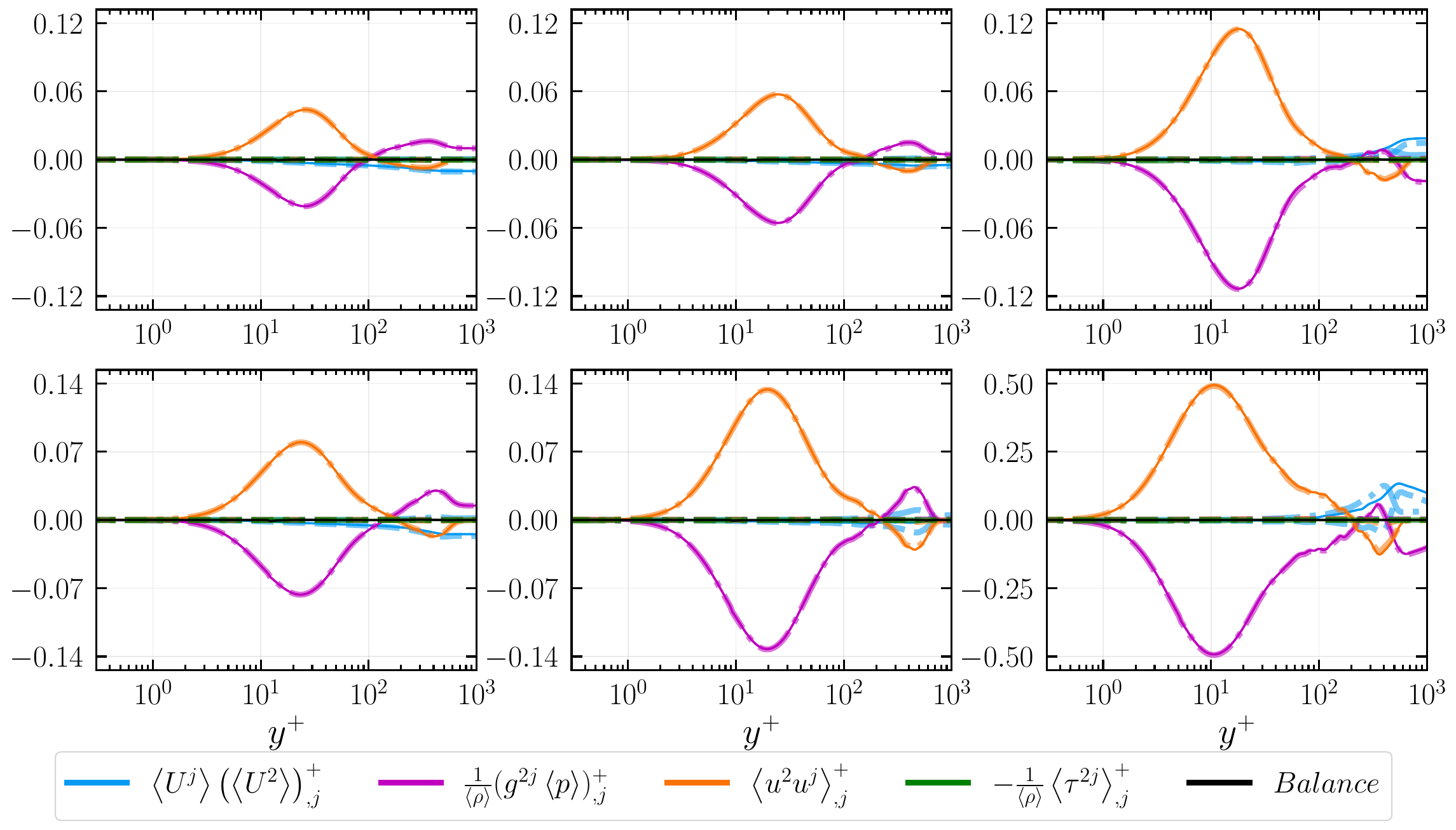}
	\put(3,57.5){(\textit{a})}
        \put(35.8,57.5){(\textit{b})}
        \put(68.55,57.5){(\textit{c})}
        \put(3,32.75){(\textit{d})}
        \put(35.8,32.75){(\textit{e})}
        \put(68.55,32.75){(\textit{f})}
  \end{overpic}        
\caption{Individual terms of the wall-normal mean momentum equation at chordwise positions (\textit{a,d}) $x_c = 0.5$, (\textit{b,e}) $0.7$, and (\textit{c,f}) $0.9$. Solid lines denote the contracted terms in equation \ref{eq:momentum_yn}, while dashed and dot-dashed lines correspond to contributions from components $j=1$ and $j=2$, respectively. Top and bottom rows show results for angles of attack $9$ and $12$ deg, respectively. All terms are non-dimensionalized by $u_\tau^3/\nu$.}
\label{fig:momentum_n}
\end{figure}

Prior to computing the balances, all terms in the equations are expressed in physical components. This procedure involves the use of  metric tensors to compute unitary covariant basis vectors \citep{aris1989}, as shown in the Appendix \ref{appA}.
The individual terms of the mean tangential momentum equation (\ref{eq:momentum_xt}) and mean wall-normal momentum equation (\ref{eq:momentum_yn}) are presented in figures \ref{fig:momentum_t} and \ref{fig:momentum_n}, respectively. In these figures, solid lines denote the contracted terms, while the dashed and dot-dashed lines correspond to quantities with $j = 1$ and $j = 2$, respectively. The top row shows the results for the $9$ deg. simulation, while the bottom row depicts those for the $12$ deg. case. The analysis is performed for chordwise positions $x_c = 0.5$, $0.7$ and $0.9$. All terms are expressed in non-dimensional form using the inner scaling $u_\tau^3/\nu$, where $u_\tau$ is the friction velocity and $\nu$ is the kinematic viscosity. For the 9 deg. case, a constant y-axis scale is used across different chordwise positions to facilitate direct comparison of the balance terms as the APG increases. For the 12 deg. case, the y-axis scale is modified at $x_c = 0.9$ to improve visualization, since most terms appearing in the balance have considerably higher values due to the inner scaling \citep{maciel2018}.

Analysis of figure \ref{fig:momentum_t} indicates that the pressure-gradient component $\left<\rho\right>^{-1}(g^{1j}\left<p\right>)_{,j}$ remains nearly constant in the wall-normal direction across both simulations and at all chordwise locations, consistent with boundary-layer theory. This pressure gradient is largely dominated by the streamwise contribution $\left<\rho\right>^{-1}(g^{11}\left<p\right>)_{,1}$, which reflects the pressure variation along the airfoil surface. At the wall, a balance is established between pressure and viscous forces. Within the buffer layer, the viscous term $\left<\tau^{1j} \right>_{,j}$, together with the pressure gradient, is counteracted by the gradient of the Reynolds stresses $\left<u^1 u^j\right>_{,j}$. As expected, both the viscous and Reynolds stresses are primarily constituted by their wall-normal components $\left<\tau^{12} \right>_{,2}$ and $\left<u^1 u^2\right>_{,2}$, respectively.
In the buffer layer and logarithmic region, particularly at downstream positions where the APG intensifies, the mean tangential convection term $\left<U^j\right>\left(\left< U^1\right>\right)_{,j}$ becomes increasingly significant in the momentum balance. Its contribution, however, is more pronounced in the outer layer, where it is primarily counterbalanced by the pressure gradient and, to a lesser extent, by the Reynolds-stress term.

The terms involving the divergence of the Reynolds stress tensor characterize the exchange of momentum between the fluctuating and mean velocity flow fields \citep{kitsios2017}. Positive values correspond to momentum transfer from the fluctuations to the mean flow, whereas negative values indicate the opposite. The predominance of the $ \left<u^1 u^2\right>_{,2}$ component highlights the central role of the Reynolds shear stress in the momentum transfer across layers. In the inner layer, the negative peak reflects intense transfer from the mean field to the fluctuations, while in the outer layer a positive peak emerges, indicating the reversed trend. The wall-normal contribution of the mean tangential convection $\left<U^2\right>\left(\left< U^1\right>\right)_{,2}$ becomes increasingly relevant in the outer layer as the APG intensifies.
Nevertheless, the streamwise component $\left<U^1\right>\left(\left< U^1\right>\right)_{,1}$, which exhibits a more pronounced peak, remains the dominant contribution. This behavior reflects the fact that boundary-layer deceleration is primarily driven by the APG, thereby shaping the overall structure of the mean tangential convection. At $x_c = 0.9$ in the $12$ deg. case, a strong APG develops, as shown in figure \ref{fig:momentum_t}(\textit{f}). At this location, the pressure-gradient term overcomes the peak of the viscous term in the buffer layer, differently from the other chordwise positions. Moreover, in the outer layer, the mean tangential convection plays a more prominent role due to both the wall-normal velocity as well as the flow deceleration, which are significantly increased in magnitude compared to all previous cases.

Figure \ref{fig:momentum_n} exhibits the individual terms of the mean momentum equation in the wall-normal direction, expressed as equation \ref{eq:momentum_yn}. It is observed that the Reynolds stress $\left<u^2 u^j\right>_{,j}$ and  pressure gradient $\left<\rho\right>^{-1}(g^{2j}\left<p\right>)_{,j}$ terms dominate the balance, particularly within the inner layer. In this region, these terms display positive and negative peaks, respectively, with the wall-normal components ($j=2$) providing the largest contribution. 
In the outer layer, a small negative peak is observed for the Reynolds stress, while a positive peak arises for the pressure-gradient component. 
The presence of the wall-normal convective term $\left<U^j\right>\left(\left< U^2\right>\right)_{,j}$ is also noticed in the outer layer. For mild APGs, this latter term is negative due to the tangential component $\left<U^1\right>\left(\left< U^2\right>\right)_{,1}$, whereas for the strong APG case shown in figure \ref{fig:momentum_n}(\textit{f}), it becomes positive due to the contributions of both tangential and wall-normal components. A wall-normal convection effect has been observed for turbulent boundary layers under strong APGs \citep{vinuesa2018}. Here, we show that two mechanisms compete: the mean flow convection exchanges energy with $\left<u^2 u^2\right>_{,2}$ and, near the edge of the boundary layer, a mild wall-normal favorable pressure gradient provides an outward acceleration of the mean flow.


The present findings are consistent with those reported by \citet{kitsios2017}, who conducted DNS of turbulent boundary layers developing over a flat plate with farfield boundary conditions set to develop different pressure gradients. Their study examined boundary layers with ZPG as well as with mild ($\beta = 1$) and strong $(\beta = 39)$ APGs, the latter closely resembling the conditions investigated in the current work for 12 deg, near the trailing edge. \citet{kitsios2017} showed that the APG increases the convective terms due the rise in mean velocity gradients, thereby enhancing the production of turbulent fluctuations. They also reported a net transfer of streamwise momentum from the mean flow to the fluctuating field in the inner layer, with the opposite trend in the outer layer, reflected by the divergence of the Reynolds stresses. To clarify the mechanisms governing this momentum transfer, we perform a quadrant analysis to isolate the contributions of ejections ($Q2$), sweeps ($Q4$), and outward and inward interactions ($Q1$ and $Q3$, respectively).
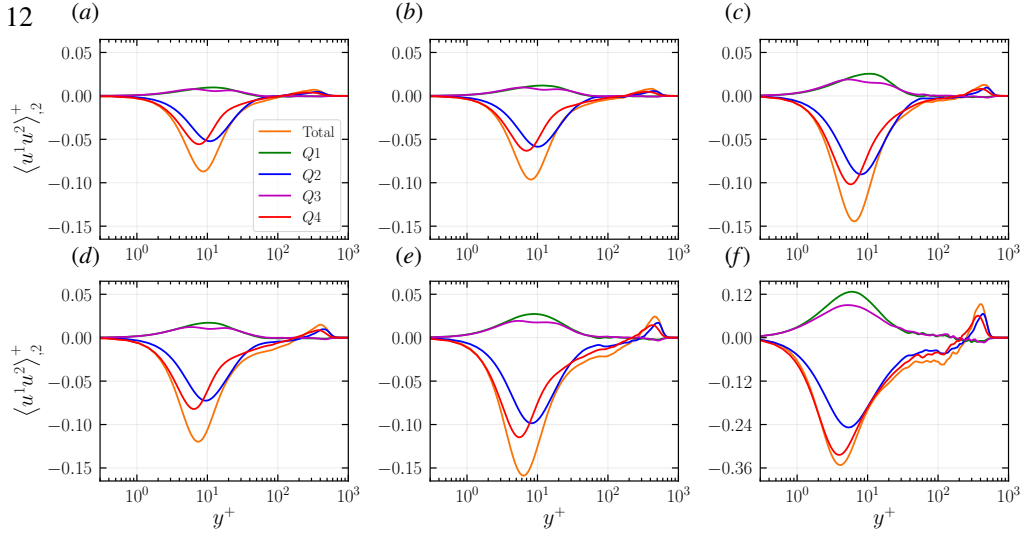
\begin{figure}
 \centering
 \begin{overpic}[width=1\textwidth]{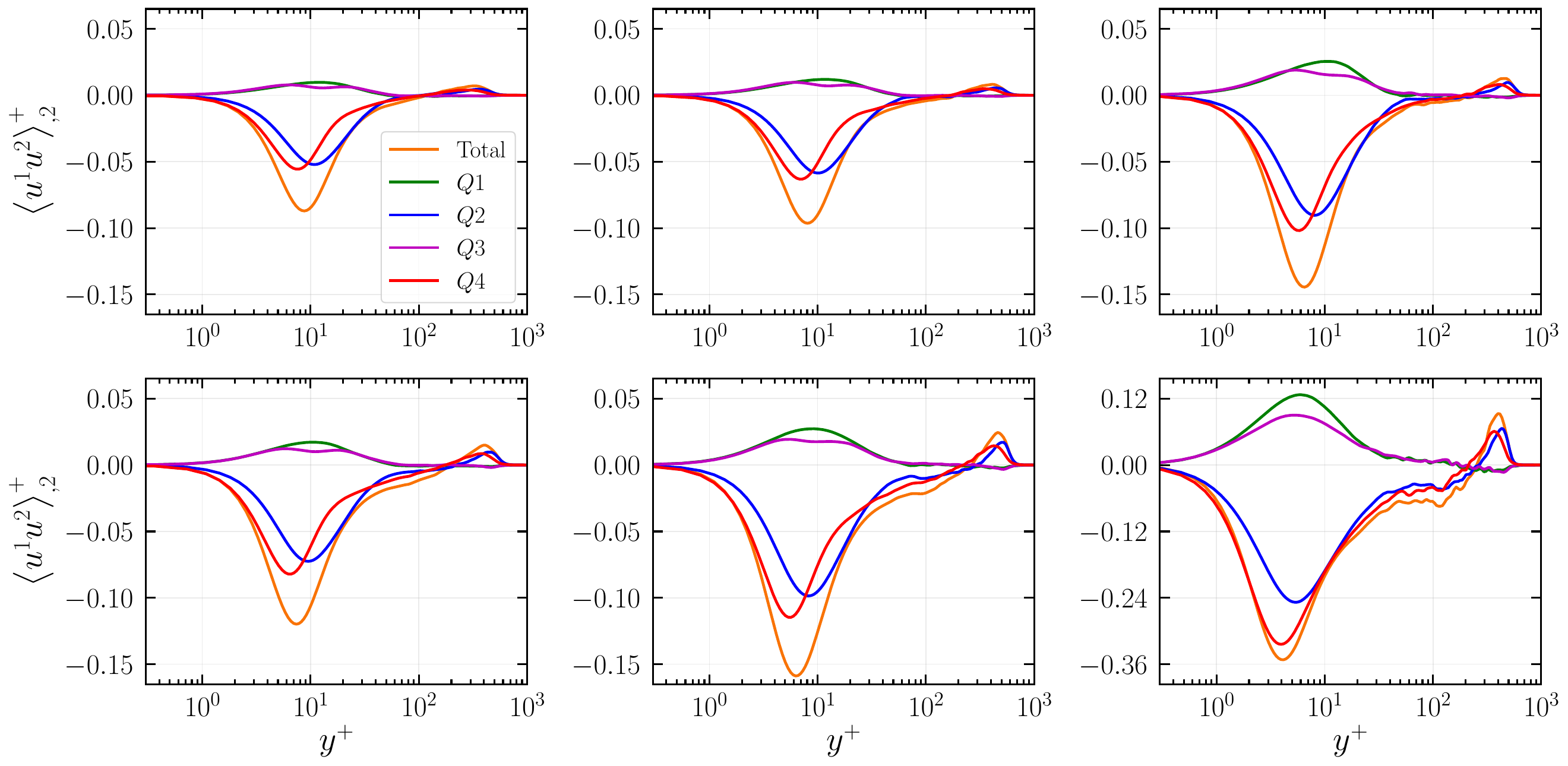}
	\put(6.5,50.25){(\textit{a})}
        \put(38.25,50.25){(\textit{b})}
        \put(70.5,50.25){(\textit{c})}
        \put(6.5,26.5){(\textit{d})}
        \put(38.25,26.5){(\textit{e})}
        \put(70.5,26.5){(\textit{f})}
  \end{overpic}        
\caption{Decomposition of the wall-normal derivative of the Reynolds shear stress into quadrant contributions at chordwise positions (\textit{a,d}) $x_c = 0.5$, (\textit{b,e}) $0.7$, and (\textit{c,f}) $0.9$. The top and bottom rows present results for $9$ and $12$ deg. angle of attack, respectively. All quantities are nondimentionalized by $u_\tau^3/\nu$.}
\label{fig:quadrants}
\end{figure}

As shown in figure \ref{eq:momentum_xt}, the dominant contributor to the divergence of the Reynolds stress tensor in the streamwise momentum balance is the shear term $\left<u^1 u^2\right>_{,2}$. Figure \ref{fig:quadrants} presents a quadrant decomposition of this term, detailing the individual contributions to momentum transfer. All quantities are nondimentionalized using the inner scaling $u_\tau^3/\nu$. The analysis is performed at the same chordwise positions considered previously: the upper panels correspond to an angle of attack of $9$ deg, and the lower panels to $12$ deg. 
Sweeps and ejections emerge as the primary mechanisms for transferring momentum from the mean flow to the fluctuating field, dominating the  total shear stress term. In contrast, inward and outward interactions contribute to momentum transfer in the opposite sense. Sweeps dominate in the viscous sublayer, while ejections prevail in the buffer layer, except for the higher pressure gradient case, where sweeps and ejections exhibit comparable influence within the buffer region. The prominence of sweeps and ejections increases with the APG, with sweep intensities near the wall rising more rapidly than those of ejections. In the outer layer, momentum transfer reverses direction, ocurring from the fluctuating to the mean field, as indicated by the positive peak in the total shear stress term. This process results from the combined action of sweeps and ejections. 
Outward and inward interactions make similar contributions along the boundary layer. They have a minor contribution compared to events in $Q2$ and $Q4$ and become non negligible only in the inner layer. The intensities of these terms also rise with the APG, and outward interactions become more pronounced in the buffer region. 

\section{Extreme wall shear stress events under adverse pressure gradients}
\label{sec:extreme}


In this section, we characterize the occurrence of negative and positive extreme wall shear stress events, and examine their connection with coherent motions within the TBL. 
As noted by \citet{farazmand2017}, extreme events leave detectable signatures that can be identified by statistical methods. We therefore begin with a statistical analysis based on probability density functions (PDFs) to characterize the occurrence of extreme wall shear stress events as a function of the APGs. Next, we compute space-time correlations to examine their spatiotemporal support, followed by conditional statistics to assess the impact of extreme positive and backflow events on the mean velocity and Reynolds shear stress profiles. Finally, we investigate the spatial and temporal evolution of the turbulent structures responsible for generating extreme wall shear stress.


\subsection{Characterization of extreme wall shear stress events}
\label{sec:characterization}

\begin{figure}
 \centering
 \begin{overpic}[width=\textwidth]{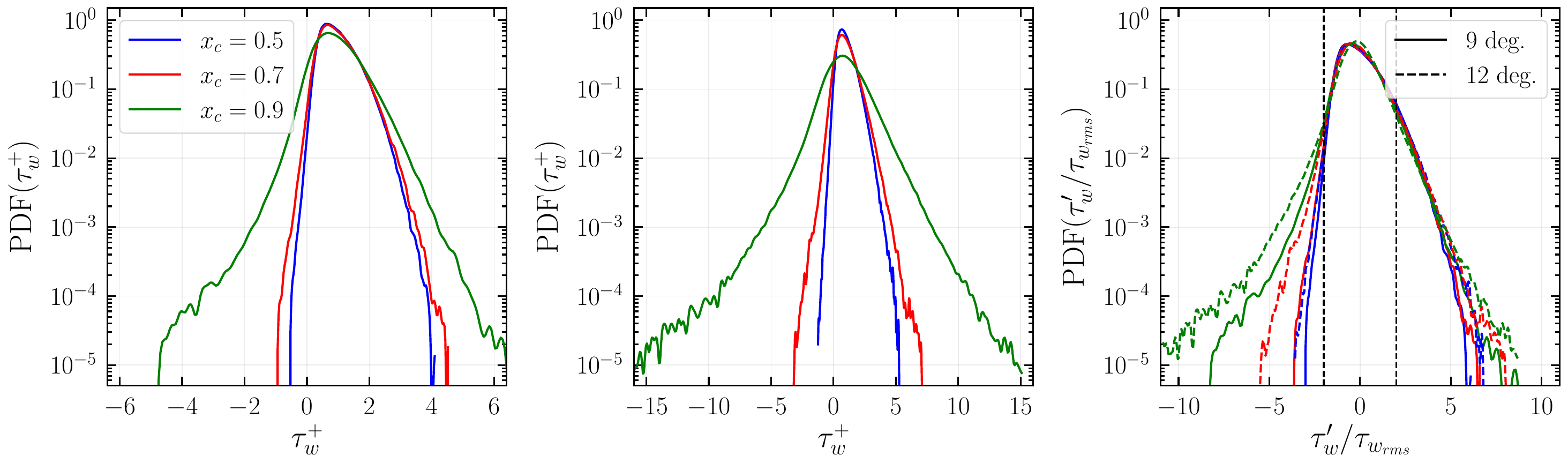}
	\put(3,30){(\textit{a})}
        \put(36.5,30){(\textit{b})}
        \put(70.5,30){(\textit{c})}
  \end{overpic}  
\caption{Probability density functions of the inner-scaled wall shear stress $\tau_w^+$ computed at chordwise positions $x=0.5$, $0.7$ and $0.9$ for (\textit{a}) $9$ and (\textit{b}) $12$ deg. angle of attack, and (\textit{c}) results of the wall shear stress fluctuations normalized by their root-mean-square value $\tau'_w/\tau_{w_{rms}}$. The vertical black dashed lines in (\textit{c}) mark $\tau_w'/\tau_{w_{\text{rms}}} = \pm 2$, which will serve as a threshold for identifying extreme positive and backflow events.}
\label{fig:pdf_tauw}
\end{figure}

Figures \ref{fig:pdf_tauw}(\textit{a}) and (\textit{b}) present PDFs of the inner-scaled wall shear stress, $\tau_w^+$, at chordwise positions $x_c = 0.5$, $0.7$, and $0.9$ for angles of attack $9$ and $12$ deg, respectively. 
It can be observed from these figures 
that the PDFs are more skewed toward positive values of $\tau_w^+$, in agreement with previous findings \citep{lenaers2012,guerrero2020}. This positive skewness is more pronounced at upstream locations, where the APG is weaker. In contrast, the emergence of a negative tail reflects intermittent backflow events, whose occurrence increases with the strength of the APG \citep{vinuesa2017}. Under the strongest APG condition investigated, at $x_c = 0.9$ and $12$ deg, the PDF becomes more symmetric about the mean, as seen in figure \ref{fig:pdf_tauw}(\textit{b}). 

To further characterize the wall shear stress under varying APGs, table \ref{tab:events} reports the mean and root-mean-square (rms) values, $\langle \tau_w \rangle$ and $\tau_{w_{\text{rms}}}$, respectively, at different chordwise positions. The skewness values, $S(\tau_w)$, are also included and corroborate the trends observed in figure \ref{fig:pdf_tauw}. Specifically, skewness decreases with increasing APG, confirming that the PDFs become progressively more symmetric. As also shown in the table, the mean wall shear stress diminishes with stronger APGs, consistent with an increased likelihood of local instantaneous flow separation. Similarly, the rms values also decrease. 
The PDFs of the wall shear stress fluctuations normalized by their respective rms values are presented in figure \ref{fig:pdf_tauw}(\textit{c}). These distributions reveal that larger negative fluctuations become more probable at chordwise positions subjected to stronger APGs, whereas positive fluctuations remain comparable up to $\tau_w'/\tau_{w_{\text{rms}}} \approx 5$. Beyond this value, stronger APGs promote larger positive deviations. The vertical black dashed lines in figure \ref{fig:pdf_tauw}(\textit{c}) mark $\tau_w'/\tau_{w_{\text{rms}}} = \pm 2$, which will serve as a threshold for identifying both extreme positive and backflow events.

\begin{table}
\begin{center}
\def~{\hphantom{0}}
\begin{tabular}{cccccccccccc}
\begin{tabular}[c]{@{}c@{}}AoA \\ (deg.)\end{tabular} &
  $x_c$ & $\beta$ &
  \begin{tabular}[c]{@{}c@{}}$\left< \tau_w \right>$\\ $(\times 10^{-5})$\end{tabular} &
  \begin{tabular}[c]{@{}c@{}}$ \tau_{w_{rms}}$ \\ $(\times 10^{-5})$\end{tabular} &
  $S(\tau_w)$ &
  $\% \mathrm{EP}_{\mathrm{S}}$ &
  $\% \mathrm{BF}_{\mathrm{S}}$ &
  $N_{\mathrm{EP}}$ &
  $N_{\mathrm{BF}}$ &
  $\% \mathrm{EP}_{\mathrm{E}}$ &
  $\% \mathrm{BF}_{\mathrm{E}}$ \\
\multirow{3}{*}{9}  & 0.5 & 1.8 & 7.574 & 3.845 & 0.838 & 4.25 & 0.12 & 5361 & 460  & 0.467 & 0.040 \\
                    & 0.7 & 3.1 & 5.711 & 3.049 & 0.817 & 4.11 & 0.28 & 4764 & 962  & 0.415 & 0.084 \\
                    & 0.9 & 9.4 & 3.382 & 2.350 & 0.563 & 3.67 & 1.07 & 4721 & 3287 & 0.411 & 0.286 \\
\multirow{3}{*}{12} & 0.5 & 5.2 & 5.402 & 3.384 & 0.889 & 4.19 & 0.29 & 9585 & 1892 & 0.416 & 0.082 \\
                    & 0.7 & 12.0 & 3.118 & 2.361 & 0.776 & 3.92 & 0.82 & 7497 & 3672 & 0.285 & 0.160 \\
                    & 0.9 & 46.2 & 1.002 & 1.624 & 0.142 & 3.01 & 1.95 & 5789 & 5245 & 0.251 & 0.228
\end{tabular}
  \caption{Statistics of wall shear stress for different angles of attack and chordwise locations, with corresponding counts and percentages of extreme events. 
  The mean, RMS and skewness of $\tau_w$ are represented by $\left< \tau_w \right>$, $ \tau_{w_{rms}}$, and $S(\tau_w)$, respectively. The percentage of samples above or below the threshold $\tau_w' = \pm 2 \tau_{w_{\text{rms}}}$ is given by $\% \mathrm{EP}_{\mathrm{S}}$ and $\% \mathrm{BF}_{\mathrm{S}}$. The parameters $N_{\mathrm{EP}}$ and $N_{\mathrm{BF}}$ represent the number of extreme positive and backflow realizations tracked in the temporal signals, as indicated by blue and red dots marked in figure \ref{fig:tauw_sig}. The percentage of extreme positive and backflow events is finally indicated by $\% \mathrm{EP}_{\mathrm{E}}$ and $\% \mathrm{BF}_{\mathrm{E}}$, respectively.} 
  \label{tab:events}
\end{center}
\end{table}

To characterize the extreme positive (EP) and backflow (BF) events, we first analyze the PDFs of $\tau_w$, together with the temporal signals obtained at each spanwise grid point for chordwise positions $x_c = 0.5$, $0.7$, and $0.9$ in both simulations. Based on these analyses, a threshold of $\pm 2 \tau_{w_{\text{rms}}}$ was selected to define extreme events. For EP samples ($\mathrm{EP}_{\mathrm{S}}$), this criterion corresponds to fewer than $5\%$ of the total samples. In the case of BF events, the threshold ensures the occurrence of negative wall shear stress even at the weakest APG location investigated, namely $x_c = 0.5$ for $9$ deg. angle of attack. As can be seen in table \ref{tab:events}, the percentage of BF samples ($\mathrm{BF}_{\mathrm{S}}$) is lower than $2\%$. 

\begin{figure}
 \centering
 \begin{overpic}[width=\textwidth]{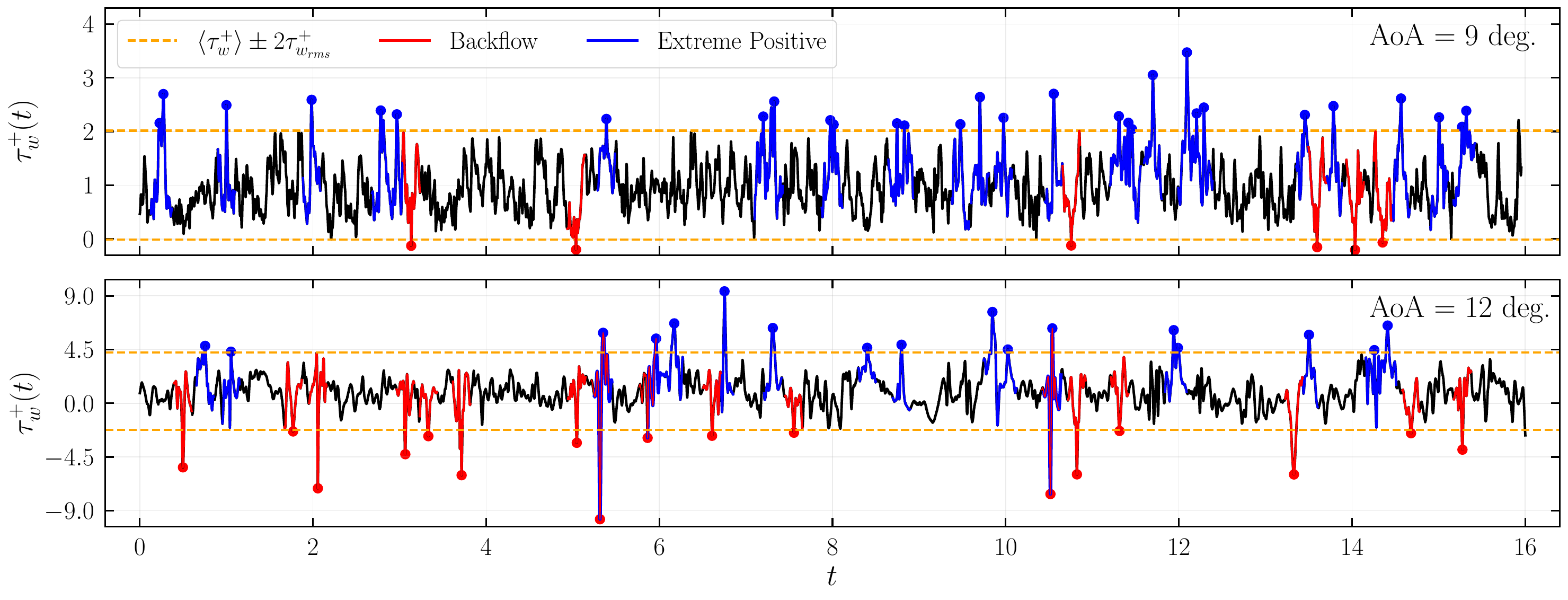}
	\put(2,38){(\textit{a})}
        \put(2,21){(\textit{b})}
  \end{overpic}  
\caption{Examples of inner-scaled wall shear stress signals over the full simulation duration, where (\textit{a}) represents a probe at $x_c = 0.5$ for $9$ deg. angle of attack and (\textit{b}) represents a probe at $x_c = 0.9$ for $12$ deg. angle of attack. Blue and red dots mark extreme positive (EP) and extreme negative (backflow, BF) events, respectively, while the corresponding colored lines represent the temporal windows of length $\Delta t = 0.202$ used to track their evolution.}
\label{fig:tauw_sig}
\end{figure}

To illustrate the temporal variation of the wall shear stress under the weakest and strongest APGs analyzed, figures \ref{fig:tauw_sig}(\textit{a}) and (\textit{b}) show the time signals of $\tau_w^+$ at $x_c = 0.5$ for the $9$ deg. angle of attack case and at $x_c = 0.9$ for the $12$ deg. case, respectively. The results are presented for a single point along the span. In these figures, blue and red dots mark extreme positive and negative (backflow) events. 
It should be noted that each dot in the figure corresponds to a single extreme event ($\mathrm{EP}_{\mathrm{E}}$ or $\mathrm{BF}_{\mathrm{E}}$) in the statistics reported in table \ref{tab:events}.
The associated colored lines delimit the temporal windows selected for each event, their length chosen to capture the full temporal evolution. Each line segment spans the dimensionless time interval $\Delta t = 0.202$, which was found to be sufficient to capture the complete lifespan of the events, from their onset to their decay. Although extreme events are indicated by the circles in figure \ref{fig:tauw_sig}, it is important to note that multiple samples of $\tau_w^+$ exceed the threshold of $\tau_w' = \pm 2 \tau_{w_{\text{rms}}}$. The proportion of such samples is reported in table \ref{tab:events} as $\% \mathrm{EP}_{\mathrm{S}}$ and $\% \mathrm{BF}_{\mathrm{S}}$, corresponding to the fractions of extreme positive and backflow samples, respectively.

After identifying the extreme events along the spanwise direction at each chordwise position, we quantify both their absolute number of occurrences and relative percentage with respect to the total number of samples collected during the simulation. The results are summarized in table \ref{tab:events} for both simulations at the chordwise positions under consideration. The percentage of BF samples increases with the APG, whereas the opposite trend is observed for EP samples. The highest percentage of BF samples occurs at $x_c = 0.9$ in the $12$ deg. case. A similar trend is found for the number of BF events ($N_{\mathrm{BF}}$) and EP events ($N_{\mathrm{EP}}$). However, the percentage of BF events ($\% \mathrm{BF}_{\mathrm{E}}$) at $x_c = 0.9$ in the $12$ deg. simulation is lower than that at the same chordwise position in the $9$ deg. case, which can be related to different factors. The first concerns the deceleration of the TBL induced by the stronger APG, which substantially reduces the mean wall shear stress $\left< \tau_w \right>$, as shown in table \ref{tab:events}. This deceleration weakens the near-wall turbulence activity \citep{skare1994,maciel2018}, which in turn results in smaller wall shear stress fluctuations, as reflected by the rms values $\tau_{w_{rms}}$ also reported in table \ref{tab:events}.

\begin{figure}
 \centering
 \begin{overpic}[width=\textwidth]{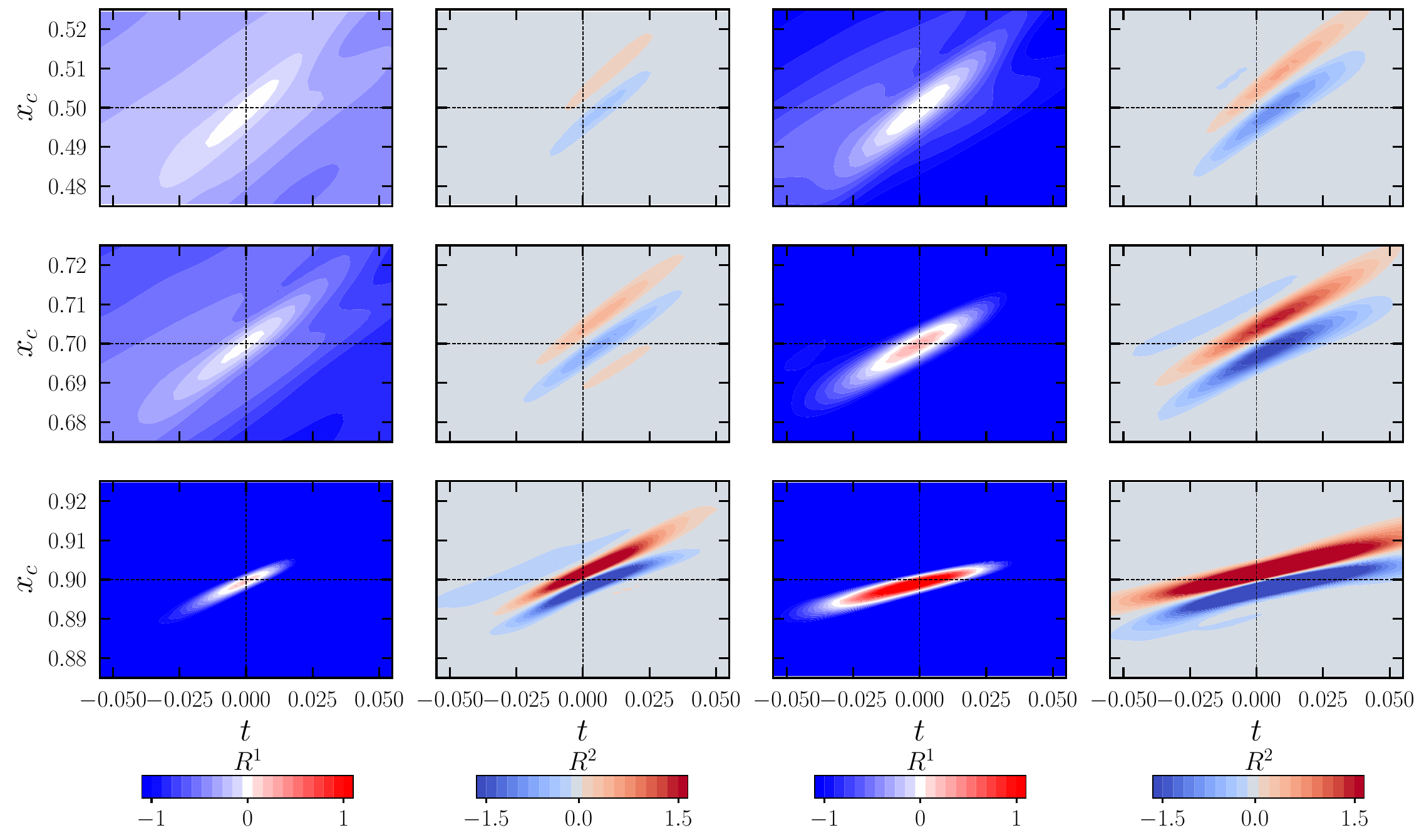}
	\put(3.5,58){(\textit{a})}
        \put(28,58){(\textit{b})}
        \put(51.5,58){(\textit{c})}
        \put(75,58){(\textit{d})}
        \put(3.5,41.5){(\textit{e})}
        \put(28,41.5){(\textit{f})}
        \put(51.5,41.5){(\textit{g})}
        \put(75,41.5){(\textit{h})}
        \put(3.5,25){(\textit{i})}
        \put(28,25){(\textit{j})}
        \put(51.5,25){(\textit{k})}
        \put(75.5,25){(\textit{l})}
  \end{overpic}  
\caption{Space-time correlations of backflow (BF) events computed with the streamwise velocity component $U^1 = U_t$ displayed in the first and third columns, and with the wall-normal velocity component $U^2=U_n$ depicted in the second and fourth columns. Results for the $9$ deg. angle of attack case are presented in the first two columns, while those for the $12$ deg. case are presented in the last two columns. The correlations shown in the first, second and third rows represent results at chordwise positions $x_c = 0.5$, $0.7$, and $0.9$, respectively.}
\label{fig:corr_bf}
\end{figure}

In addition, we observe that BF events exhibit longer persistence at locations subject to stronger APGs. Consequently, although the percentage of backflow events identified at $x_c = 0.9$ is lower in the $12$ deg. case compared to the $9$ deg. case, the absolute number of samples exceeding the threshold parameter is larger. This finding is further supported by the space-time correlation analysis between the instantaneous wall shear stress $\tau_w$ and the instantaneous velocity components $U^i$, both computed considering the threshold defining extreme events, described by:
\begin{equation}
  R^i(x_c, r, \Delta t_0) = \frac{\left< \tau_w(x_c, t_0) U^i(x_c + r, t_0 + \Delta t_0)\right>}{\tau_{w_{rms}}(x_c, t_0)U^i_{rms}(x_c + r, t_0 + \Delta t_0)} \mbox{ .}
  \label{eq:correlation}
\end{equation}
Here, $r$ denotes the spatial offset defined to vary within $-0.025 \leq r \leq 0.025$, $x_c$ is the chordwise position under analysis, $t_0$ is the time instant of the selected event, and $\Delta t_0$ is the time window centered on the event. Figure \ref{fig:corr_bf} presents the space-time correlations for the BF events. The first and second columns correspond to the $9$ deg. angle of attack case, whereas the third and fourth columns show the results for the $12$ deg. case. The first, second and third lines display results at the chordwise positions $x_c = 0.5$, $0.7$, and $0.9$, respectively. 

The correlations $R^1$ and $R^2$ are computed for the instantaneous streamwise velocity component $U^1 = U_t$ (first and third columns), and the wall-normal velocity component $U^2 = U_n$ (second and fourth columns), respectively. These components are sampled at a wall-normal position corresponding to $y^+ = 2.5$. It is important to note that the correlations are performed using the instantaneous, rather than the fluctuating, components of the variables. As a result, the magnitude of the correlation $|R^i(x_c, r, \Delta t)|$ may exceed unity. The slope of the correlation reflects the mean convection velocity of the events. As the APG increases, the associated timescales become longer, consistent with the flow deceleration. Furthermore, the correlation amplitudes increase with the APG, which can be attributed to the reduction in rms values.

%

To investigate the dynamics of BF events, it is necessary to examine the tangential ($U_t$) and wall-normal ($U_n$) velocity components simultaneously. For a broader perspective, attention is focused on the strongest APG condition, at $x_c = 0.9$ in the $12$ deg. angle of attack case, which corresponds to the most pronounced BF events. The corresponding results are shown in figures \ref{fig:corr_bf}(\textit{k}) and (\textit{l}). In these calculations, the wall shear stress $\tau_w$ is fixed; during BF events, $\tau_w$ assumes negative values. Following the horizontal dashed line at $x_c = 0.9$ in figure \ref{fig:corr_bf}(\textit{k}), the negative correlation indicates that $U_t$ is positive in the streamwise direction, consistent with $\tau_w < 0$. At $t \approx -0.025$, a positive correlation emerges in $U_n$ as shown in figure \ref{fig:corr_bf}(\textit{l}), corresponding to a velocity directed towards the wall. The combined action of these two components produces a sweep motion that drives fluid into the near-wall region. 

As time advances, the correlation $R^1$ increases, reflecting a reversal in its orientation and signaling the onset of backflow. 
At $t = 0.0$, the correlation $R^1$ reaches its maximum, while $R^2$ vanishes, indicating motion in the opposite direction of the flow. Subsequently, the correlation $R^2$ becomes negative, implying motion away from the wall, while the correlation $R^1$ remains positive.
This combination reveals that the backflow event is followed by an ejection. With further time progression, the correlation $R^1$ decreases and becomes negative, marking the reorientation of tangential velocity back toward the streamwise direction, while $R^2$ remains negative, sustaining an outward motion. A similar inspection for the other cases at different chordwise positions under different APG strengths demonstrates that the same fundamental mechanism persists. The primary differences lie in the characteristic timescales, which increase with stronger APGs.

\begin{figure}
 \centering
 \begin{overpic}[width=\textwidth]{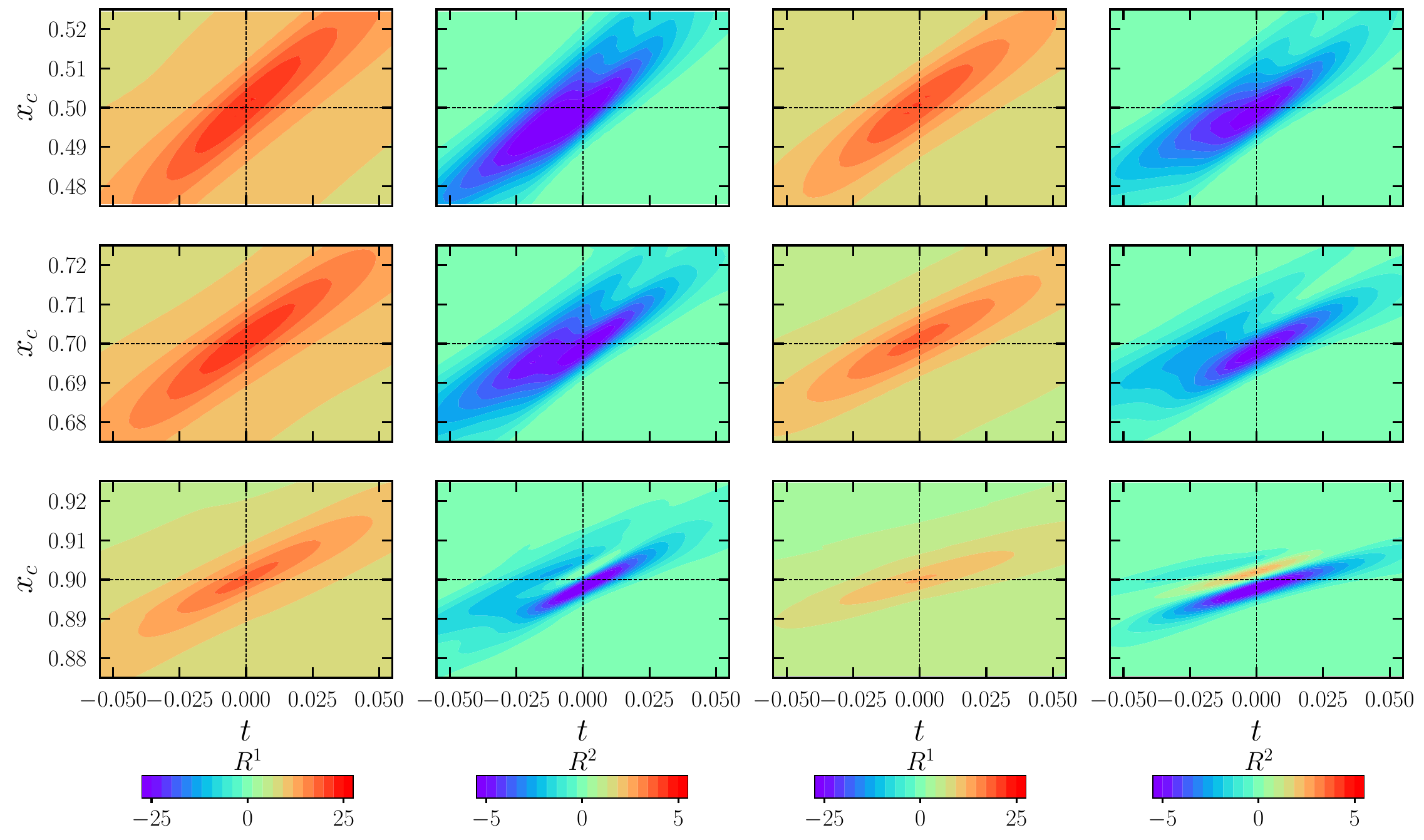}
	\put(3.5,58){(\textit{a})}
        \put(28,58){(\textit{b})}
        \put(51.5,58){(\textit{c})}
        \put(75,58){(\textit{d})}
        \put(3.5,41.5){(\textit{e})}
        \put(28,41.5){(\textit{f})}
        \put(51.5,41.5){(\textit{g})}
        \put(75,41.5){(\textit{h})}
        \put(3.5,25){(\textit{i})}
        \put(28,25){(\textit{j})}
        \put(51.5,25){(\textit{k})}
        \put(75.5,25){(\textit{l})}
  \end{overpic}  
\caption{Space-time correlations of extreme positive (EP) events computed with the streamwise velocity component ($U^1 = U_t$) displayed in the first and third columns, and with the wall-normal velocity component ($U^2=U_n$) depicted in the second and fourth columns. Results for the $9$ deg. angle of attack case are presented in the first two columns, while those for the $12$ deg. case are presented in the last two columns. The correlations shown in the first, second and third rows represent results at chordwise positions $x_c = 0.5$, $0.7$, and $0.9$, respectively.}
\label{fig:corr_ep}
\end{figure}

The same analysis is carried out for the EP case, with the results presented in figure \ref{fig:corr_ep}, following the same format as in the previous figure. In EP events, the wall shear stress $\tau_w$ is positive. For a more comprehensive view, attention is directed to  position $x_c = 0.5$ of the $9$ deg. angle of attack case, shown in figures \ref{fig:corr_ep}(\textit{a}) and (\textit{b}). At this location, the APG is weaker, making EP events more frequent (see table \ref{tab:events}). Following the dashed horizontal line at $x_c = 0.5$, the tangential velocity $U_t$ exhibits a positive correlation with $\tau_w$, indicating that the flow is oriented in the streamwise direction. As the time progresses, this correlation increases, reflecting acceleration of $U_t$. At the same time, the correlation $R^2$ decreases and becomes negative, corresponding to a velocity directed toward the wall. The combined behavior of these two components characterizes a sweep motion originating in the near-wall region. The correlation $R^1$ continues to grow, reaching its maximum at $t = 0.0$, after which it declines, marking the weakening of the event. This onset mechanism of EP events is observed consistently across the other cases analyzed, as shown in figure \ref{fig:corr_ep}, although the APG reduces both their spatial and temporal extent. Moreover, at $x_c = 0.9$ in both simulations, the correlations $R^2$ show small positive regions preceding the event (figures \ref{fig:corr_ep}(\textit{j}) and (\textit{l})). This suggests the presence of a weak outward interaction that anticipates the sweep motion. 

\subsection{Conditional statistics of extreme events}
\label{sec:conditional}

This section examines the temporal evolution of the wall-tangential $U^1$ and wall-normal $U^2$ velocity profiles, together with the Reynolds shear stress $\langle u^1 u^2 \rangle$, during extreme positive and backflow events. The profiles are obtained through conditional averaging of the events using the characteristic time scales extracted from the space-time correlations presented in figures \ref{fig:corr_bf} and \ref{fig:corr_ep}. Results are reported in nondimensional form based on inner scaling.
The conditional averaging procedure is defined as
\begin{align}
  \widetilde{f}_{EP}(x_c,y_n,t_0 \pm \Delta t_0) &= \left< f(x_c,y_n,\Delta z_{EP},t_0 + \Delta t_0) \quad | \quad \tau_w' \geq 2\tau_{w_{rms}} \right>  \mbox{ , and}
  \label{eq:conditional_ep} \\
  \widetilde{f}_{BF}(x_c,y_n,t_0 \pm \Delta t_0) &= \left< f(x_c,y_n,\Delta z_{BF},t_0 + \Delta t_0) \quad | \quad \tau_w' \leq -2\tau_{w_{rms}} \right>  \mbox{ ,}
  \label{eq:conditional_bf}
\end{align}
for extreme positive and backflow events, respectively. Here, $\widetilde{(\cdot)}$ denotes the conditional mean of the variable $f$ at a fixed chordwise position $x_c$, calculated along the wall-normal direction $y_n$. The averaging is carried out in the spanwise direction over the extent of the extreme event, $\Delta z$. To capture temporal evolution, the conditional mean is computed as an ensemble average at different time lags $\Delta t_0$ relative to the reference event time $t_0$.
The positions $x_c = 0.5$ for the 9 deg. angle of attack and $x_c = 0.9$ of the 12 deg. case are selected to represent the results of the analysis, as they correspond to the weakest and strongest APG effects, respectively (see table \ref{tab:events}). Inspection of other chordwise positions (presented as supplementary material) revealed qualitatively similar behavior for both EP and BF events.

\begin{figure}
 \centering
 \begin{overpic}[width=\textwidth]{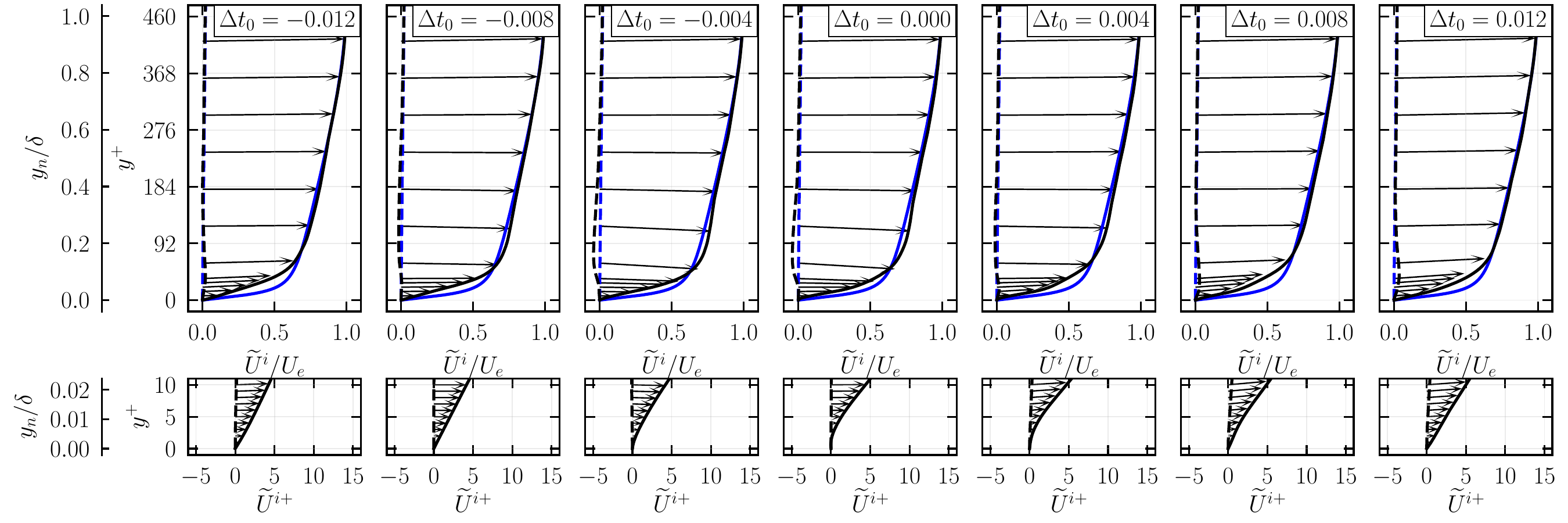}
	\put(3.5,34){(\textit{a})}
  \end{overpic} \\[2mm]
  \begin{overpic}[width=\textwidth]{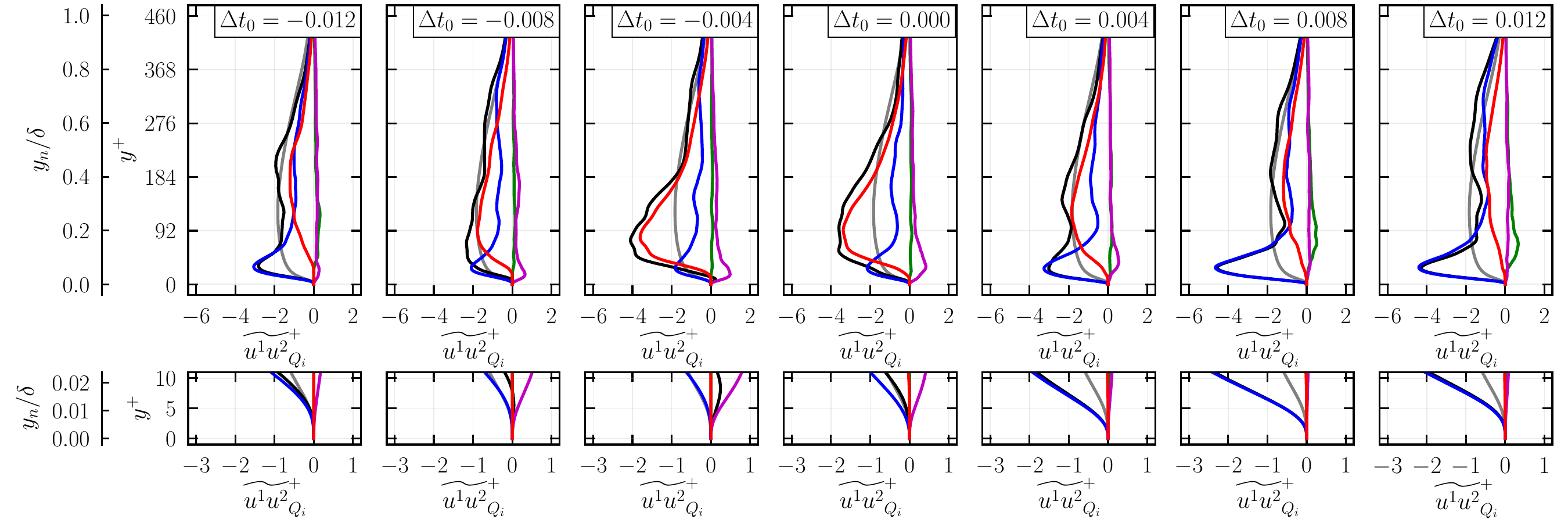}
	\put(3.5,34){(\textit{b})}
  \end{overpic} \\[2mm] 
  \begin{overpic}[width=\textwidth]{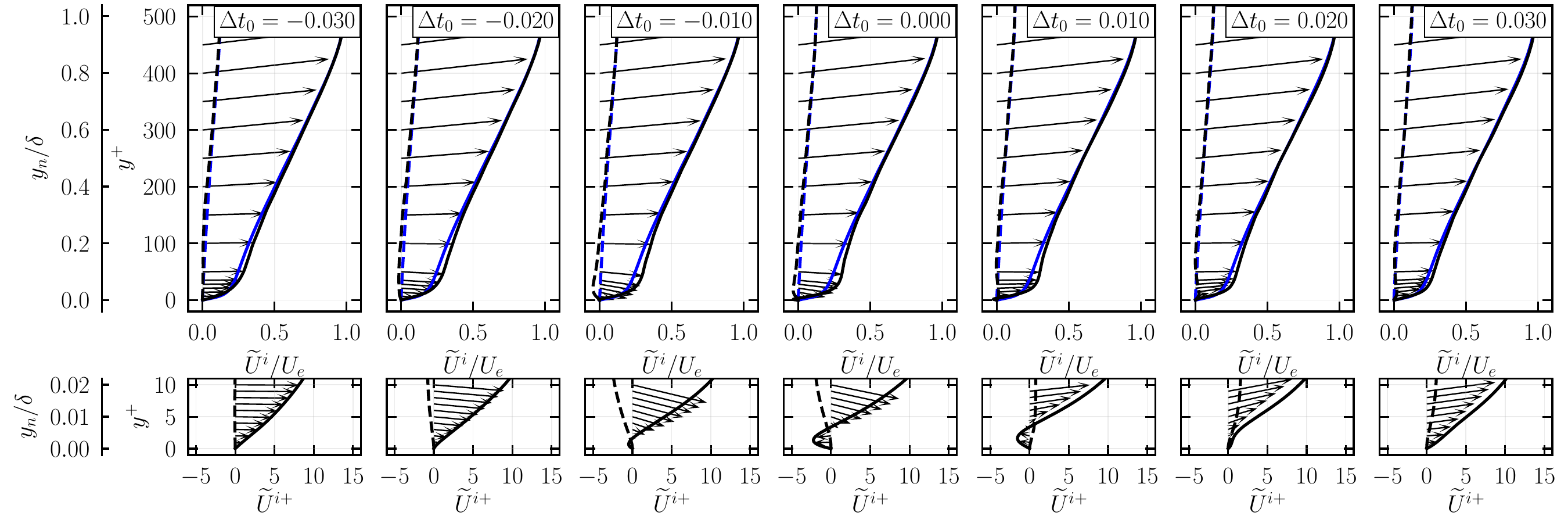}
	\put(3.5,34){(\textit{c})}
  \end{overpic} \\[2mm] 
  \begin{overpic}[width=\textwidth]{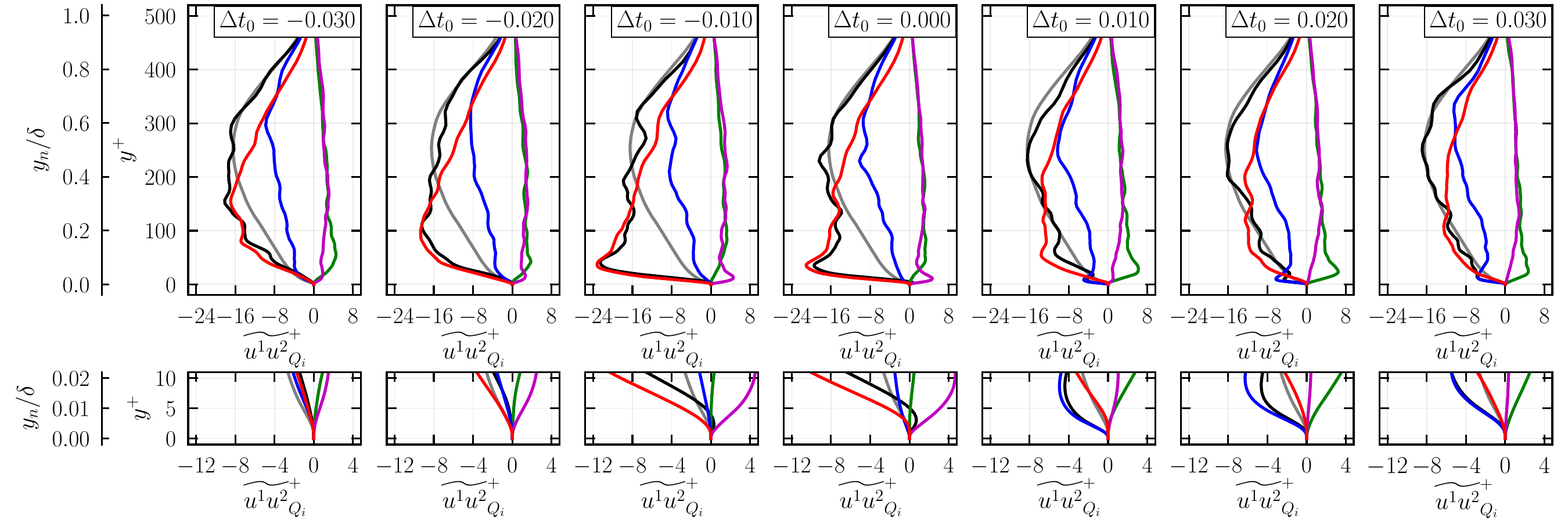}
	\put(3.5,34){(\textit{d})}
  \end{overpic}  
\caption{Temporal evolution of conditional statistics for backflow (BF) events. (\textit{a,c}) Conditional mean velocity profiles of the wall-tangential ($\widetilde{U}^1$, solid black line) and wall-normal ($\widetilde{U}^2$, dashed black line) components, compared with the corresponding total mean profiles (blue solid and dashed lines). (\textit{b,d}) Conditional Reynolds shear-stress profiles ($\widetilde{u^1u^2}$, black), together with the quadrant decomposition contributions: outward interactions ($Q1$, green), ejections ($Q2$, blue), inward interactions ($Q3$, magenta), and sweeps ($Q4$, red). The total mean Reynolds shear stress is shown in gray. Results in figures (\textit{a}, \textit{b}) correspond to the chordwise position $x_c = 0.5$ of the 9 deg. angle of attack case, while figures (\textit{c}, \textit{d}) correspond to $x_c = 0.9$ of the $12$ deg. case.}
\label{fig:bf_mean}
\end{figure}

Figures \ref{fig:bf_mean}(\textit{a}) and (\textit{c}) show the evolution of the wall-tangential and wall-normal conditional velocity profiles for BF events, plotted as solid and dashed black lines, respectively. For reference, the mean flow velocity profiles, obtained through temporal and spanwise averaging of the entire dataset, are shown in blue. This comparison highlights the differences between the conditional profiles of the extreme events and the mean flow field. In the top panels, the velocity profiles are normalized by the velocity at the edge of the boundary layer, $U_e$, while the bottom panels provide a detailed view of the near-wall region normalized by the friction velocity, $u_\tau$. Together, these figures capture the conditional mean velocity dynamics during BF events across both the inner and outer layers.
In a similar manner, figures \ref{fig:bf_mean}(\textit{b}) and (\textit{d}) present the evolution of the conditional Reynolds shear stress during BF events, shown by black lines. For reference, the mean Reynolds shear stress profile, obtained from the full dataset, is depicted by the gray line. To further characterize the dynamics, a quadrant splitting analysis  \citep{wallace1972,willmarth1972} is applied to the conditional averages, enabling the identification of the relative contributions of sweeps ($Q4$), ejections($Q2$), inward ($Q3$) and outward ($Q1$) interactions throughout the BF event evolution. 

It can be observed from figure \ref{fig:bf_mean}(\textit{a}) that, for the $9$ deg. angle of attack case at chordwise position $x_c = 0.5$, the backflow event begins with the streamwise velocity component $\widetilde{U}^1$ being lower than the local mean in the inner layer.
Additionally, a small positive value of the wall-normal component $\widetilde{U}^2$ is present, indicating a slight ejection. This interpretation is supported by figure \ref{fig:bf_mean}(\textit{b}), which shows that the Reynolds shear stress $\widetilde{u^1u^2}$ in the inner layer is initially dominated by ejections, although their intensity diminishes over time. At the time lag $\Delta t_0 = -0.008$, figure \ref{fig:bf_mean}(\textit{b}) indicates the onset of a sweep around $y^+ \approx 90$, characterized by a small negative $\widetilde{U}^2$ component. This implies a downward transport of fluid toward the wall, as also indicated by the directional arrows in figure \ref{fig:bf_mean}(\textit{a}). Notably, an inward interaction $Q3$ event becomes apparent in the near-wall region, as shown in the lower panel of figure \ref{fig:bf_mean}(\textit{b}). 
This suggests that the extreme BF event is driven by a combination of downward motion and flow acceleration, as reflected by the increasing tangential component $\widetilde{U}^1$, although it remains below the total mean near the wall. 
This mechanism intensifies up to the moment of backflow occurrence at $\Delta t_0 = 0.000$. Following this point, an ejection emerges in the near-wall region, surpassing the sweep, which eventually weakens. Simultaneously, a small outward interaction develops, positioned above the ejection peak. The wall-normal velocity component $\widetilde{U}^2$, which was negative before the event, progressively transitions to small positive values in the inner layer, indicating a reversal in the direction of vertical transport. 

The results for the $12$ deg. angle of attack case at chordwise position $x_c = 0.9$ are shown in figures \ref{fig:bf_mean}(\textit{c}) and (\textit{d}). Under this strong APG condition, the backflow event enhances the conditional streamwise velocity $\widetilde{U}^1$ near the wall, as can be seen from the comparison between the solid black and blue lines in the top panels of figure \ref{fig:bf_mean}(\textit{c}). This increase arises from the sweep motion across the boundary layer, which dominates the Reynolds shear stress profiles in figure \ref{fig:bf_mean}(\textit{d}) as the BF event develops.
For negative time lags, as the extreme event intensifies, sweeps become increasingly dominant near the wall, while inward interactions also strengthen within the viscous sublayer. Their influence is reflected in the wall-normal velocity component $\widetilde{U}^2$, which becomes progressively more negative in the inner layer, and in the tangential component $\widetilde{U}^1$ in the viscous sublayer (bottom panels of figure \ref{fig:bf_mean}(\textit{c})). This trend culminates at the onset of backflow ($\Delta t_0 = 0.000$), when $\widetilde{U}^1$ reaches its most negative value at $y^+ \approx 2.5$, leading to a more intense BF event compared to the lower APG case previously analyzed. After the event, the Reynolds shear stress near the wall is governed mainly by ejections and outward interactions, replacing the earlier combination of sweeps and inward interactions. This transition indicates that, for positive time lags, the flow moves away from the wall. Figure \ref{fig:bf_mean}(\textit{c}) confirms this outward motion where $\widetilde{U}^2$ presents positive values in the inner layer, and $\widetilde{U}^1$ exceeds the total mean profile. The outward transport is further emphasized by the arrows in figure \ref{fig:bf_mean}(\textit{c}), which point away from the wall. Subsequently, the flow gradually returns to the mean state.

From a broad perspective encompassing both APG cases, a BF event is initiated by a sweep motion originating in the inner layer, which transports fluid toward the wall. This motion is accompanied by inward interactions very close to the wall, as shown in the bottom panels of figures \ref{fig:bf_mean}(\textit{b}) and (\textit{d}), preceding the onset of backflow. The backflow itself produces a local deceleration in the viscous sublayer, visible in the bottom panels of figures \ref{fig:bf_mean}(\textit{a}) and (\textit{c}) at $y^+<5$, until the extreme event reaches full development. Subsequently, the wall-directed interactions are replaced by ejections and outward motions for $y^+<10$, driving fluid away from the wall. In both cases, the largest departures from the total mean profiles are confined to the inner layer, indicating that the mechanisms governing BF events, as well as their impact on the TBL, are localized within this region, consistent with the observations of \citet{hutchins2011}. 
The main difference between the two APG conditions lies in the Reynolds shear stress quadrant dynamics, aside from the distinct intensities of the tangential velocity during the event. In the 9 deg. case, a strong ejection arises at the onset of the BF, briefly exceeding the sweep contribution. During the event, the sweeps dominate the Reynolds shear stress in the inner layer, but are later overtaken by ejections, which surpass sweeps by a factor of four (top panels of figure \ref{fig:bf_mean}(\textit{b})). In contrast, for the 12 deg. case, sweeps remain dominant throughout the entire BF event along the inner layer. Only after the event do ejections become slightly more pronounced, but this effect is confined to the region very close to the wall.


\begin{figure}
 \centering
 \begin{overpic}[width=\textwidth]{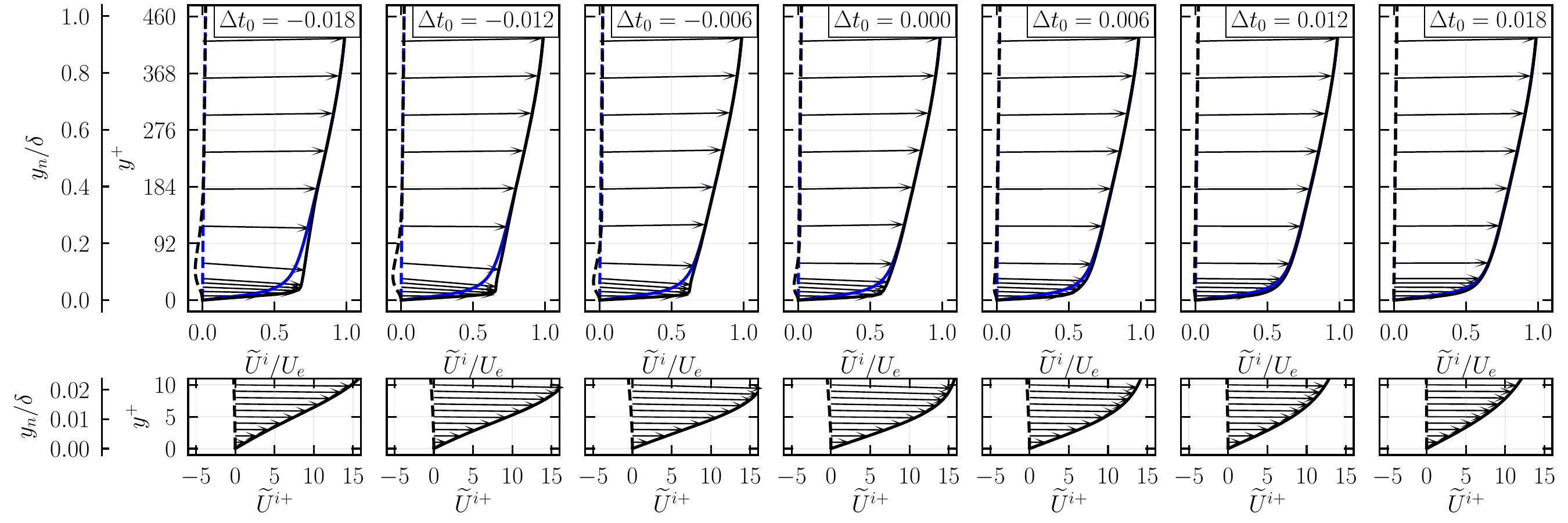}
        \put(3.5,34){(\textit{a})}
  \end{overpic} \\[2mm]
  \begin{overpic}[width=\textwidth]{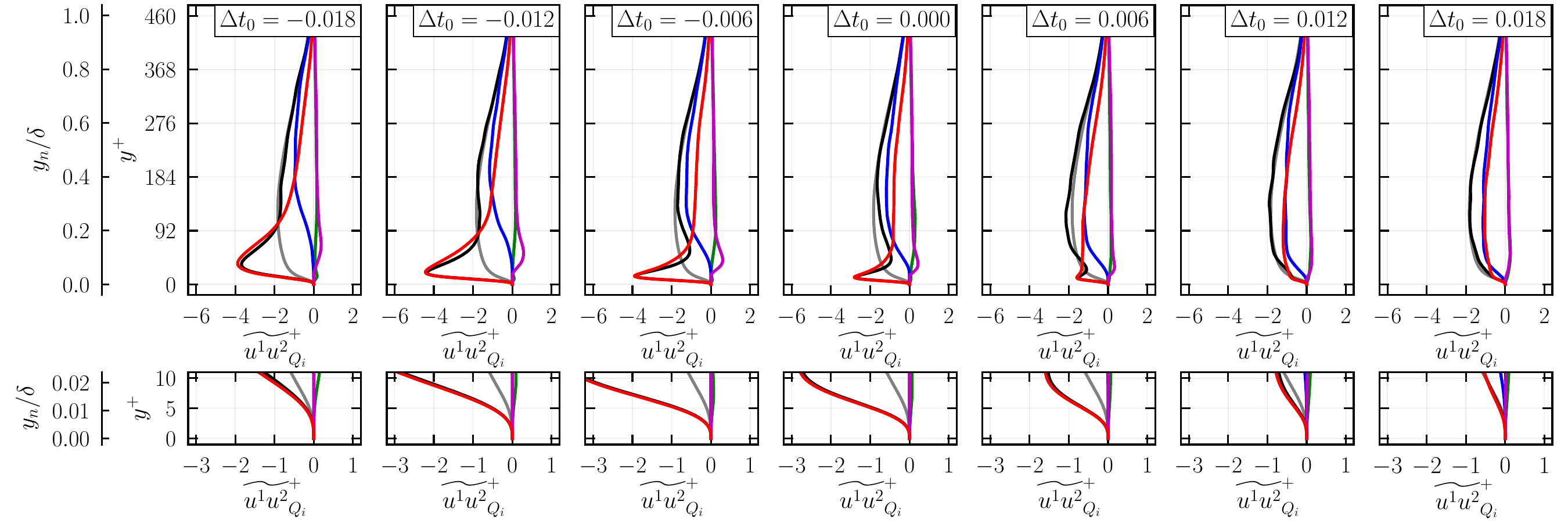}
	\put(3.5,34){(\textit{b})}
  \end{overpic} \\[2mm]
  \begin{overpic}[width=\textwidth]{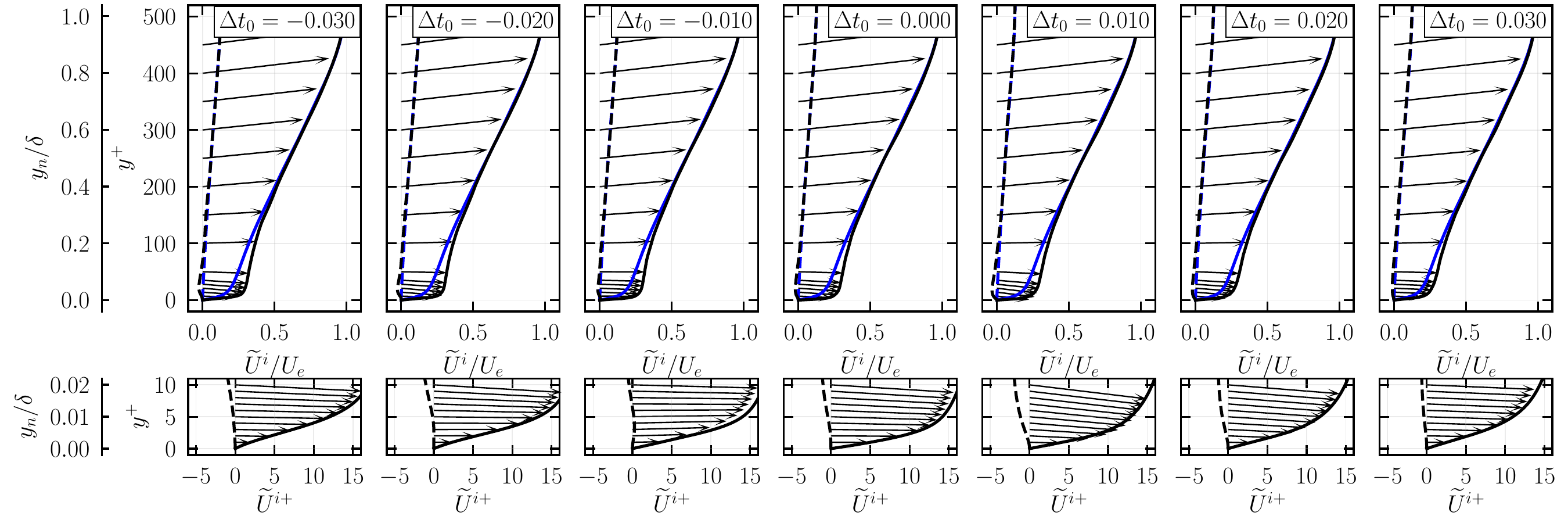}
	\put(3.5,34){(\textit{c})}
  \end{overpic} \\[2mm] 
  \begin{overpic}[width=\textwidth]{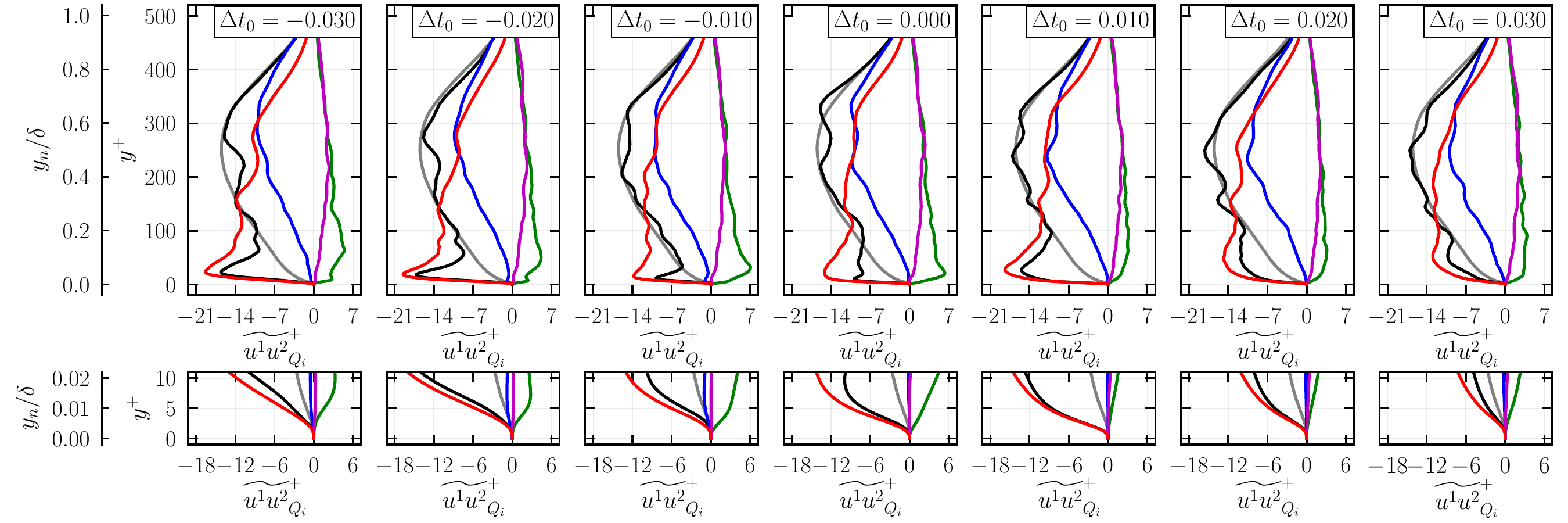}
	\put(3.5,34){(\textit{d})}
  \end{overpic}  
\caption{Temporal evolution of conditional statistics for extreme positive (EP) events. (\textit{a,c}) Conditional mean velocity profiles of the wall-tangential ($\widetilde{U}^1$, solid black line) and wall-normal ($\widetilde{U}^2$, dashed black line) components, compared with the corresponding total mean profiles (blue solid and dashed lines). (\textit{b,d}) Conditional Reynolds shear-stress profiles ($\widetilde{u^1u^2}$, black), together with the quadrant decomposition contributions: outward interactions ($Q1$, green), ejections ($Q2$, blue), inward interactions ($Q3$, magenta), and sweeps ($Q4$, red). The total mean Reynolds shear stress is shown in gray. Results in figures (\textit{a}, \textit{b}) correspond to the chordwise position $x_c = 0.5$ of the 9 deg. angle of attack case, while figures (\textit{c}, \textit{d}) correspond to $x_c = 0.9$ of the $12$ deg. case.}
\label{fig:ep_mean}
\end{figure}

The statistics of extreme positive wall shear stress events are also examined in relation to the temporal evolution of velocity profiles and Reynolds shear stress. The corresponding results are presented in figure \ref{fig:ep_mean}, following the same procedure applied to the backflow cases. Figures \ref{fig:ep_mean}(\textit{a}) and (\textit{b}) show the evolution of statistics for the $9$ deg. angle of attack case at the chordwise position $x_c = 0.5$. The event is initiated by a negative wall-normal velocity component $\widetilde{U}^2$ in the inner layer and an enhanced tangential velocity $\widetilde{U}^1$ relative to the total mean profiles indicated by the blue lines. This behavior suggests that the fluid is being accelerated toward the wall, as illustrated by the arrows in figure \ref{fig:ep_mean}(\textit{a}), which is consistent with the sweep motion observed in the Reynolds shear stress profile in figure \ref{fig:ep_mean}(\textit{b}). In addition to the dominant sweep contribution, a small inward interaction $Q3$ is also present, although its influence remains significantly weaker within the same region. As time progresses, the sweep motion becomes increasingly confined to the near-wall region, further accelerating the flow near the wall until the extreme event occurs at $\Delta t_0=0.000$. Following the event, the sweep intensity diminishes and approaches that of the ejections along the inner layer. This evolution is reflected in the straightening of the velocity profile shown in figure \ref{fig:ep_mean}(\textit{a}), which gradually returns to its mean value.

Figures \ref{fig:ep_mean}(\textit{c}) and (\textit{d}) show the statistics of EP events for the $12$ deg. case at the chordwise position $x_c = 0.9$. Similar to the previous case with weaker APG, the event initiates with a sweep motion originating in the near-wall region, as depicted in figure \ref{fig:ep_mean}(\textit{d}).
However, in contrast to the $9$ deg. case at $x_c = 0.5$, a substantial contribution from outward interactions $Q1$ is also observed along the inner layer. As shown in figure \ref{fig:ep_mean}(\textit{c}), the tangential velocity component $\widetilde{U}^1$ exceeds the total mean profile at $y^+<100$, while a small negative wall-normal velocity component $\widetilde{U}^2$ appears within the same region, 
indicating a high-speed fluid motion directed toward the wall. Although sweeps remain the dominant mechanism (approximately three times stronger than outward interactions), the interplay between these two motions influences the intensity of the EP events. 
After the event, the flow gradually relaxes toward the mean state and the sweep weakens near the wall as shown in the bottom panels of figure \ref{fig:ep_mean}(\textit{d}),  yet it continues to dominate across the inner layer. This evolution is reflected in figure \ref{fig:ep_mean}(\textit{c}), where $\widetilde{U}^2$ progressively approaches its mean value while maintaining a downward flow direction, and $\widetilde{U}^1$ decelerates but also remains above the mean profile. 
In both configurations analyzed, the initiation, development, and decay of EP events occur within a confined region in the inner layer.


\subsection{Inspection of coherent turbulent structures associated with extreme events} \label{sec:coherent_structures}

\begin{figure}
 \centering
 \begin{overpic}[width=0.475\textwidth,trim={0.25cm 6cm 7cm 17.5cm},clip]{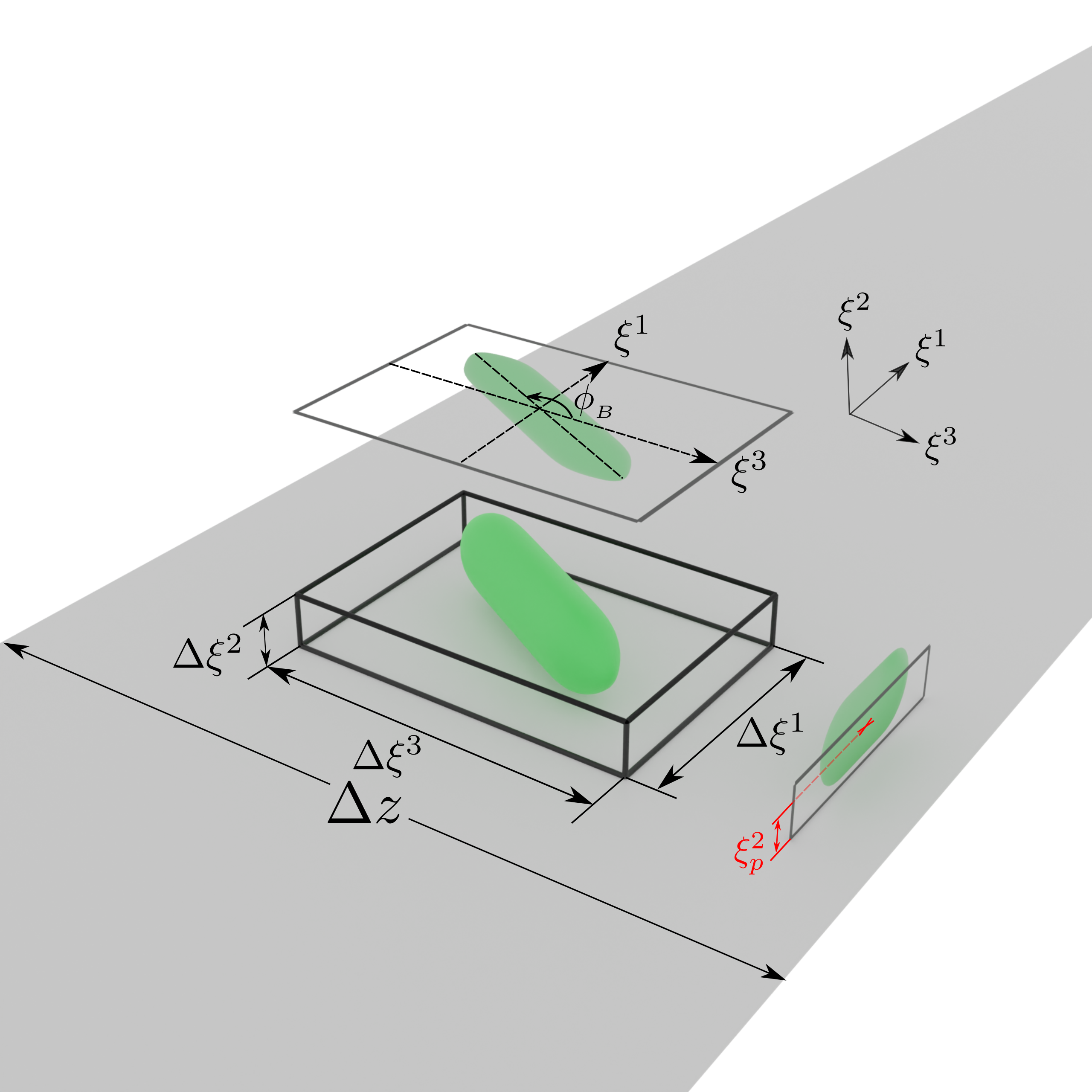}
	\put(1,69){(\textit{a})}
  \end{overpic}
  \begin{overpic}[width=0.475\textwidth,trim={0.25cm 6cm 7cm 17.5cm},clip]{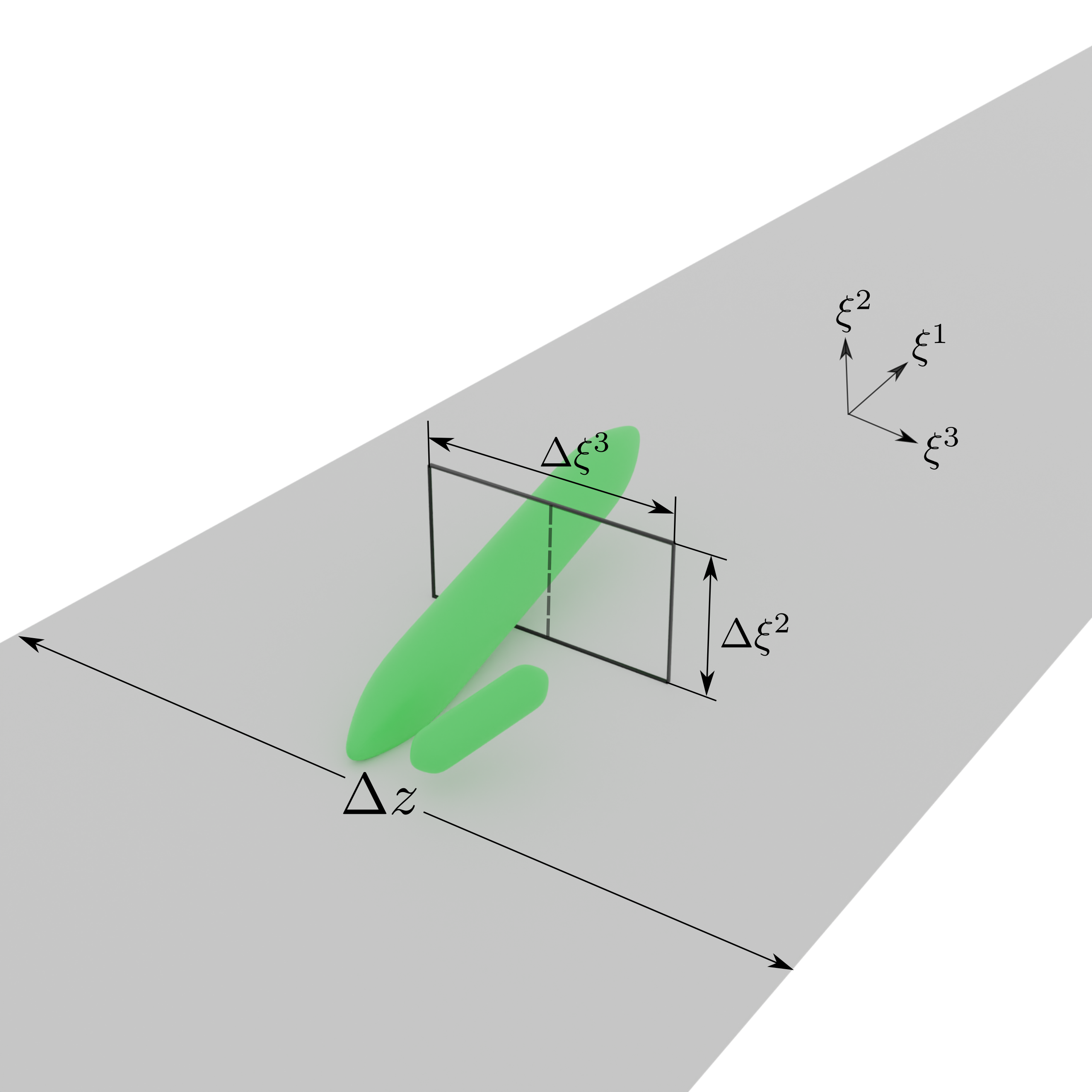}
	\put(1,69){(\textit{b})}
  \end{overpic}
\caption{Schematic representation of vortical structures used for the definition of (\textit{a}) backflow and (\textit{b}) extreme positive events.}
\label{fig:event_definition}
\end{figure}

To investigate the turbulence structures responsible for the extreme wall shear stress events and their mutual interactions, we perform a conditional analysis similar to that described in the previous section, but extended here to the three-dimensional space. Once a specific event is identified, it is centered within a domain of spanwise extent of $\Delta z = 0.034$. This value was chosen to encompass at least three times the maximum width of the wall shear stress events observed across all cases analyzed. Previous studies by \citet{sheng2009} and \citet{guerrero2020} have shown that extreme positive and backflow events either originate from or generate flow structures exhibiting a preferential side orientation. Such asymmetry could be distorted by conditional averaging if not properly accounted for, and the influence of this effect on the conditional averages is discussed in the Appendix \ref{appB}. Following these considerations, the backflow and extreme positive events are classified according to their defining characteristics, as schematically illustrated in figure \ref{fig:event_definition}.

Figure \ref{fig:event_definition}(\textit{a}) presents a schematic representation of the backflow event definition. The isometric view illustrates a vortical structure defined by the $Q$-criterion, which emerges during the evolution of a backflow event \citep{guerrero2022}. To track this structure, a $\xi^1 - \xi^2$ plane is positioned at the event center with a spatial window of size $\Delta \xi^1 \times \Delta \xi^{2^+} \approx 0.025 \times 20$.
Within this window, the location of the maximum positive $Q$-criterion is identified, as indicated by the red line marking the position $\xi^2_p$ in figure \ref{fig:event_definition}(\textit{a}). Once this reference point is determined, a 
constant $\xi^2$ plane coincident with $\xi^2_p$ is defined, with dimensions $\Delta \xi^1 \times \Delta \xi^3 \approx 0.025 \times 0.013$. This plane provides a top view of the $Q$-criterion structure, from which the orientation angle $\phi_B$ is measured relative to the $\xi^3$ axis passing through the event center. Turbulent structures with orientation angles $0 < \phi \leq \pi/2$ are classified as right-side structures, while those with $\pi/2 < \phi \leq \pi$, such as the one shown in figure \ref{fig:event_definition}(\textit{a}), are categorized as left-side structures.

The definition of extreme positive events is represented schematically in figure \ref{fig:event_definition}(\textit{b}). This event is predominantly associated with a single streamwise vortex oriented either to the left or right of the event center, or less frequently, with a pair of asymmetric counter-rotating streamwise vortices \citep{sheng2009,guerrero2020}. To characterize this turbulent structure, a spatial window is defined on a 
constant $\xi^1$ plane centered at the event location, with dimensions $\Delta \xi^{2^+} \times \Delta \xi^3 \approx 60 \times 0.013$. Within this window, the structures located to the left and right of the centerline are separated and conditionally averaged. In addition, for both event types, a region of positive $Q$-criterion with a negative pressure condition is required in the selected $\xi^1$ plane to ensure that the identified flow feature corresponds to a vortex.

After applying this procedure, the identified structures are conditionally averaged according to their respective characteristics. For the backflow events, the orientation angle of each structure is also computed to provide insights into its evolution. Table \ref{tab:3d_events} summarizes the total number of three-dimensional events ($N_{\mathrm{B}_{3D}}$, $N_{\mathrm{P}_{3D}}$) identified using the present methodology, along with the number of events classified on the left ($N_\mathrm{LB}$, $N_\mathrm{LP}$) and right ($N_\mathrm{RB}$, $N_\mathrm{RP}$) sides. Here, the subscripts $(B)$ and $(P)$ denote backflow and extreme positive events, respectively, and the mean orientation angle of the backflow events, $\phi_B$, is also reported in the table. As can be noticed, approximately $20\%$ of the detected events do not meet the classification criteria. A closer inspection of these unclassified events revealed that they correspond either to structures slightly displaced from the spatial window or to complex interactions between outer-layer structures that directly influence the inner layer in the case of backflow events. For extreme positive events, the unclassified cases were similarly associated with interactions involving outer-layer structures that affect the near-wall region, or in some cases, with absence of a negative pressure fluctuation at the vortex core. Moreover, table \ref{tab:3d_events} shows that the mean orientation angle tends to decrease with an increasing APG. This trend indicates that the turbulent structures originated from the evolution of backflow events become more two-dimensional, or spanwise-aligned, under stronger APG conditions, reflecting enhanced interactions among turbulence structures directly influenced by the pressure gradient.
\begin{table}
\begin{center}
\def~{\hphantom{0}}
\begin{tabular}{cccccccccc}
  AoA (deg.) &
  $x_c$ &
  $N_{\mathrm{B}_{3D}}$ &
  $N_{\mathrm{LB}}$ &
  $N_{\mathrm{RB}}$ &
  $\phi_{\mathrm{LB}}$ (deg.) &
  $\phi_{\mathrm{RB}}$ (deg.) &
  $N_{\mathrm{P}_{3D}}$ &
  $N_\mathrm{LE}$ &
  $N_\mathrm{RE}$ \\
\multirow{3}{*}{9}  & 0.5 & 297  & 113  & 117  & 106.4 & 77.1 & 3393 & 1367 & 1306 \\
                    & 0.7 & 554  & 227  & 249  & 109.0 & 72.0 & 3069 & 1147 & 1266 \\
                    & 0.9 & 1808 & 915  & 845  & 132.8 & 46.9 & 3055 & 1209 & 1239 \\
\multirow{3}{*}{12} & 0.5 & 1042 & 350  & 350  & 106.2 & 73.5 & 6107 & 2354 & 2452 \\
                    & 0.7 & 1905 & 559  & 658  & 114.6 & 66.8 & 4786 & 1799 & 1822 \\
                    & 0.9 & 3111 & 1409 & 1313 & 141.4 & 38.9 & 3724 & 1398 & 1404
\end{tabular}
  \caption{Statistics of three-dimensional extreme wall shear stress events, including the total number of events ($N_{\mathrm{B}_{3D}}$, $N_{\mathrm{P}_{3D}}$), and the number of events classified on the left ($N_\mathrm{LB}$,$N_\mathrm{LP}$) and right ($N_\mathrm{RB}$,$N_\mathrm{RP}$) sides. Subscripts B and P denote backflow and extreme positive events, respectively. The mean orientation angle $\phi_B$ of the structures generated during the backflow event evolution is also reported.} 
  \label{tab:3d_events}
\end{center}
\end{table}

\subsubsection{Dynamics of coherent structures associated with backflow events}\label{sec:bf_struc}

\begin{figure}
 \centering  
   \begin{overpic}[width=0.32\textwidth]{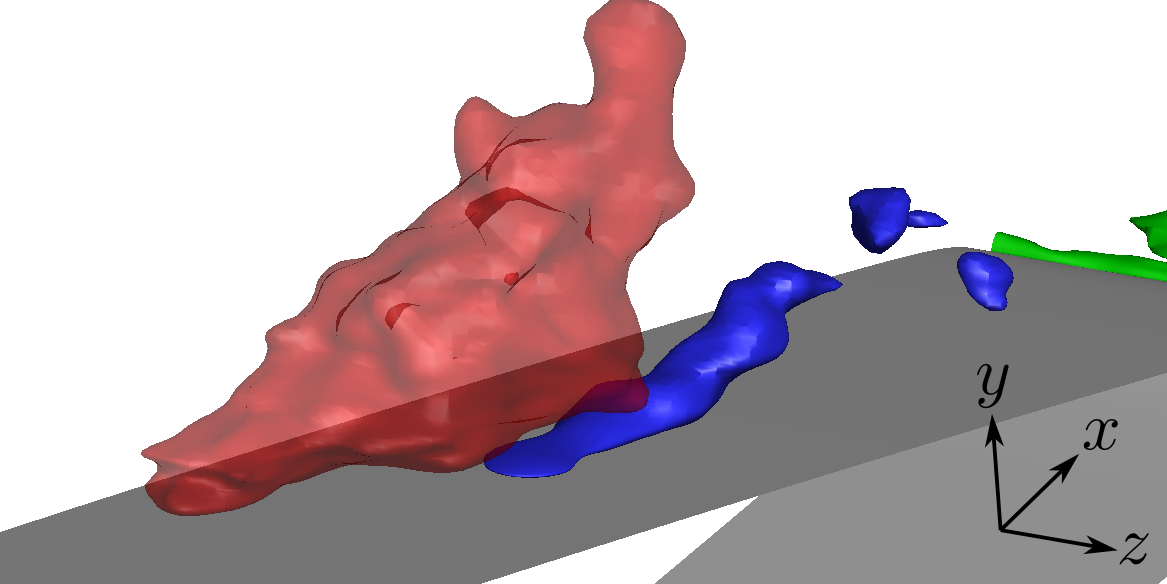}
	\put(1,43){(\textit{a})}
  \end{overpic}
  \begin{overpic}[width=0.32\textwidth]{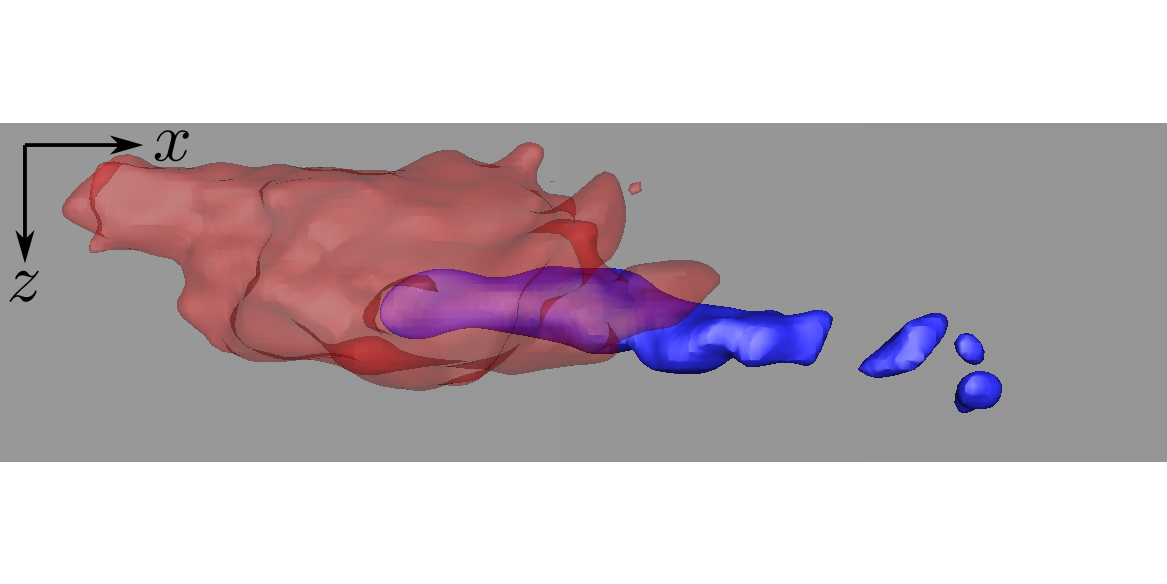}
        \put(27.5,42){$\Delta t_0 = -0.080$}
  \end{overpic}
    \begin{overpic}[width=0.32\textwidth]{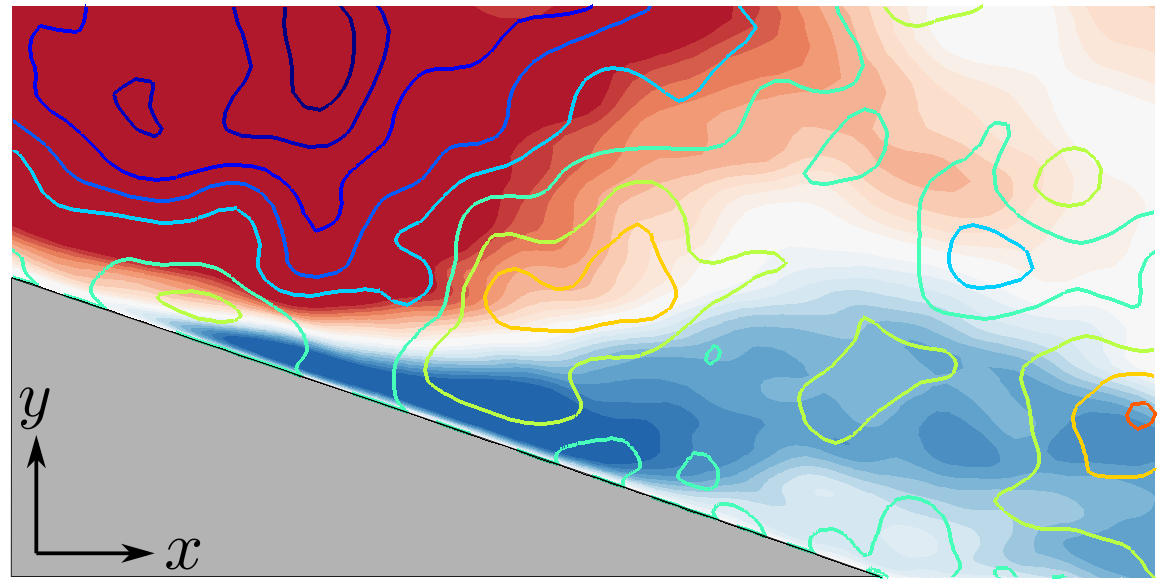}
  \end{overpic}
   \begin{overpic}[width=0.32\textwidth]{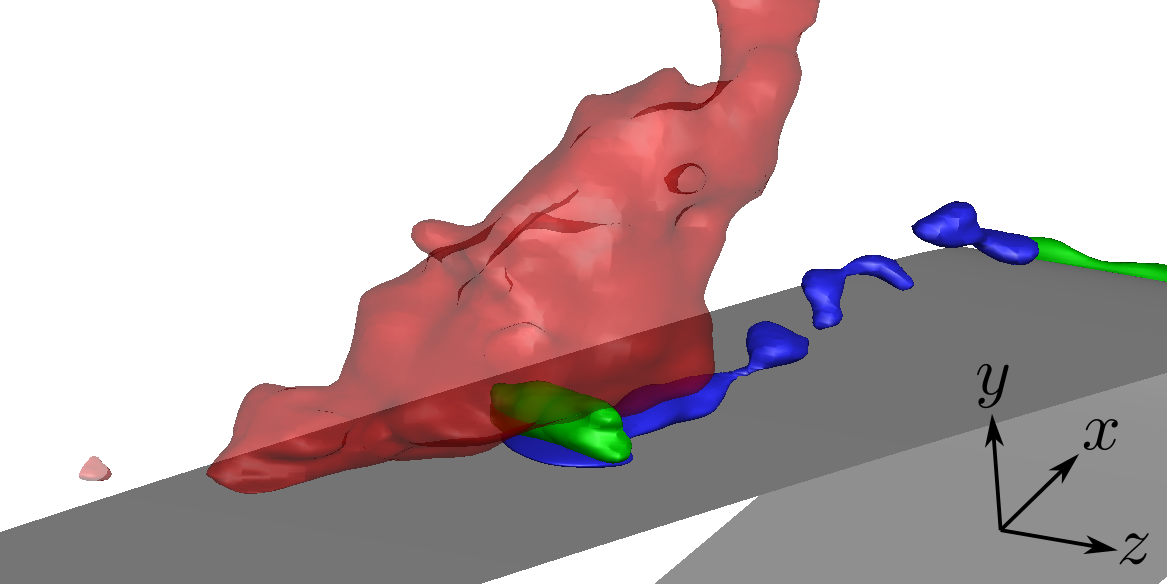}
	\put(1,43){(\textit{b})}
  \end{overpic}
  \begin{overpic}[width=0.32\textwidth]{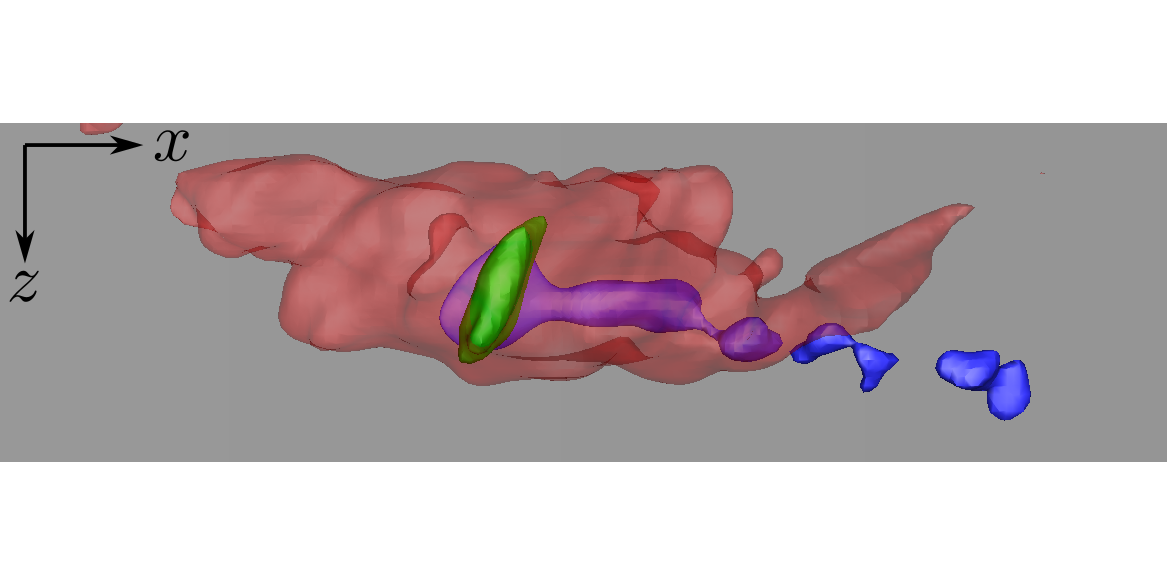}
        \put(27.5,42){$\Delta t_0 = -0.040$}
  \end{overpic}
    \begin{overpic}[width=0.32\textwidth]{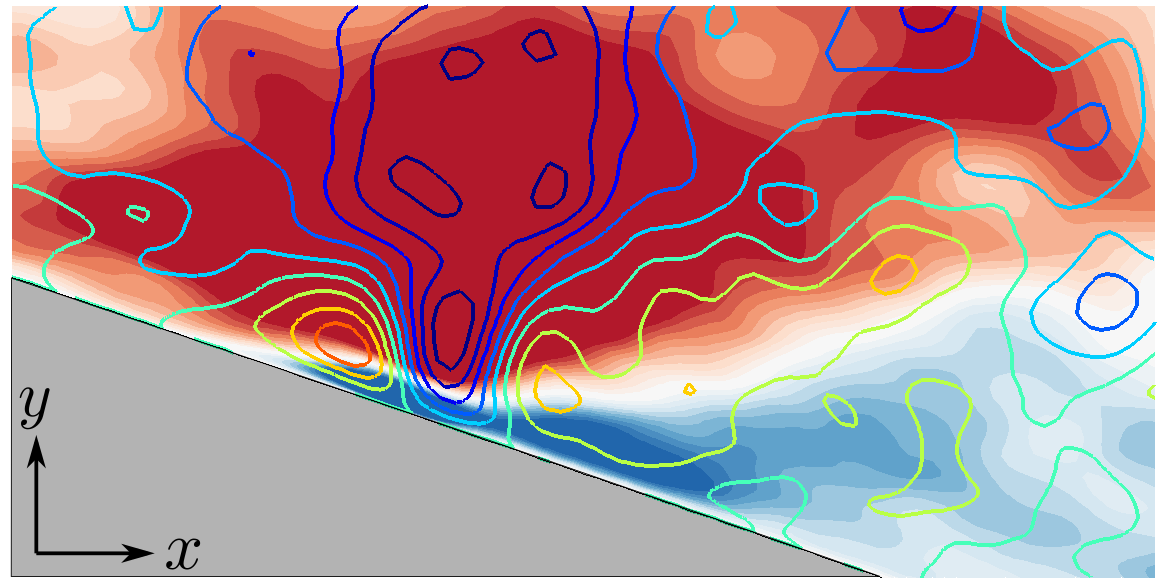}
  \end{overpic}
   \begin{overpic}[width=0.32\textwidth]{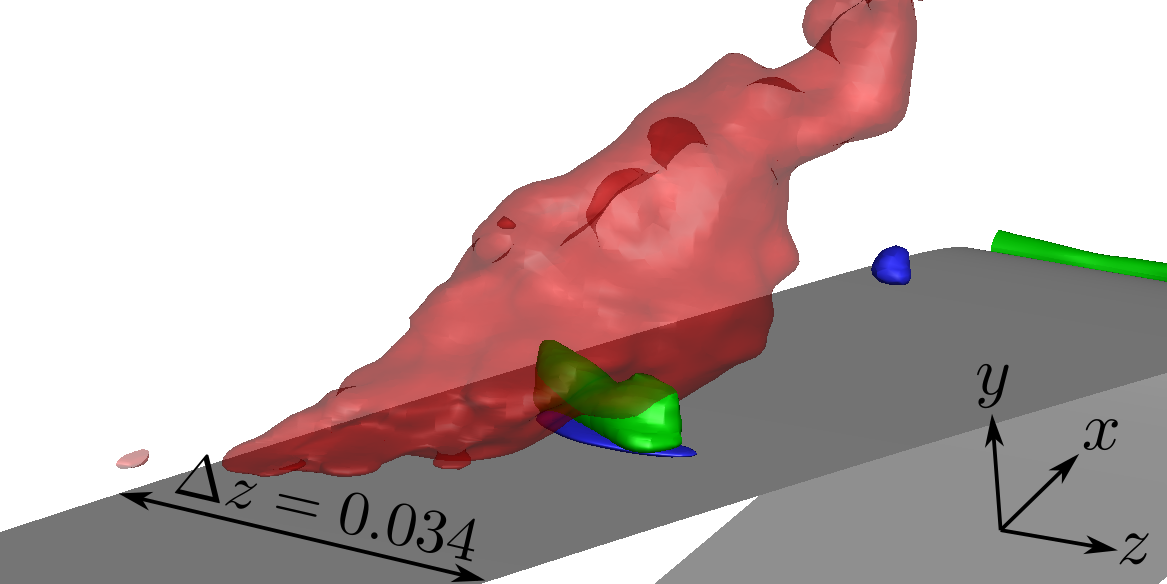}
	\put(1,43){(\textit{c})}
  \end{overpic}
  \begin{overpic}[width=0.32\textwidth]{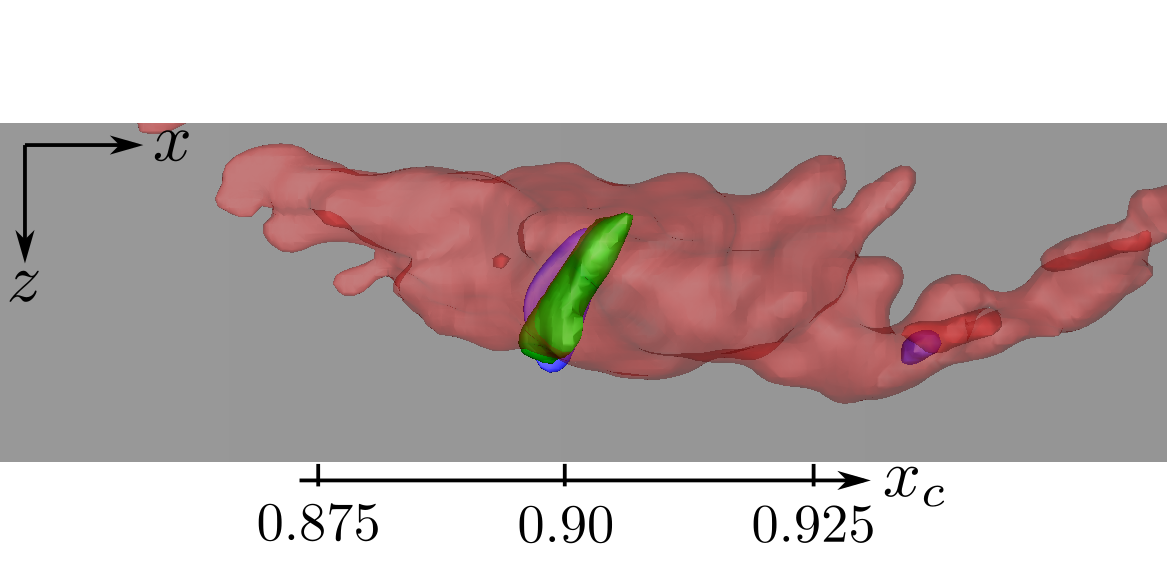}
        \put(27.5,42){$\Delta t_0 = 0.000$}
  \end{overpic}
    \begin{overpic}[width=0.32\textwidth]{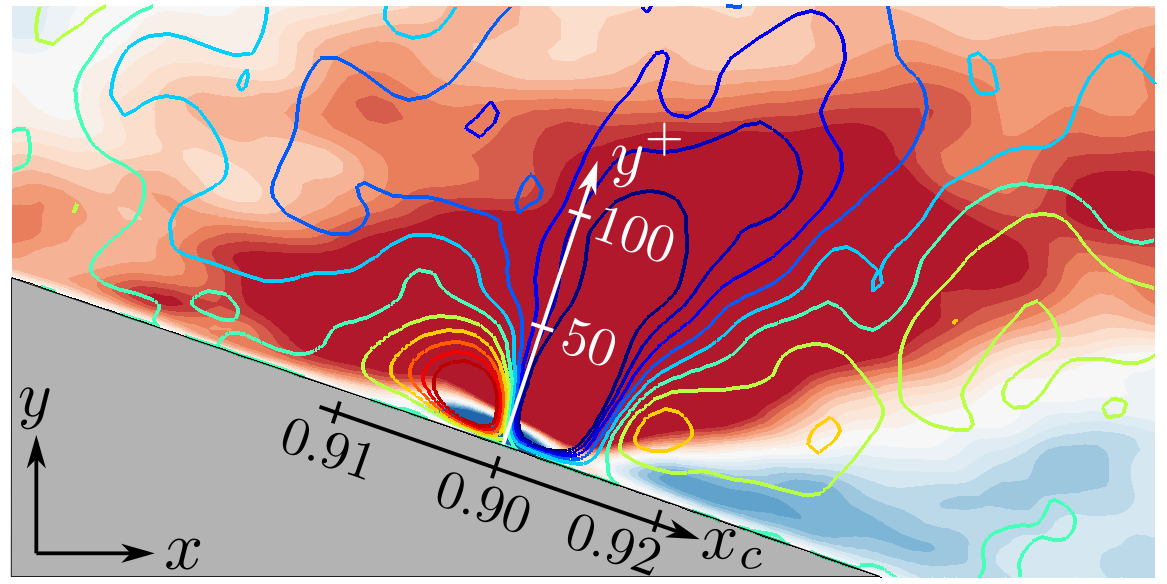}
  \end{overpic}
   \begin{overpic}[width=0.32\textwidth]{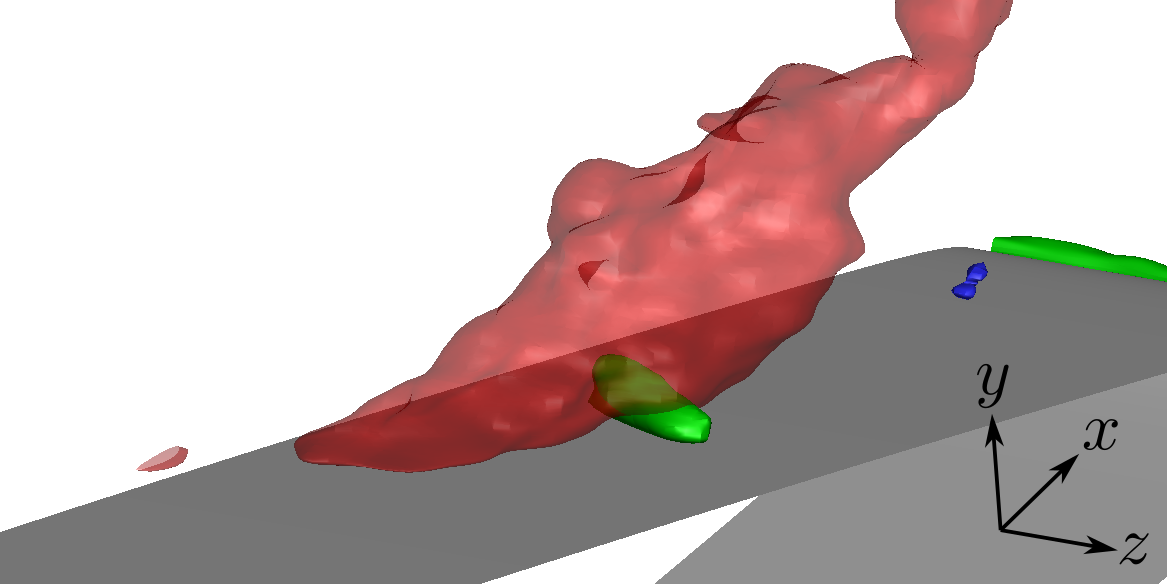}
	\put(1,43){(\textit{d})}
  \end{overpic}
  \begin{overpic}[width=0.32\textwidth]{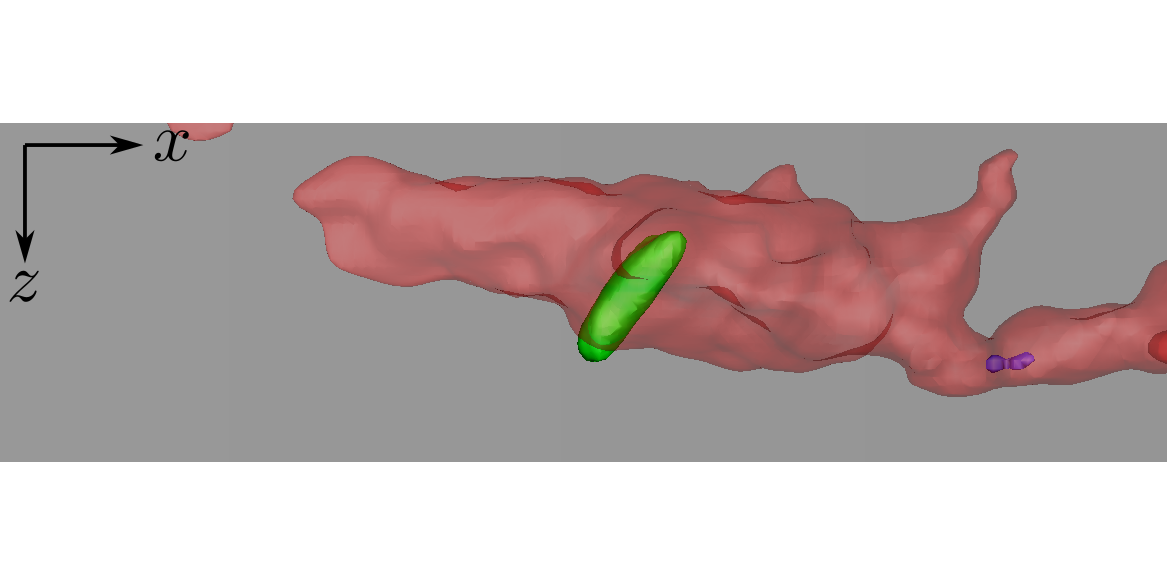}
        \put(27.5,42){$\Delta t_0 = 0.040$}
  \end{overpic}
    \begin{overpic}[width=0.32\textwidth]{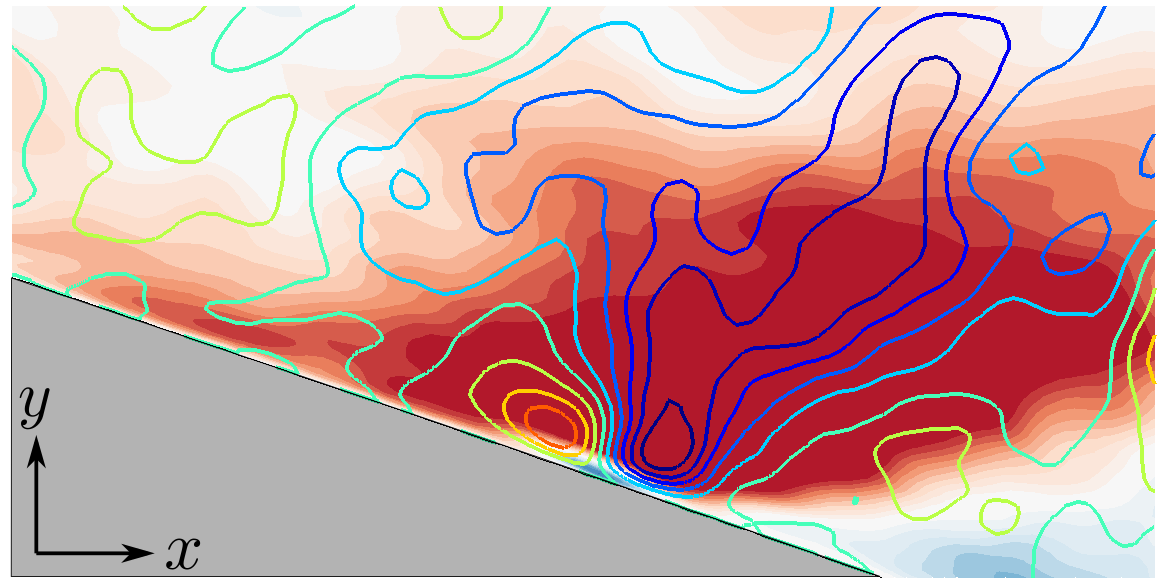}
  \end{overpic}
   \begin{overpic}[width=0.32\textwidth]{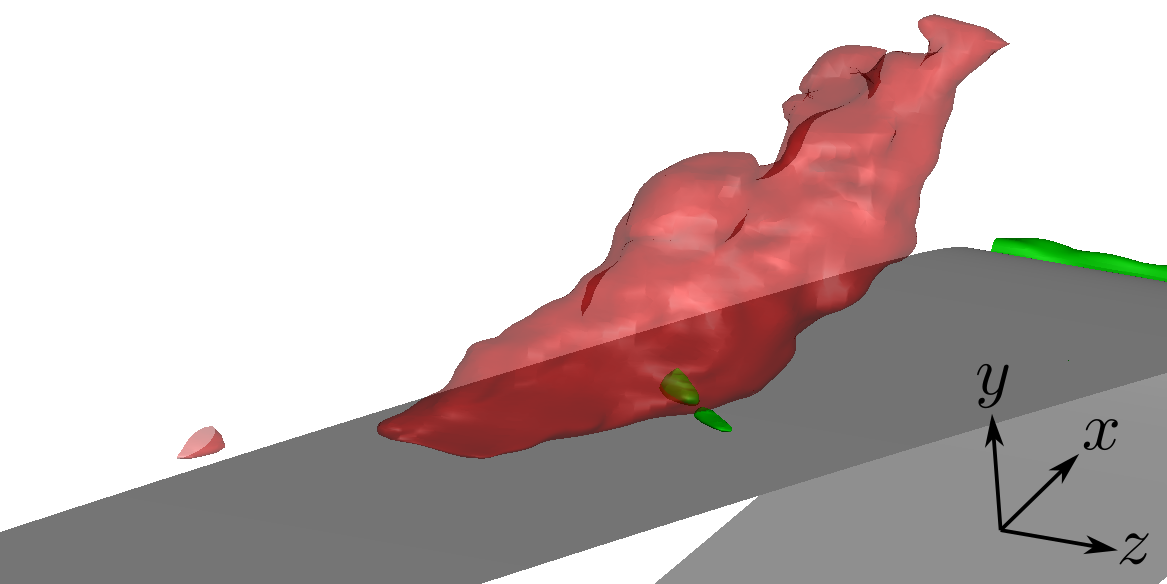}
	\put(1,43){(\textit{e})}
  \end{overpic}
  \begin{overpic}[width=0.32\textwidth]{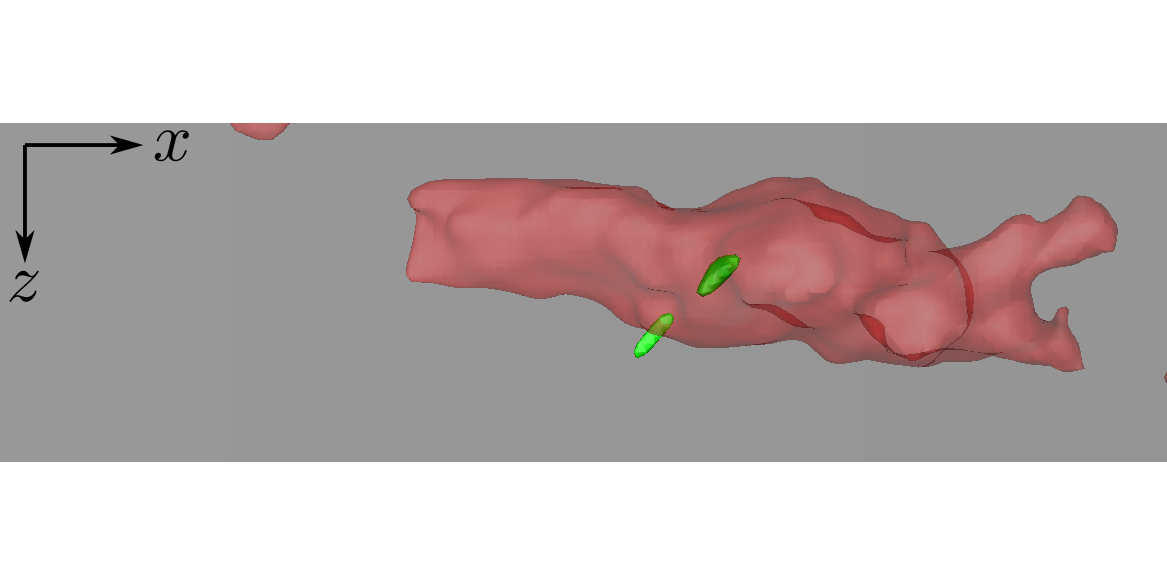}
        \put(27.5,42){$\Delta t_0 = 0.080$}
  \end{overpic}
  \begin{overpic}[width=0.32\textwidth]{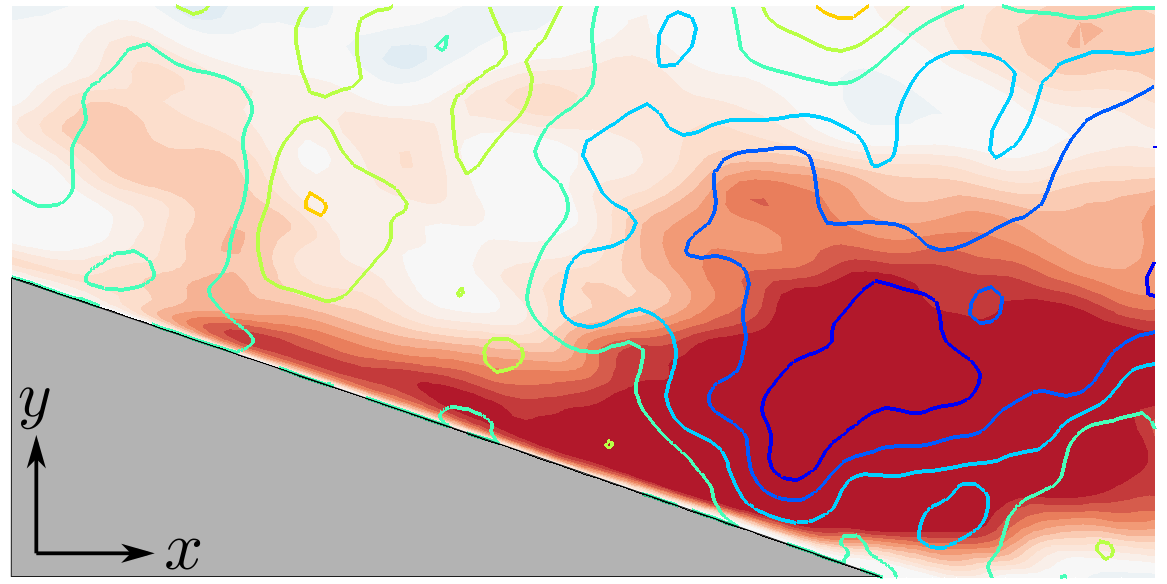}
  \end{overpic}
  \begin{overpic}[width=0.15\textwidth]{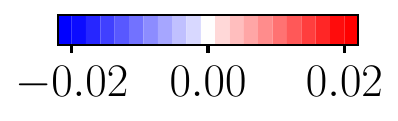}
         \put(-30,17.5){$\widetilde{u}_t/U_\infty$}
  \end{overpic}
  \hspace{1.5cm}
  \begin{overpic}[width=0.15\textwidth]{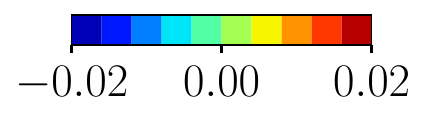}
         \put(-29.5,17.5){$\widetilde{u}_n/U_\infty$}
  \end{overpic}
\caption{Three-dimensional evolution of the coherent structures associated with the backflow event for the $12$ deg. angle of attack case at chordwise position $x_c=0.9$. The first column shows a three-dimensional perspective, while the second column presents a top view of the event dynamics. Red and blue isosurfaces correspond to regions of tangential velocity fluctuations of $\widetilde{u}_t=0.02$ and $\widetilde{u}_t=-0.02$, respectively, highlighting a high-speed large-scale structure approaching a low-speed streak. A vortex structure is visualized using a $Q$-criterion isosurface colored in green. The third column shows a constant $z$ plane passing through the event center, where the background contours represent the tangential velocity fluctuations $\widetilde{u}_t$ and the overlaid isolines denote the wall-normal velocity fluctuations $\widetilde{u}_n$. Movie 1 is provided as supplementary material showing the evolution of coherent structures and their interactions for this case. 
}
\label{fig:bk_3d_12}
\end{figure}

To further examine the turbulence structures involved in the evolution of backflow events, figure \ref{fig:bk_3d_12} presents results for the $12$ deg. angle of attack case at chordwise position $x_c = 0.9$, where the boundary layer is subject to a strong APG. Five representative time instants are shown to illustrate the complete temporal development of the event. Isosurfaces of constant tangential velocity fluctuation are plotted to identify the coherent structures, and all quantities are normalized by the freestream velocity. The first column provides a perspective view of the three-dimensional structures, while the second column shows their corresponding top view. In these panels, the red isosurface is plotted with $\widetilde{u}_t = 0.02$, showing a high-speed large-scale structure (LSS), whereas the blue isosurface, plotted with $\widetilde{u}_t = -0.02$, highlights a low-speed streak.
A vortical structure generated during the event is visualized using the $Q$-criterion, rendered in green. The third column displays a constant $z$ plane located at the mid-span of the event, where the background colors represent the tangential velocity fluctuation $\widetilde{u}_t$, and the overlaid isolines correspond to the wall-normal velocity fluctuation $\widetilde{u}_n$.

In figure \ref{fig:bk_3d_12}(\textit{a}), the initiation of the event is shown, where a high-speed LSS approaches a low-speed streak located near the wall. The inspection of the constant $z$ plane reveals that this LSS also has a negative wall-normal velocity fluctuation, indicating a sweep motion directed toward the low-speed streak. Subsequently, in figure \ref{fig:bk_3d_12}(\textit{b}), an inclined vortex structure emerges as a result of the interaction between these two turbulence structures. The sweep motion associated with the LSS continues to push the flow toward the wall, enhancing its interaction with the low-speed streak. In the rear portion of the streak, a positive wall-normal velocity fluctuation develops, demonstrating that this interaction also induces an ejection motion. At the time instant corresponding to the maximum backflow magnitude ($\Delta t_0 = 0.000$), shown in figure \ref{fig:bk_3d_12}(\textit{c}), the LSS disrupts the tail of the low-speed streak, while both sweep and ejection motions intensify, as evidenced by the $u_n$ isolines. The vortex structure also exhibits an inclination toward the streamwise direction, consistent with the observations of \citet{guerrero2022}. Following this stage, the low-speed streak and the sweep motion of the LSS weaken in figure \ref{fig:bk_3d_12}(\textit{d}), while the vortex structure continues to tilt in the streamwise direction before dissipating, as shown in figure \ref{fig:bk_3d_12}(\textit{e}).

\begin{figure}
 \centering
 \begin{overpic}[width=0.32\textwidth]{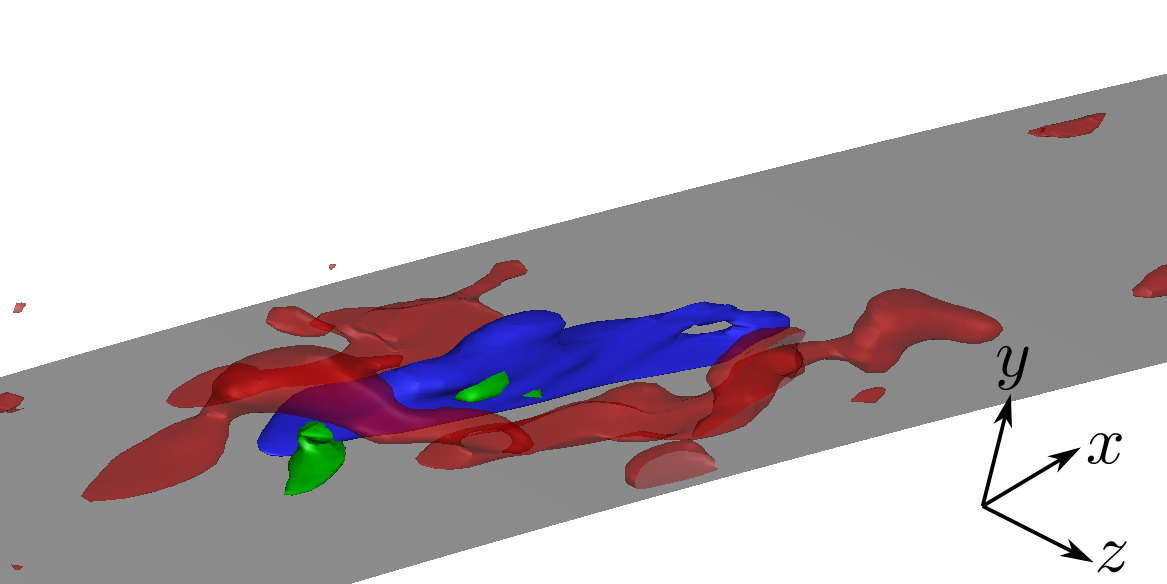}
	\put(1,43){(\textit{a})}
  \end{overpic}
  \begin{overpic}[width=0.32\textwidth]{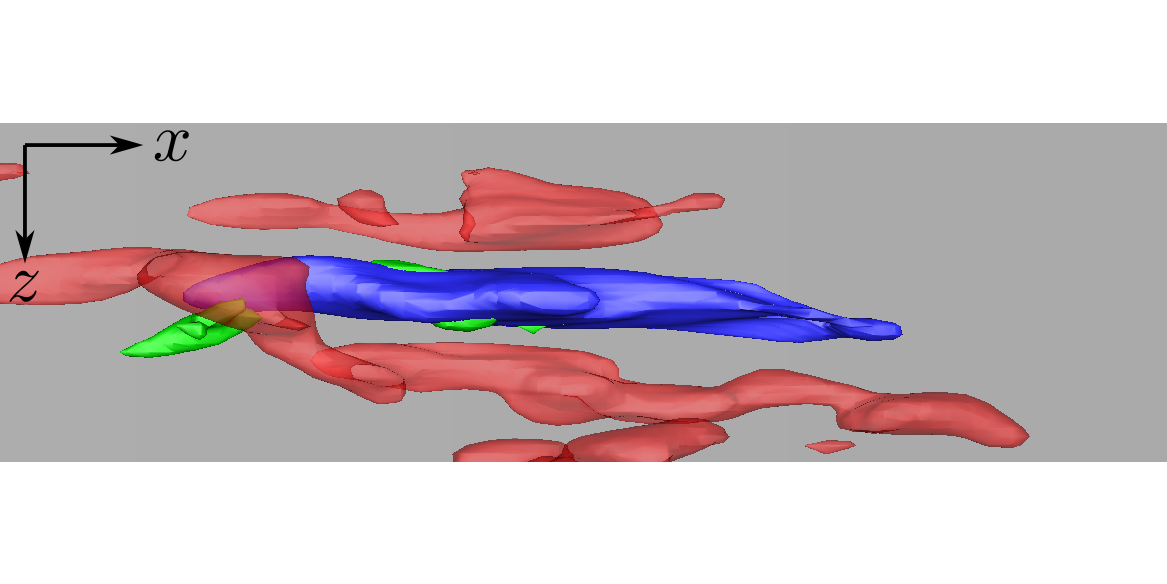}
          \put(27.5,42){$\Delta t_0 = -0.048$}
  \end{overpic}
    \begin{overpic}[width=0.32\textwidth]{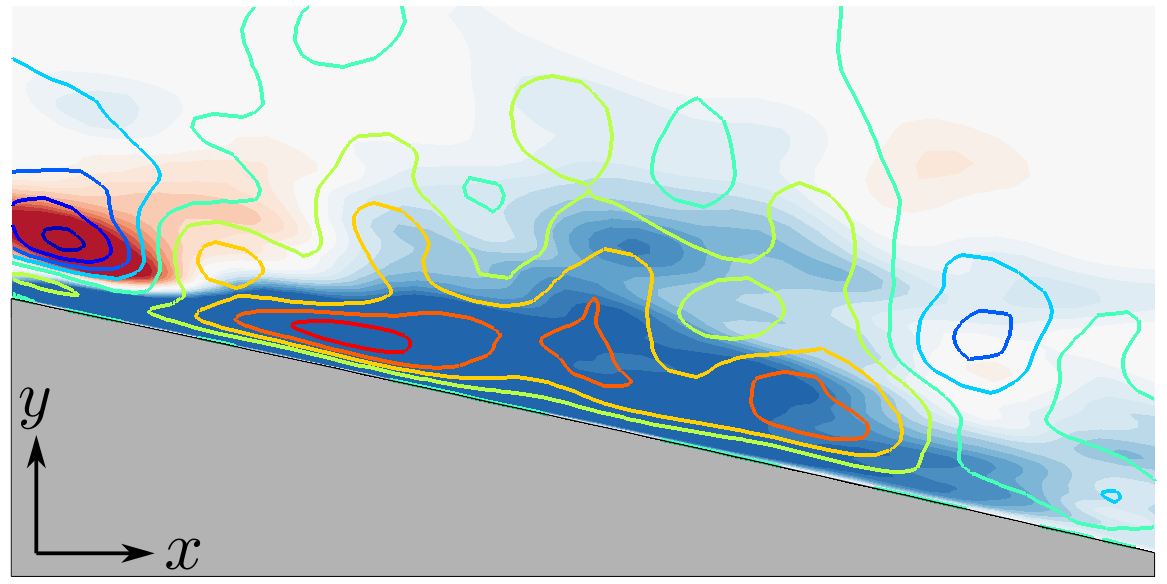}
  \end{overpic}
   \begin{overpic}[width=0.32\textwidth]{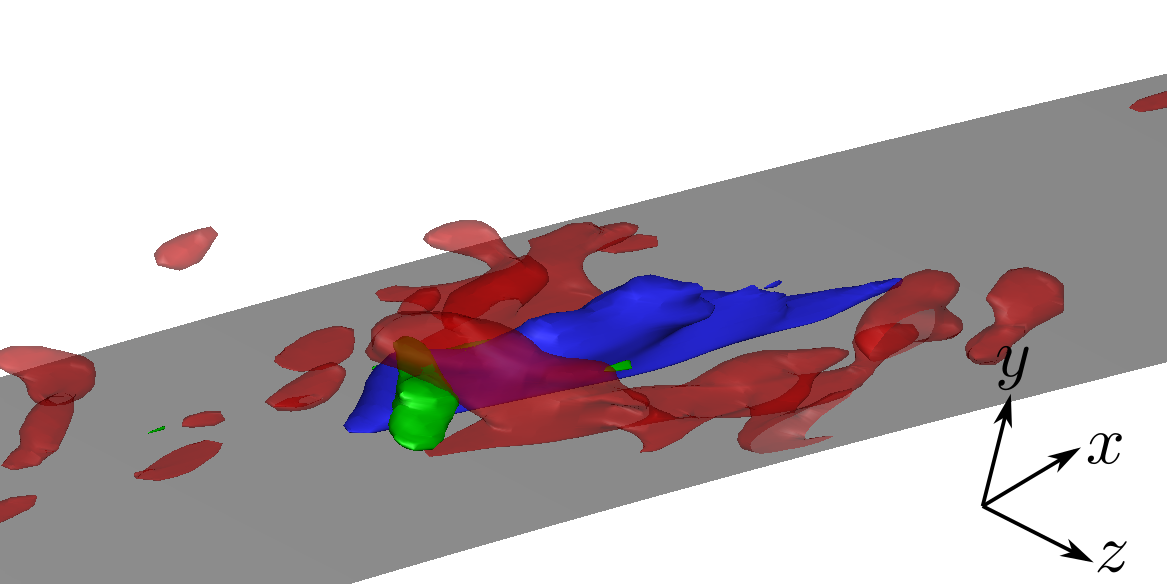}
	\put(1,43){(\textit{b})}
  \end{overpic}
  \begin{overpic}[width=0.32\textwidth]{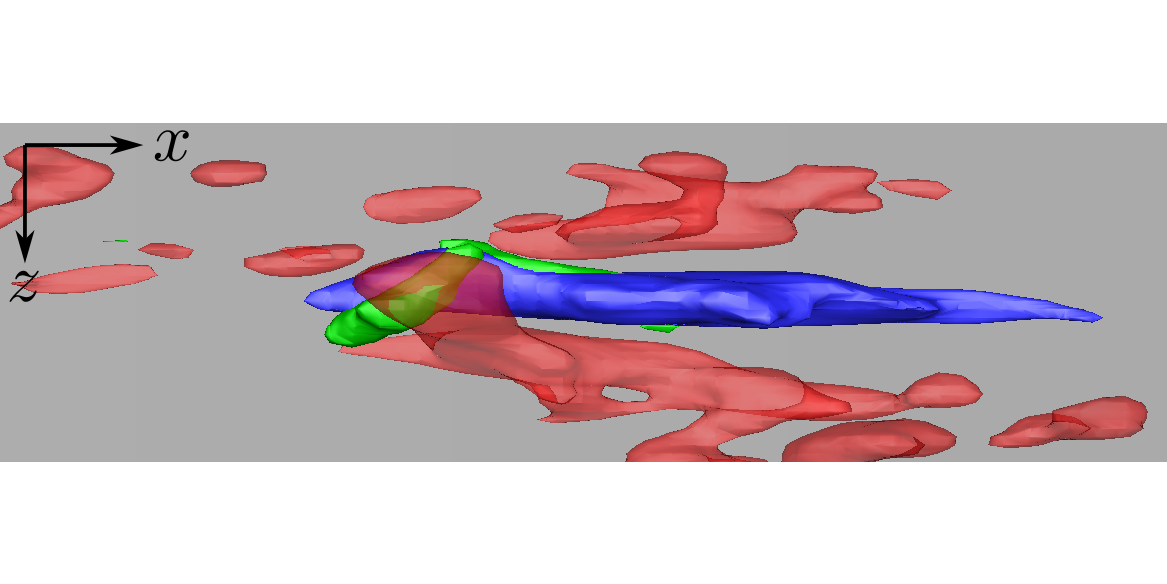}
          \put(27.5,42){$\Delta t_0 = -0.024$}
  \end{overpic}
    \begin{overpic}[width=0.32\textwidth]{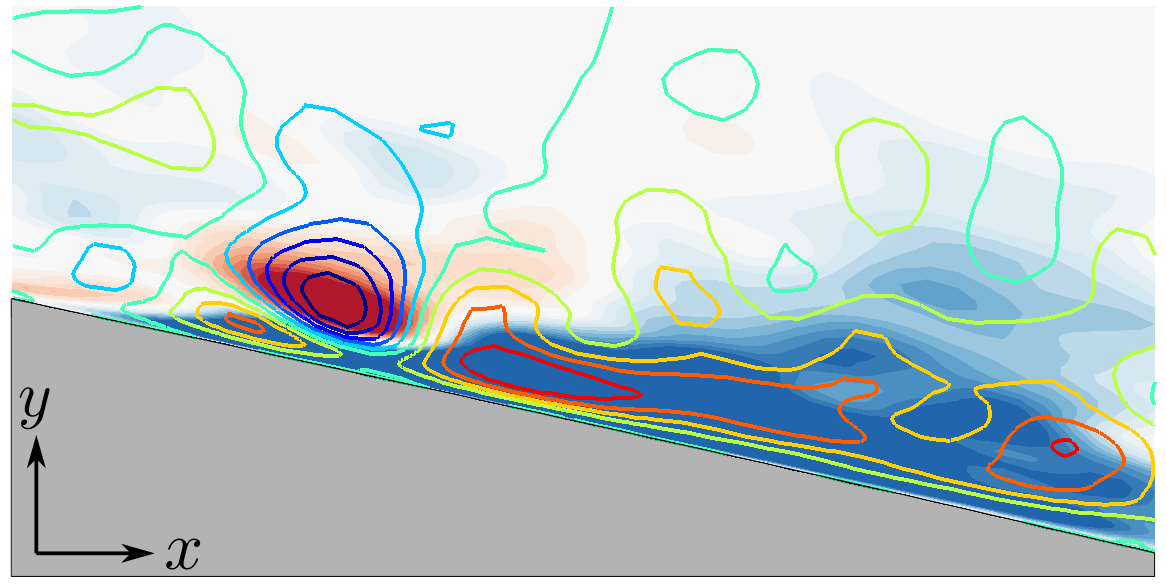}
  \end{overpic}
   \begin{overpic}[width=0.32\textwidth]{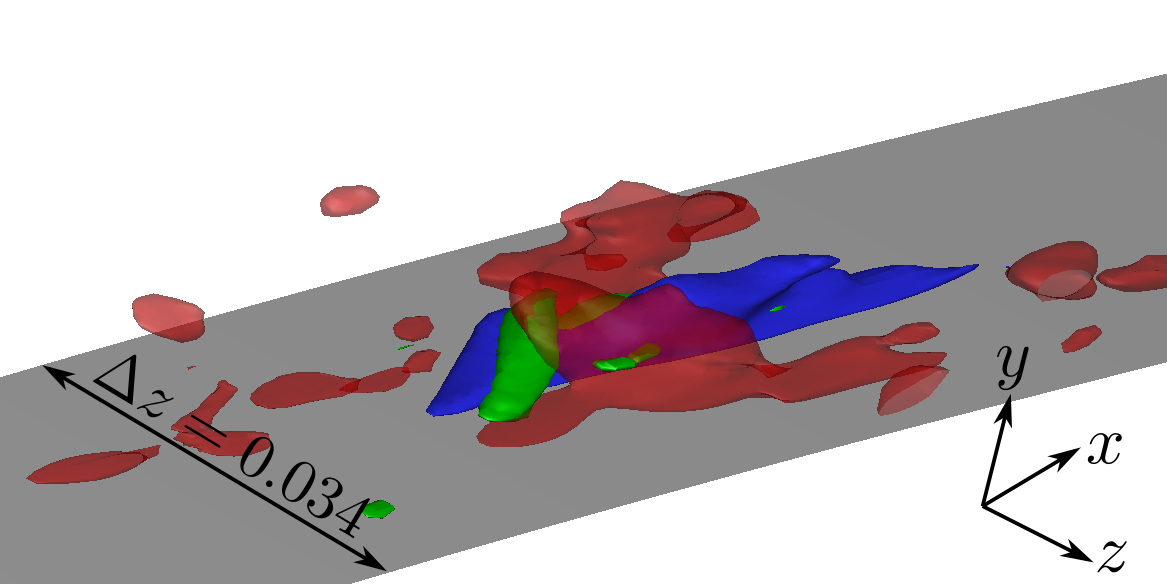}
	\put(1,43){(\textit{c})}
  \end{overpic}
  \begin{overpic}[width=0.32\textwidth]{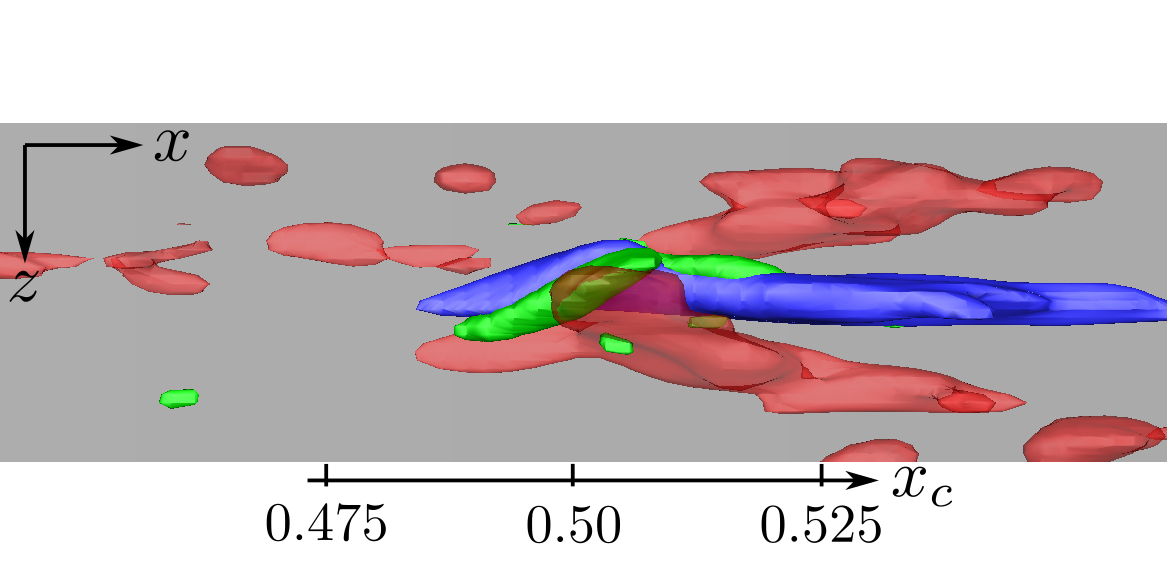}
          \put(27.5,42){$\Delta t_0 = 0.000$}
  \end{overpic}
    \begin{overpic}[width=0.32\textwidth]{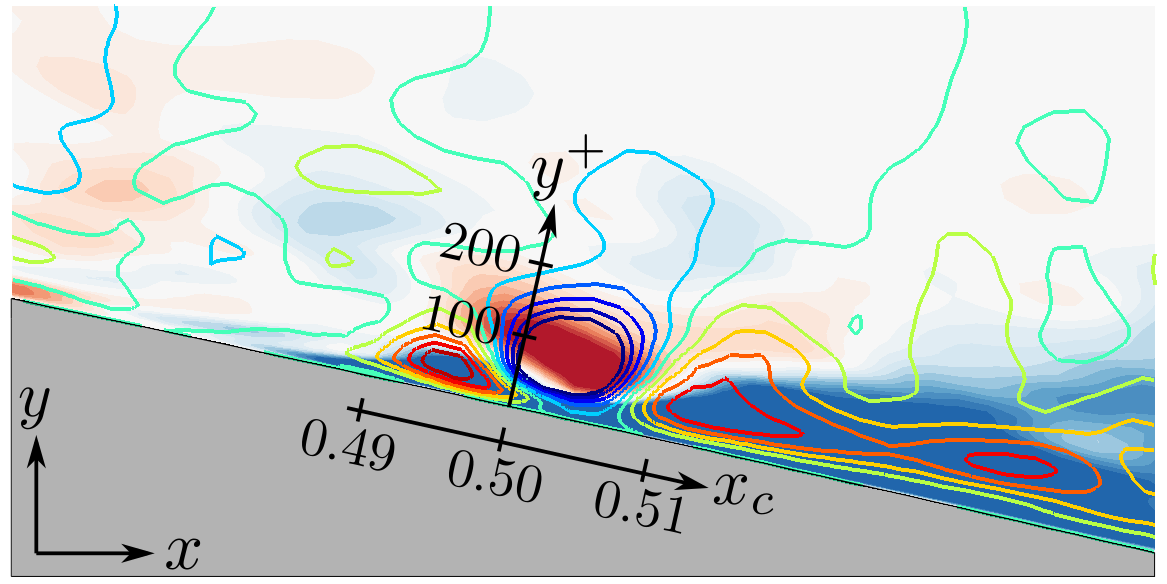}
  \end{overpic}
   \begin{overpic}[width=0.32\textwidth]{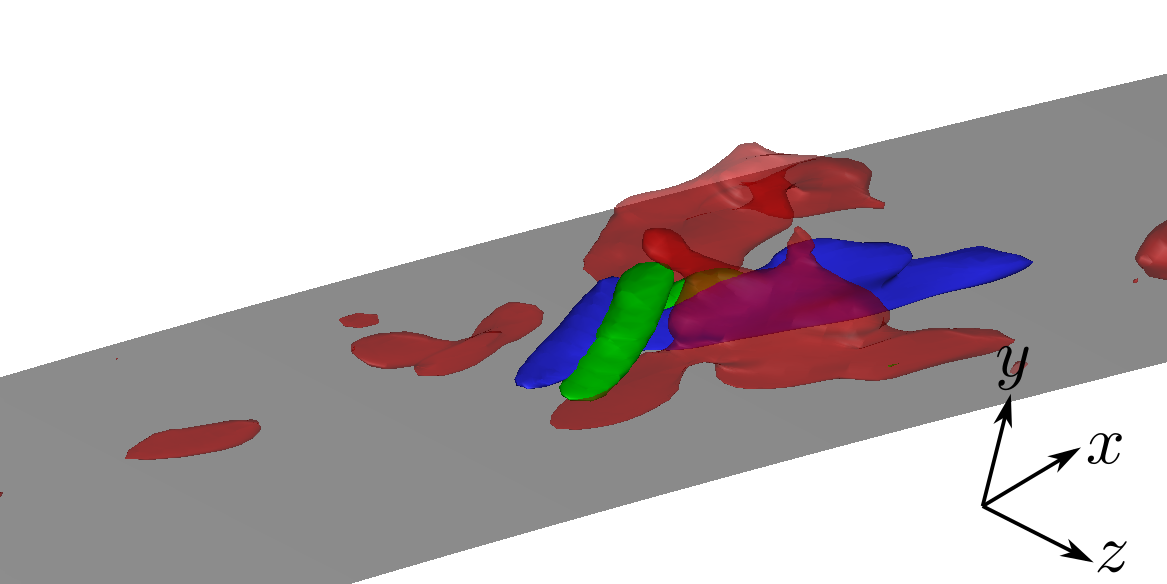}
	\put(1,43){(\textit{d})}
  \end{overpic}
  \begin{overpic}[width=0.32\textwidth]{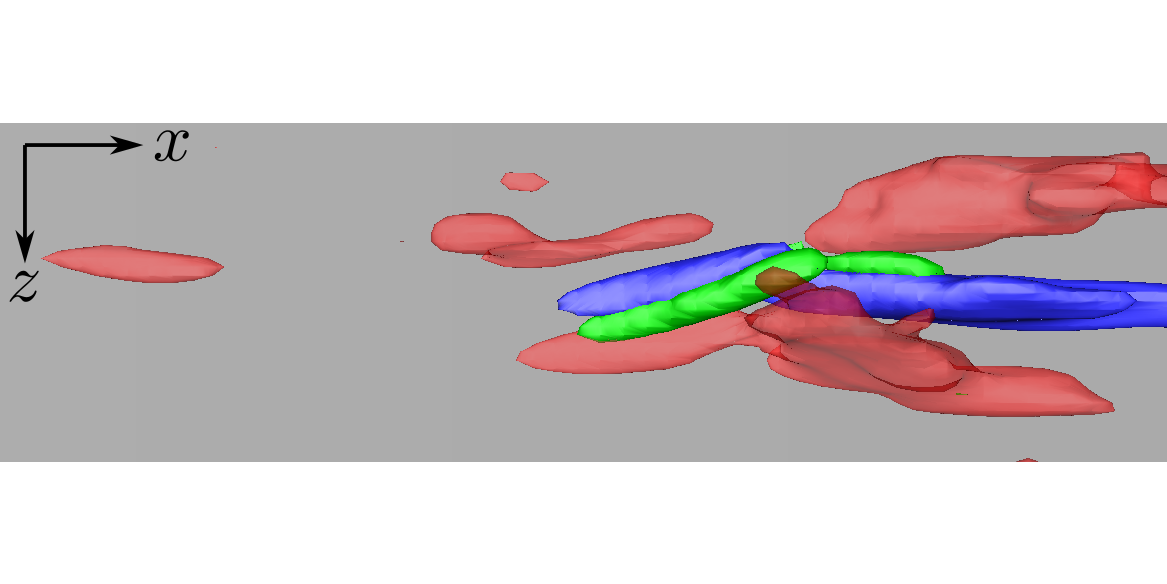}
          \put(27.5,42){$\Delta t_0 = 0.024$}
  \end{overpic}
    \begin{overpic}[width=0.32\textwidth]{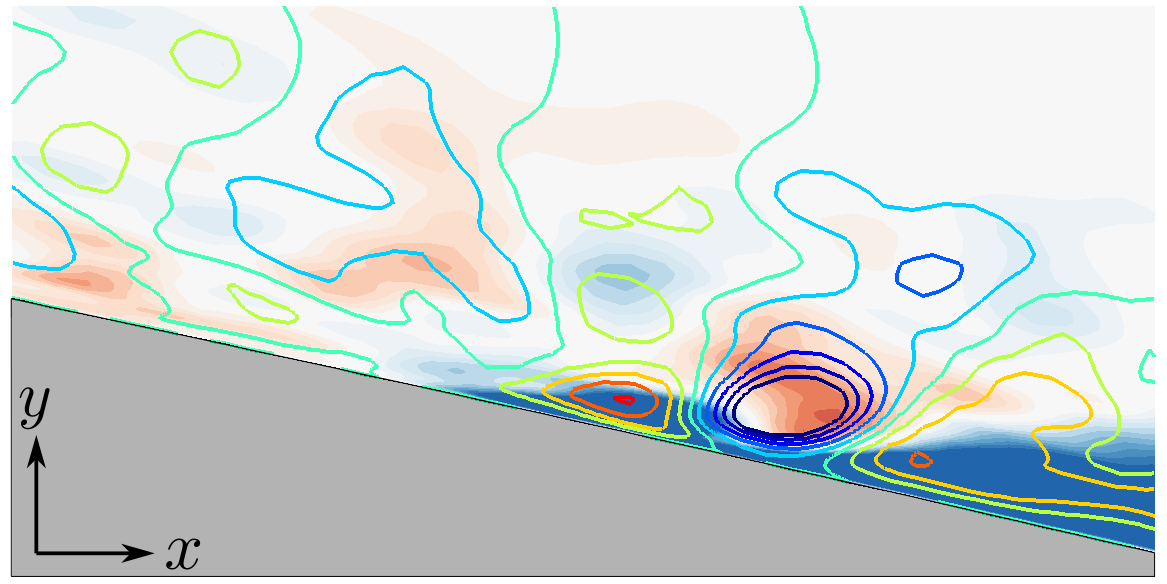}
  \end{overpic}
   \begin{overpic}[width=0.32\textwidth]{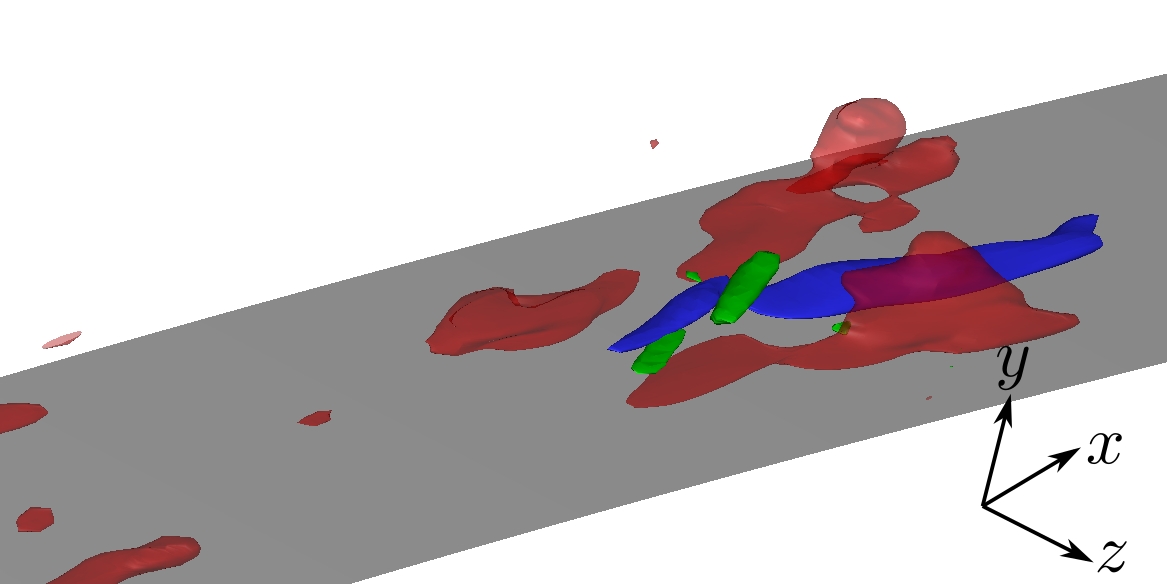}
	\put(1,43){(\textit{e})}
  \end{overpic}
  \begin{overpic}[width=0.32\textwidth]{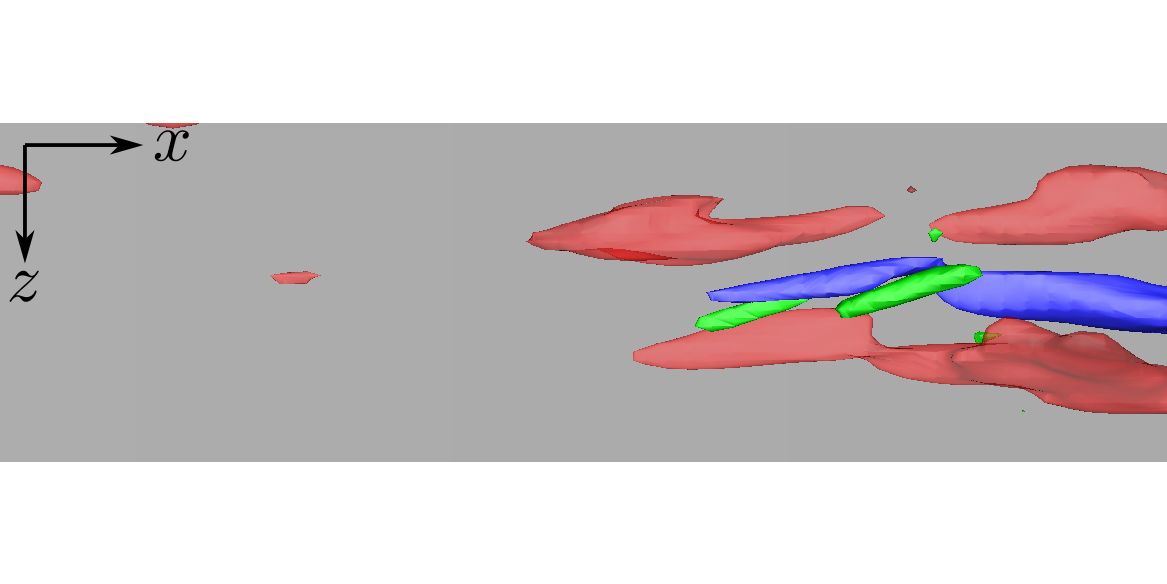}
        \put(27.5,42){$\Delta t_0 = 0.048$}
  \end{overpic}
    \begin{overpic}[width=0.32\textwidth]{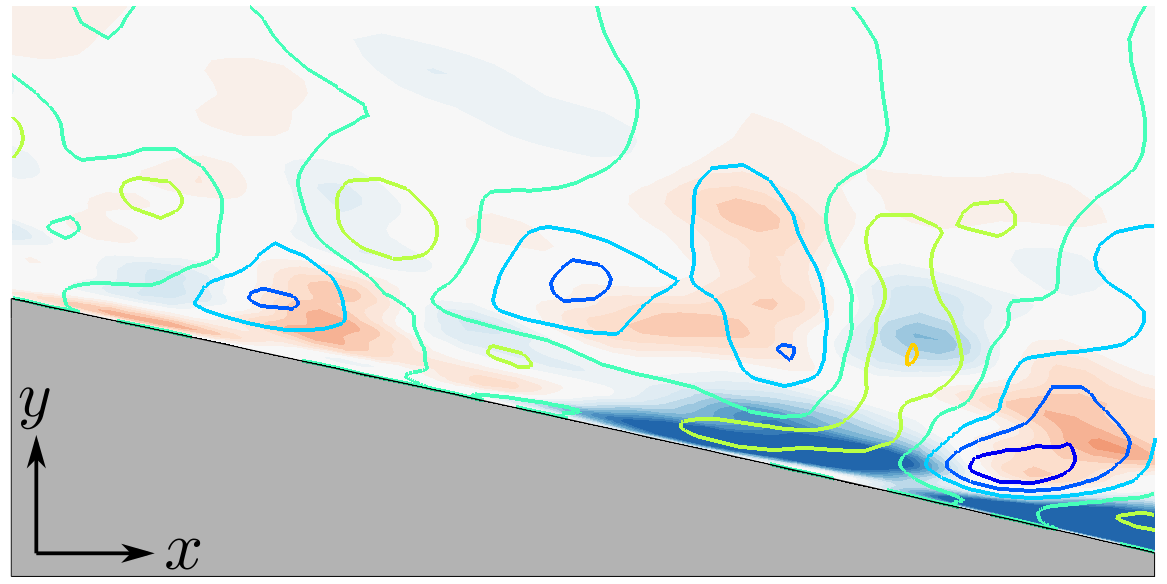}
  \end{overpic}
\begin{overpic}[width=0.15\textwidth]{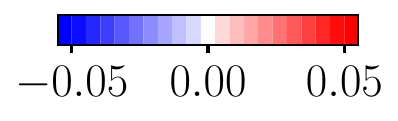}
         \put(-30,17.5){$\widetilde{u}_t/U_\infty$}
  \end{overpic}
  \hspace{1.5cm}
  \begin{overpic}[width=0.15\textwidth]{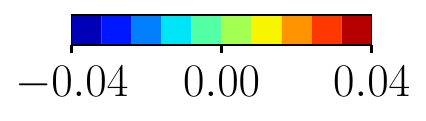}
         \put(-29.5,17.5){$\widetilde{u}_n/U_\infty$}
  \end{overpic}
\caption{Three-dimensional evolution of the coherent structures associated with the backflow event for the $9$ deg. angle of attack case at chordwise position $x_c=0.5$. The first column shows a three-dimensional perspective, while the second column presents a top view of the event dynamics. Red and blue isosurfaces correspond to regions of tangential velocity fluctuations of $\widetilde{u}_t=0.035$ and $\widetilde{u}_t=-0.07$, respectively, highlighting a local sweep motion approaching a low-speed streak. A vortex structure is visualized using a $Q$-criterion isosurface colored in green. The third column shows a constant $z$ plane passing through the event center, where the background contours represent the tangential velocity fluctuations $\widetilde{u}_t$ and the ovelaid isolines denote the wall-normal velocity fluctuations $\widetilde{u}_t$. Movie 2 is provided as supplementary material showing the evolution of coherent structures and their interactions for this case.}
\label{fig:bk_3d_9}
\end{figure}

The results for $9$ deg. angle of attack are presented in figure \ref{fig:bk_3d_9} for the airfoil chord position $x_c=0.5$, which corresponds to a location with a low APG magnitude. The visualization follows the same representation used in figure \ref{fig:bk_3d_12}, but with the velocity fluctuation thresholds for the red and blue structures set to $\widetilde{u}_t = 0.035$ and $\widetilde{u}_t = -0.07$, respectively. At the onset of the event, shown in figure \ref{fig:bk_3d_9}(\textit{a}), a large-scale low-speed streak is flanked on both sides by high-speed streaks. In addition, a high-speed structure appears above the streak region. Although smaller than in the previous case, this structure also exhibits a negative wall-normal velocity fluctuation, indicating a sweep motion toward the wall, as confirmed by the constant $z$ plane visualization. In the following instant, shown in figure \ref{fig:bk_3d_9}(\textit{b}), the high-speed structure interacts with the low-speed streak, generating a vortex structure inclined in the streamwise direction, highlighted by the green $Q$-criterion isosurface. At $\Delta t_0 = 0.000$, figure \ref{fig:bk_3d_9}(\textit{c}) shows that the high-speed structure penetrates deeper into the low-speed streak, producing a distinct ejection on its tail, as seen in the constant $z$ plane. Meanwhile, the vortex structure becomes further inclined and stretched in the streamwise direction, a process that continues to be accentuated in the following stages (figure \ref{fig:bk_3d_9}(\textit{d})). Eventually, the high-speed structure disrupts the low-speed streak and becomes weaker, as shown in figure \ref{fig:bk_3d_9}(\textit{e}).

\begin{figure}
 \centering
 \begin{overpic}[width=0.475\textwidth]{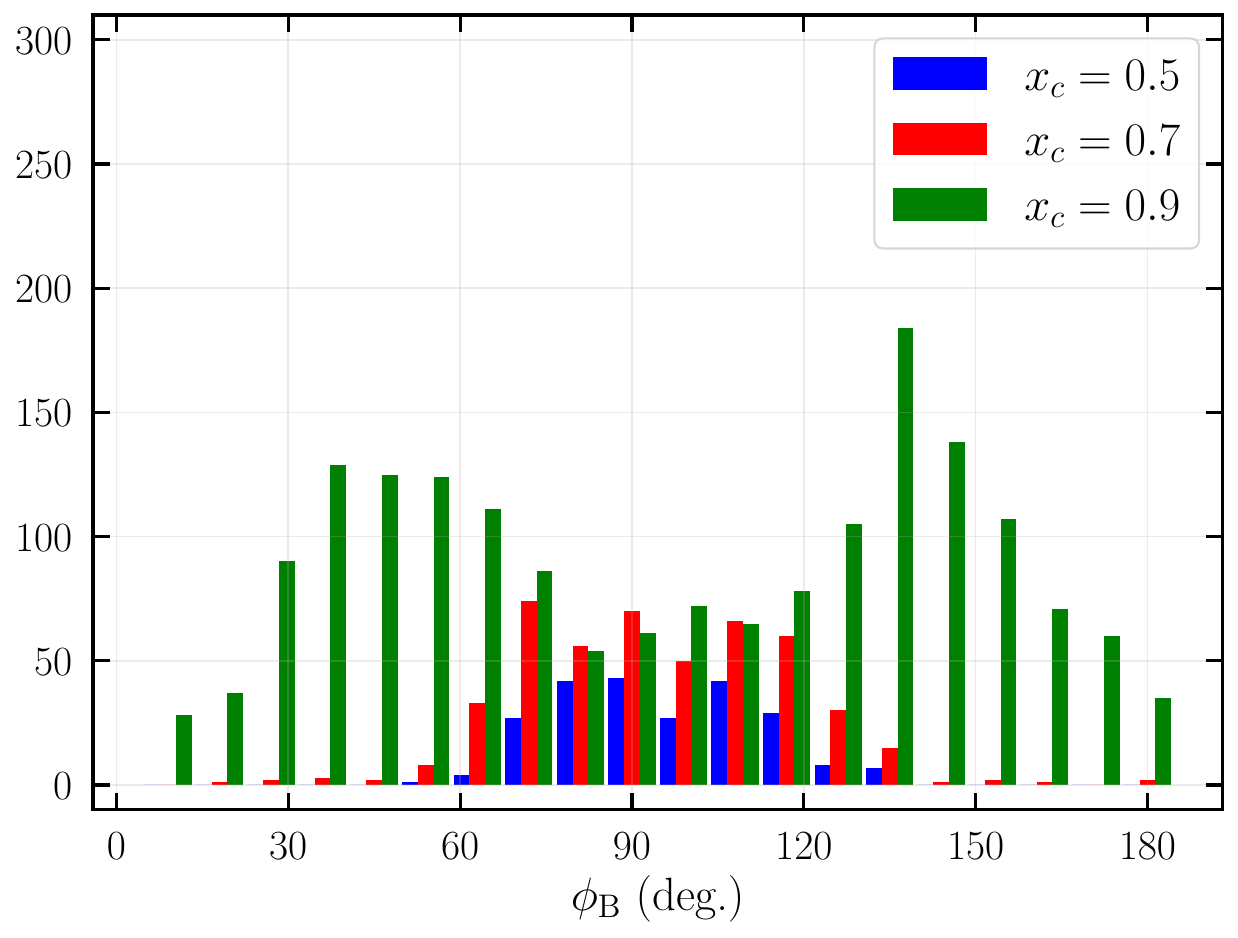}
	\put(1,77.5){(\textit{a})}
  \end{overpic}
  \begin{overpic}[width=0.475\textwidth]{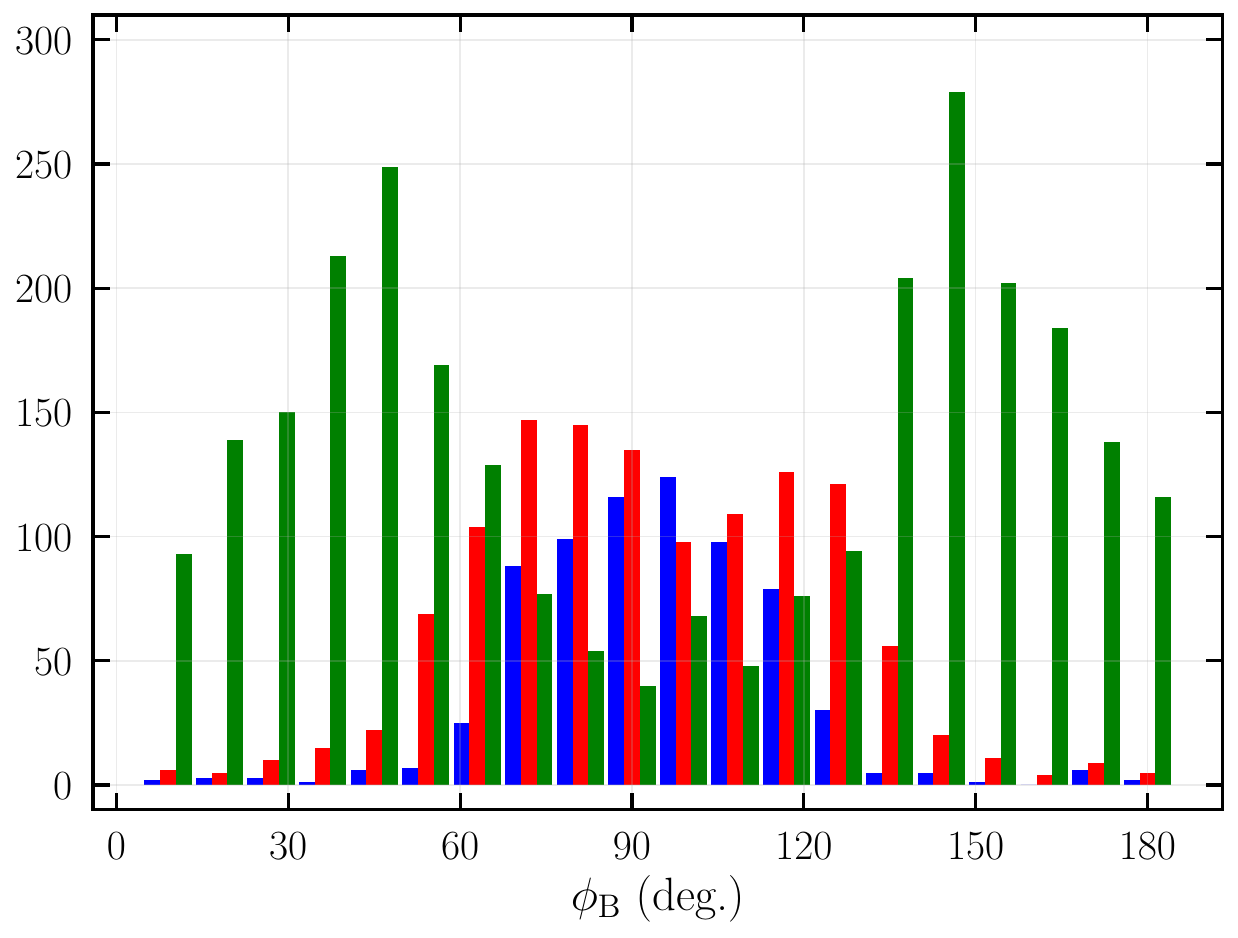}
	\put(1,77.5){(\textit{b})}
  \end{overpic}
\caption{Histogram of the vortex orientation angle generated during backflow events at chordwise positions $x_c =0.5$, $0.7$, and $0.9$. Results are shown for angles of attack of (\textit{a}) $9$ deg. and (\textit{b}) $12$ deg, respectively.}
\label{fig:histograma}
\end{figure}

The interaction between high-speed structures and low-speed streaks was investigated by \citet{brandt2008}, who analyzed both symmetric and asymmetric collisions of these structures. They observed that in the symmetric case, a lambda-vortex formed within the interaction region due to the  shear layer induced by both structures. In contrast, the asymmetric collision generated a quasi-streamwise vortex along the interaction region, also driven by the strong shear layer. This latter mechanism closely resembles our observations, where an intense shear zone produced by the interaction between a high-speed structure and a low-speed streak results in a quasi-streamwise vortex that becomes stretched and tilted as the event evolves. In the context of backflow events, a similar mechanism was also reported by \citet{guerrero2022} in a turbulent pipe flow, and the resulting vortical structures were also identified by \citet{lenaers2012} and \citet{cardesa2019} in a turbulent channel flows. Notably, the mechanism responsible for the backflow event does not appear to be affected by the presence of the APG. However, the probability of the event increases with higher APG magnitudes. Moreover, the orientation angle of the quasi-streamwise vortex generated during the event is strongly influenced by the APG magnitude. 
To assess this effect, the orientation angle is computed for all cases analyzed, and the results are presented in figure \ref{fig:histograma}. As the APG magnitude increases, the vortical structures tend to further align with the spanwise direction. This behavior can be attributed to the size differences between the high-speed structure and the low-speed streak under varying APG conditions. As shown in figure \ref{fig:bk_3d_12}, the high-speed structure in the stronger APG case is larger compared with the weaker APG case shown in figure \ref{fig:bk_3d_9}; conversely, the low-speed streak exhibits the opposite trend. Inspection of other chordwise positions for both angles of attack (summarized in the Appendix \ref{appB}) confirms that this tendency persists with an increasing APG magnitude. The resulting size disparity between the interacting structures likely alters their interplay dynamics, thereby affecting the vortex orientation angle. 

\subsubsection{Dynamics of coherent structures associated with extreme positive events}\label{sec:ep_struc}

\begin{figure}
 \centering
 \begin{overpic}[width=0.32\textwidth]{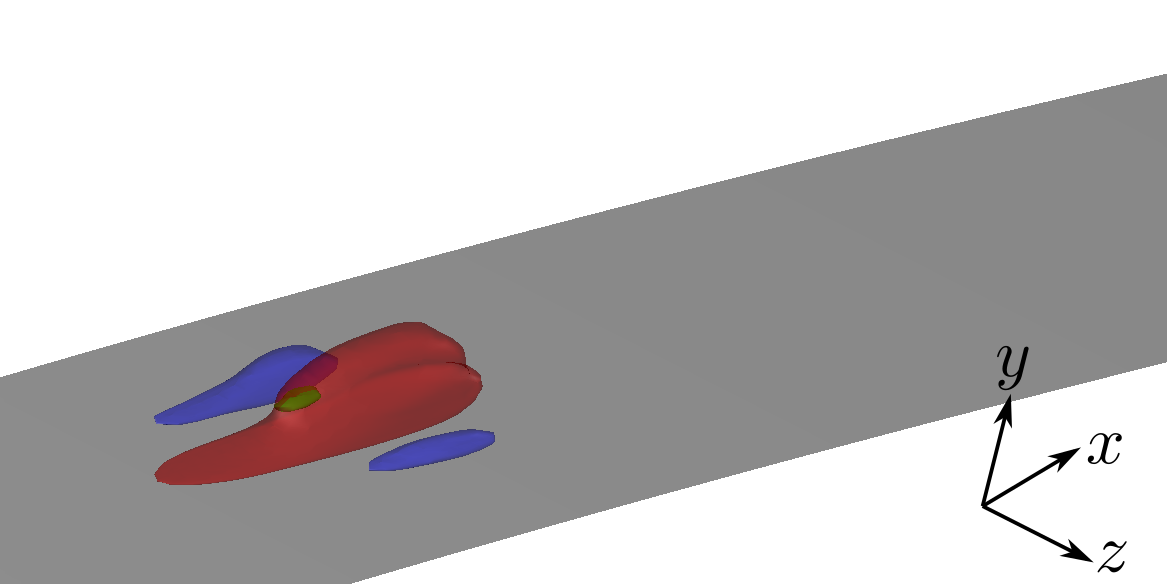}
	\put(1,43){(\textit{a})}
  \end{overpic}
  \begin{overpic}[width=0.32\textwidth]{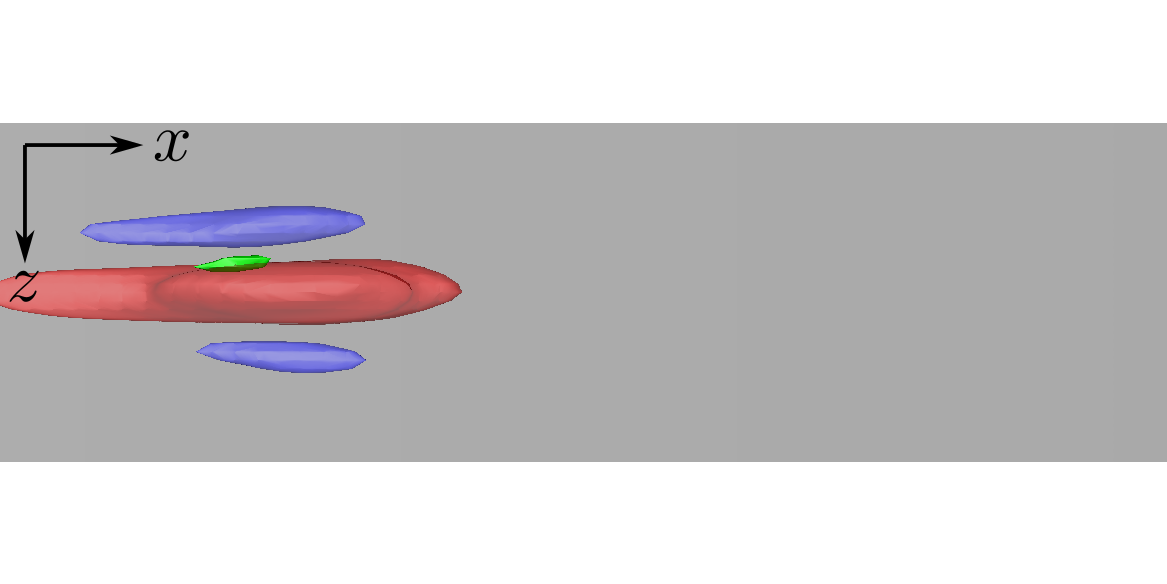}
          \put(27.5,42){$\Delta t_0 = -0.061$}
  \end{overpic}
    \begin{overpic}[width=0.32\textwidth]{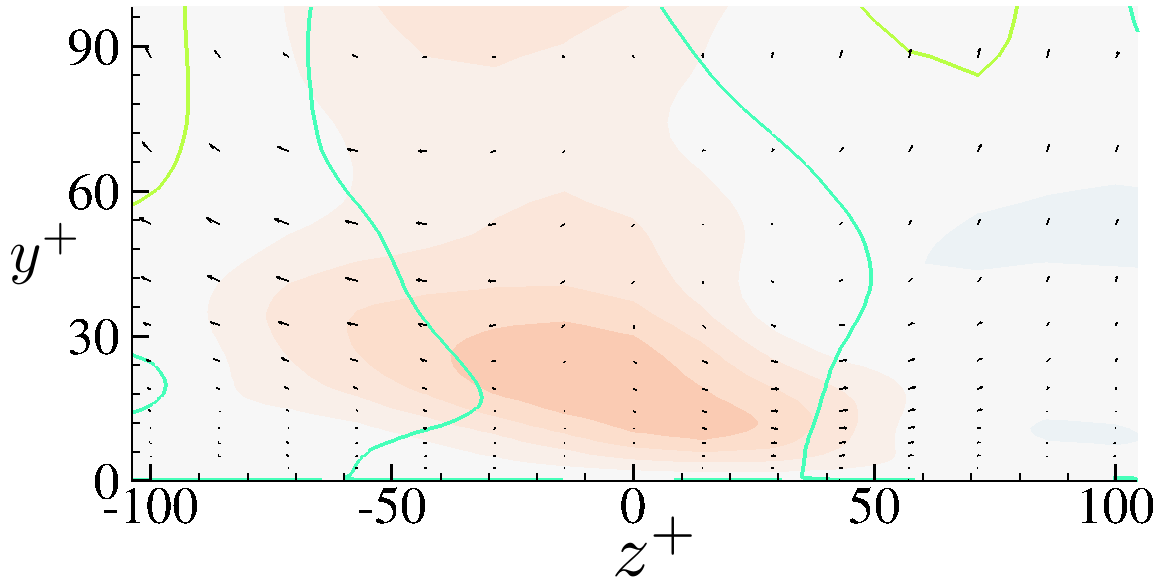}
  \end{overpic}
   \begin{overpic}[width=0.32\textwidth]{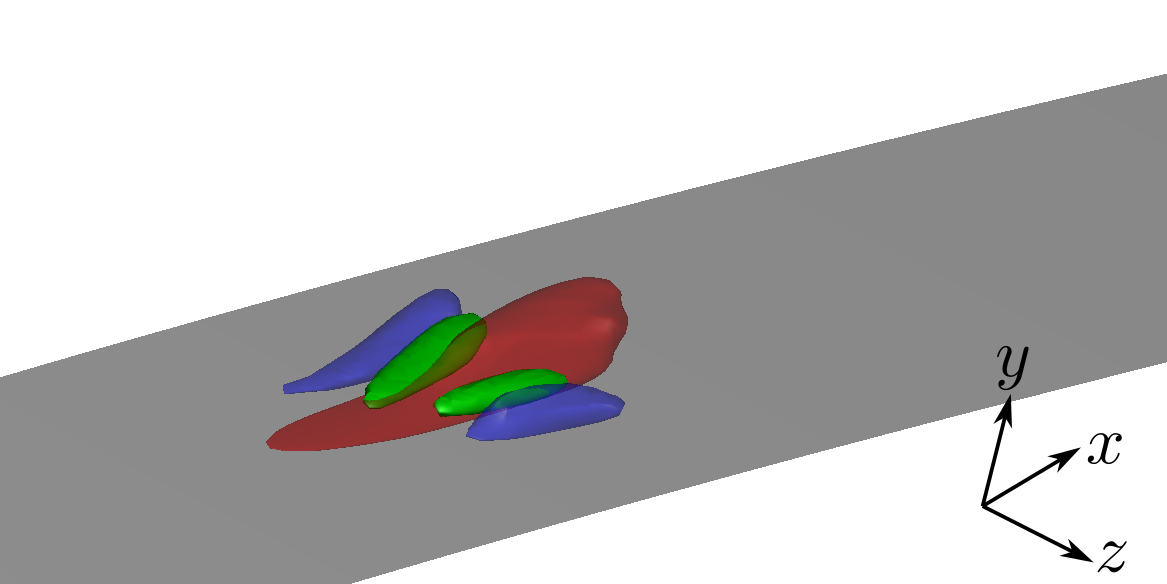}
	\put(1,43){(\textit{b})}
  \end{overpic}
  \begin{overpic}[width=0.32\textwidth]{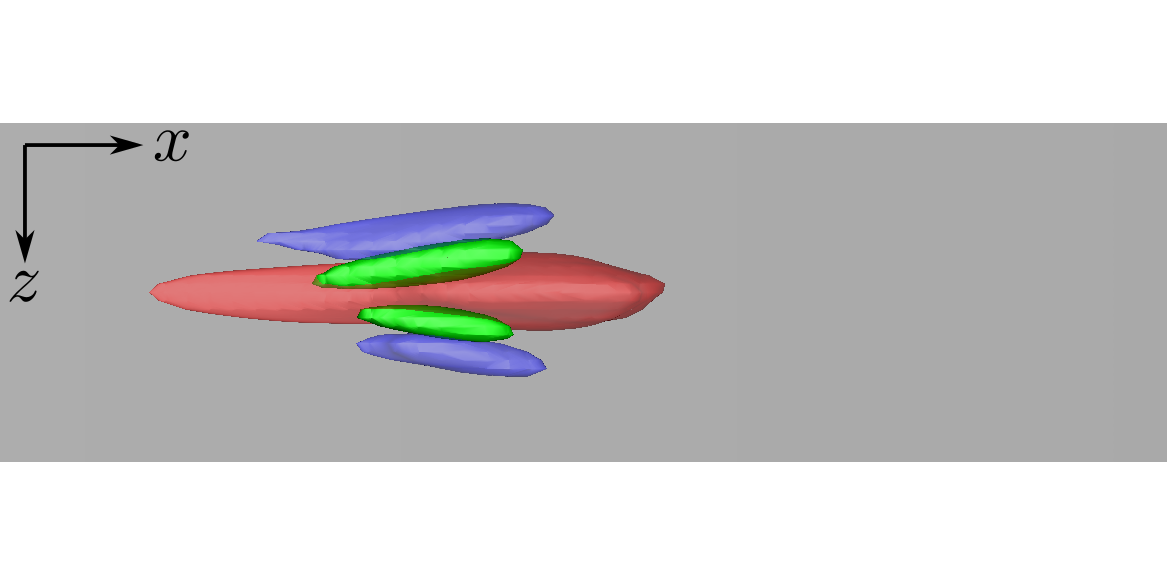}
          \put(27.5,42){$\Delta t_0 = -0.030$}
  \end{overpic}
    \begin{overpic}[width=0.32\textwidth]{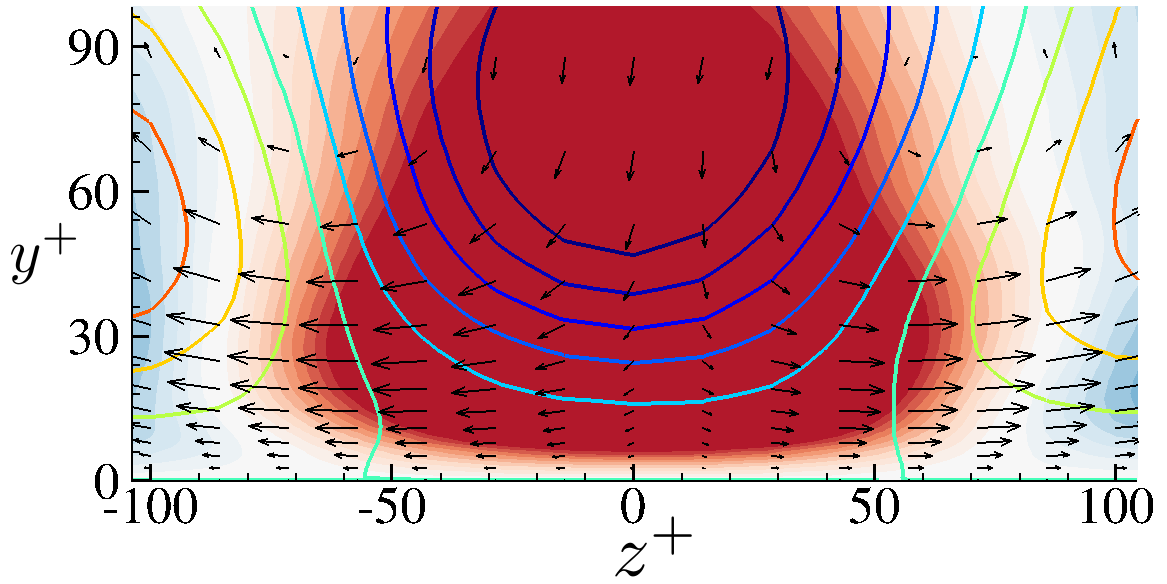}
  \end{overpic}
   \begin{overpic}[width=0.32\textwidth]{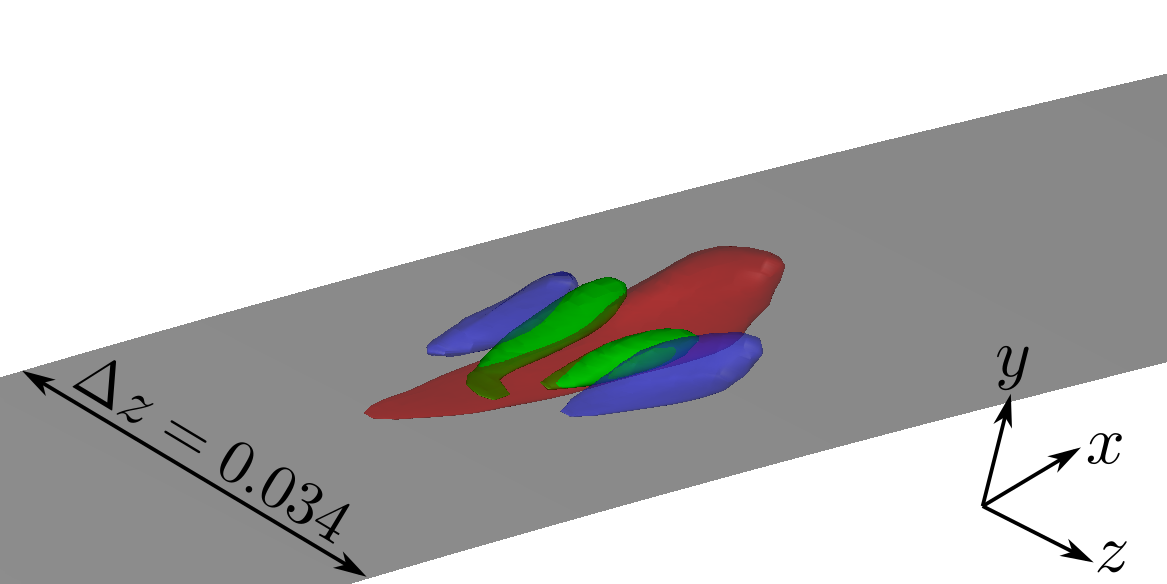}
	\put(1,43){(\textit{c})}
  \end{overpic}
  \begin{overpic}[width=0.32\textwidth]{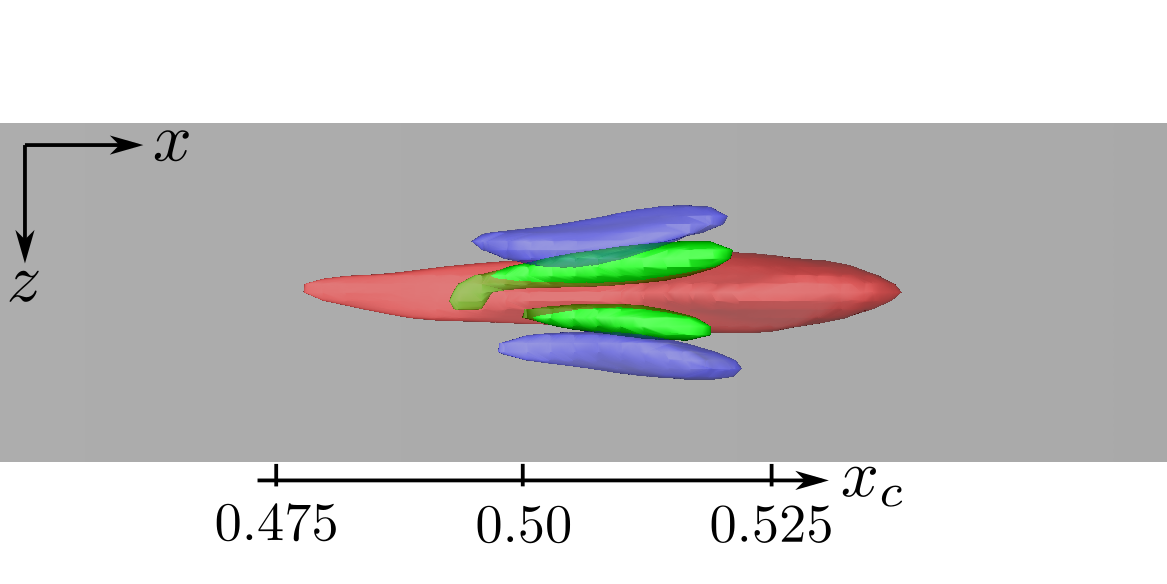}
          \put(27.5,42){$\Delta t_0 = 0.000$}
  \end{overpic}
    \begin{overpic}[width=0.32\textwidth]{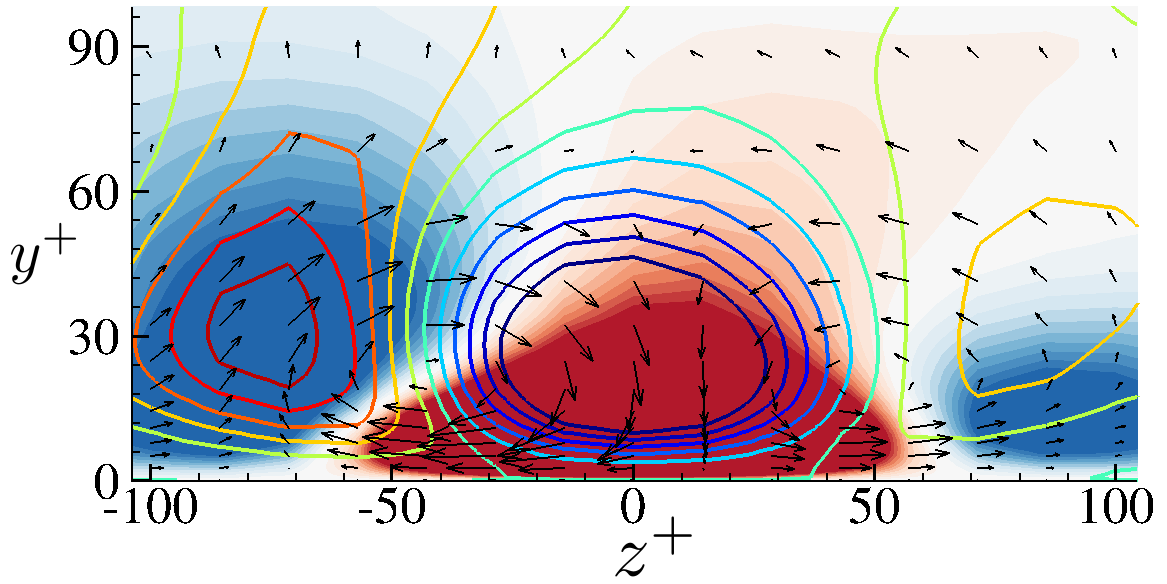}
  \end{overpic}
   \begin{overpic}[width=0.32\textwidth]{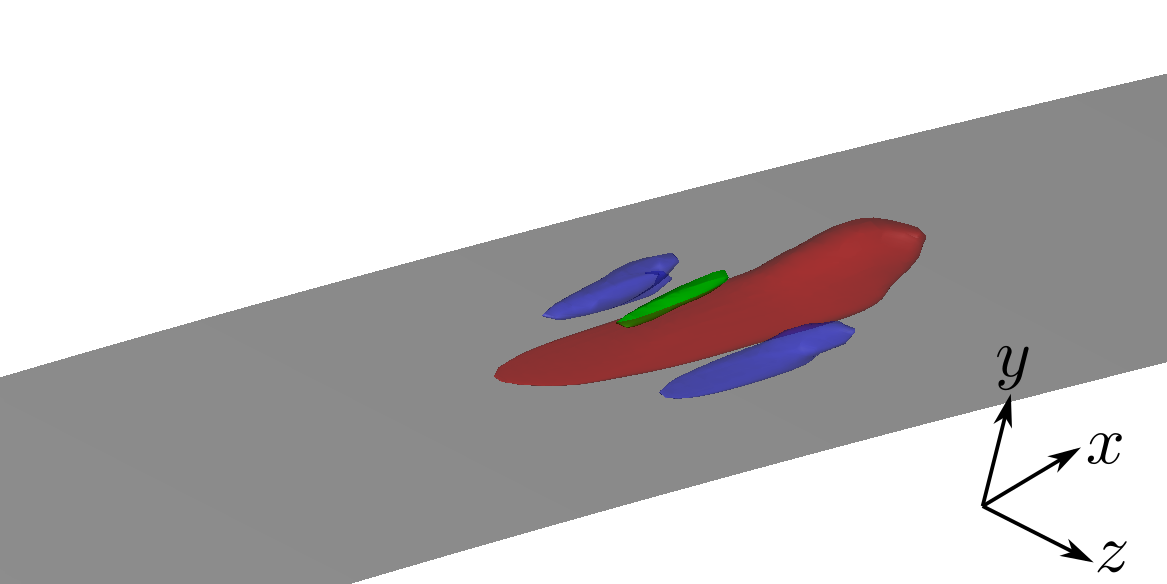}
	\put(1,43){(\textit{d})}
  \end{overpic}
  \begin{overpic}[width=0.32\textwidth]{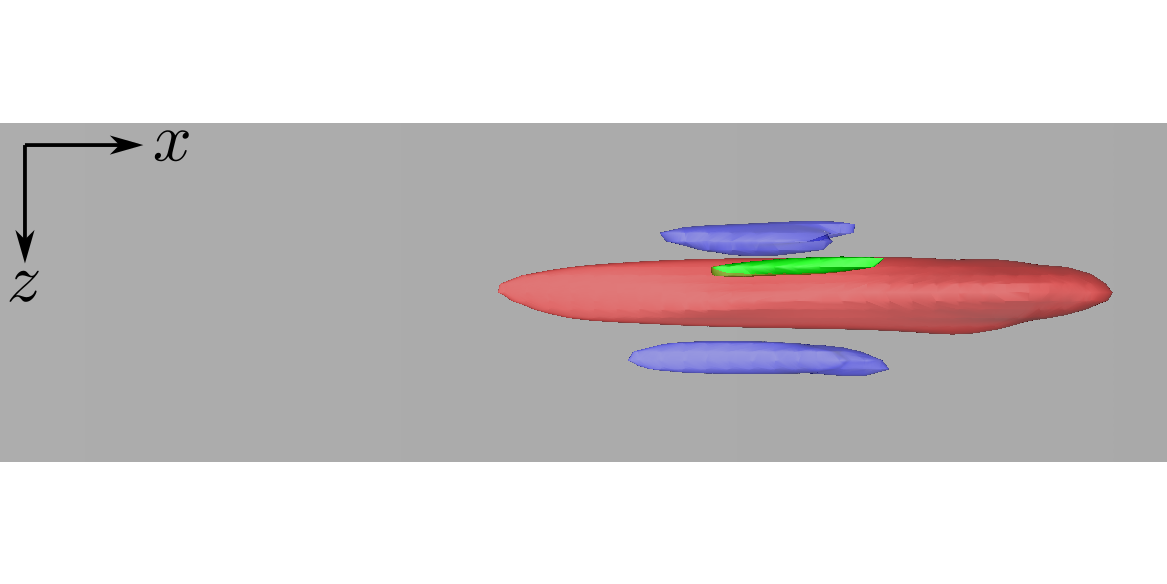}
          \put(27.5,42){$\Delta t_0 = 0.030$}
  \end{overpic}
    \begin{overpic}[width=0.32\textwidth]{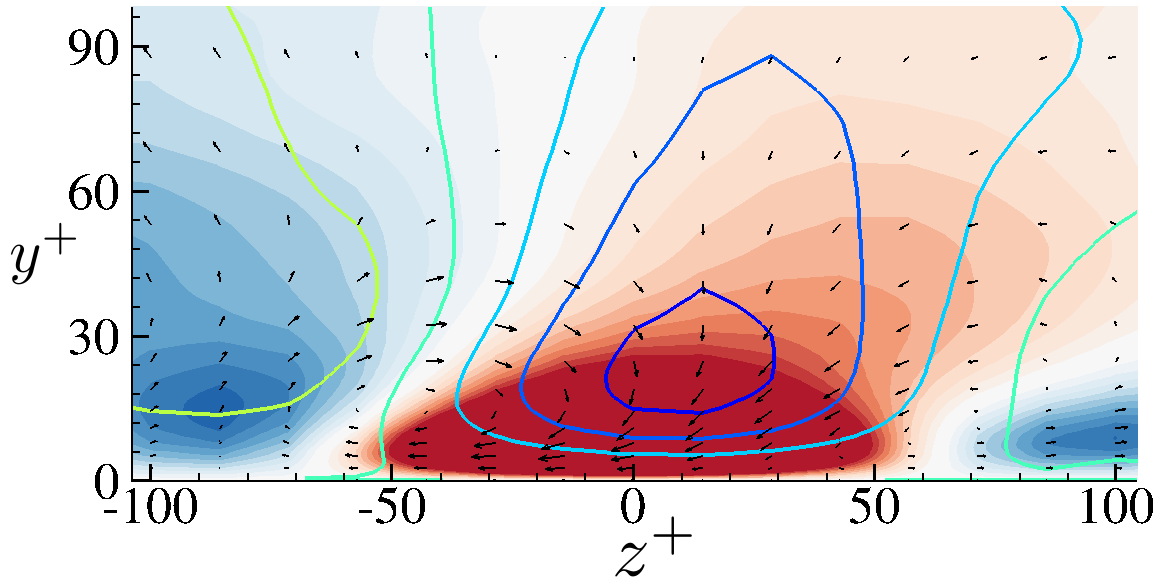}
  \end{overpic}
   \begin{overpic}[width=0.32\textwidth]{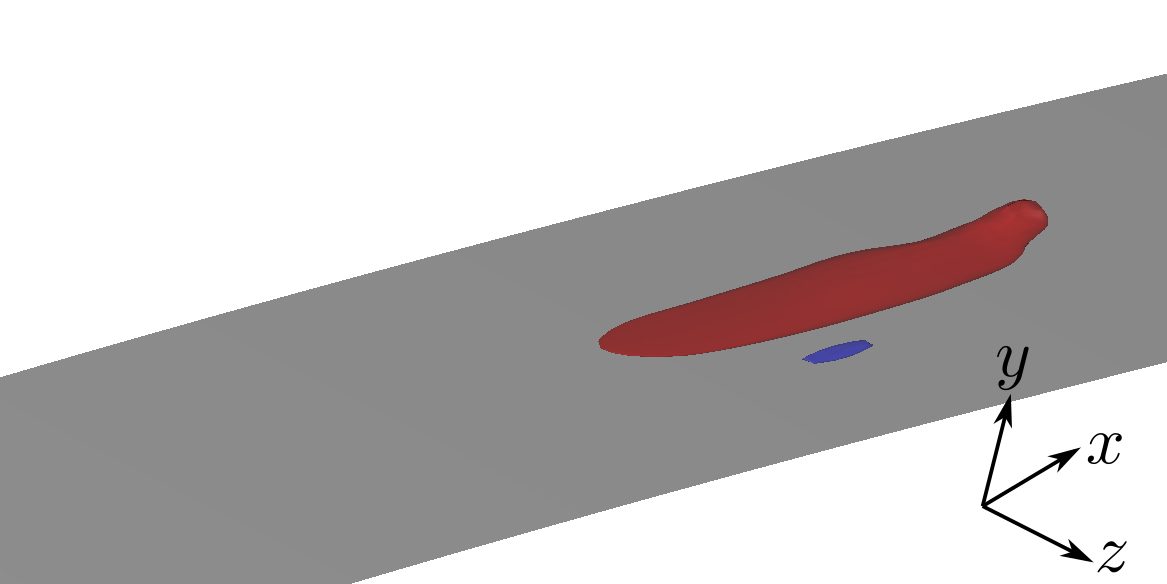}
	\put(1,43){(\textit{e})}
  \end{overpic}
  \begin{overpic}[width=0.32\textwidth]{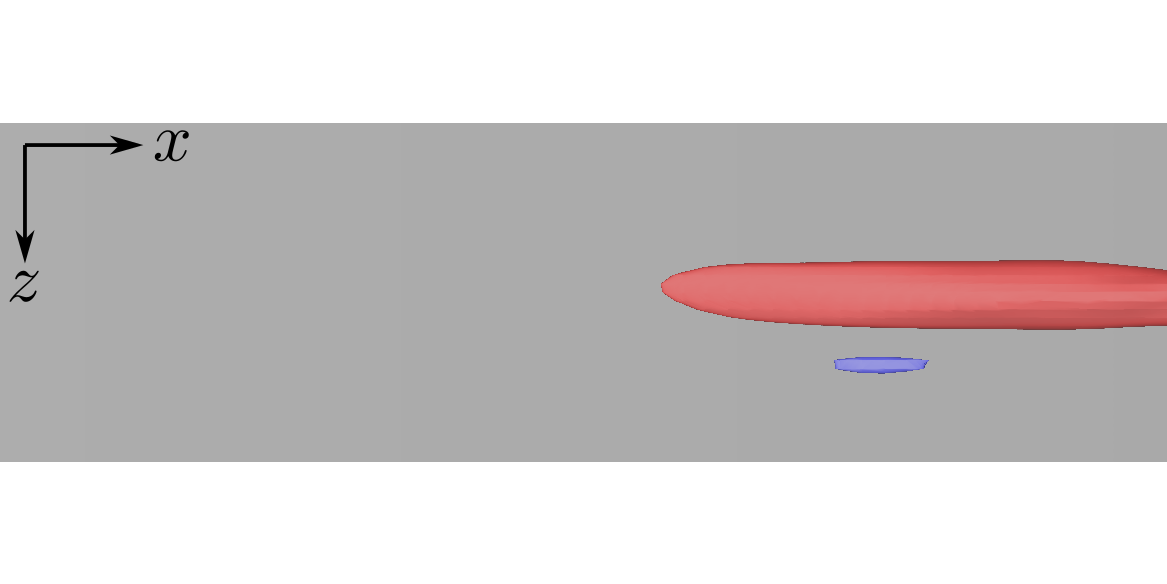}
        \put(27.5,42){$\Delta t_0 = 0.061$}
  \end{overpic}
    \begin{overpic}[width=0.32\textwidth]{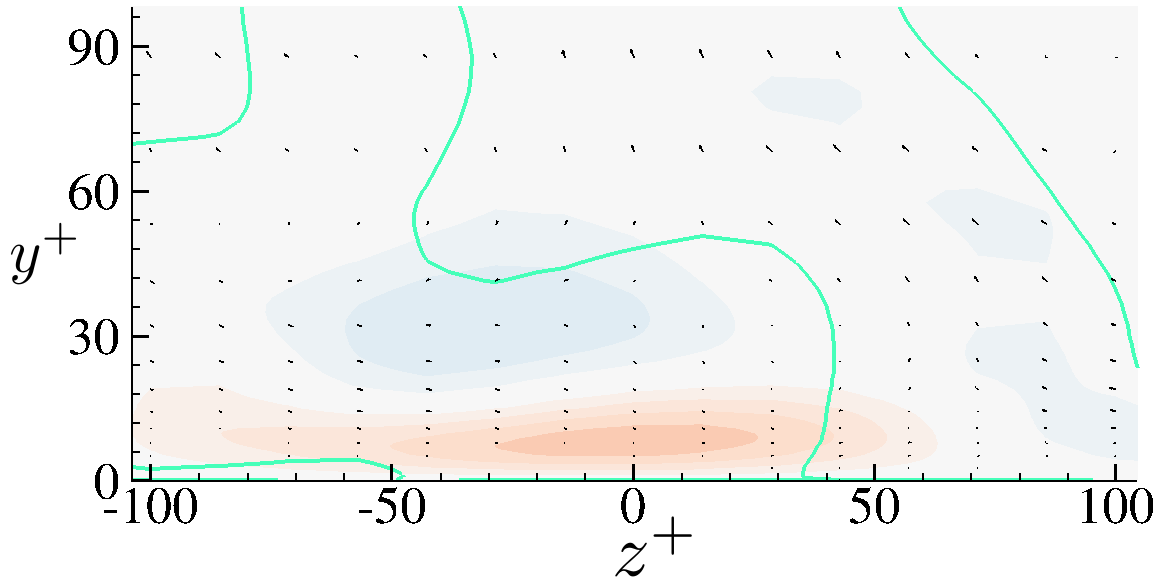}
  \end{overpic}
\begin{overpic}[width=0.15\textwidth]{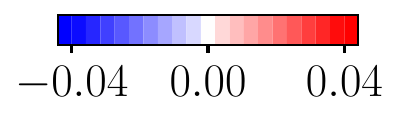}
         \put(-30,17.5){$\widetilde{u}_t/U_\infty$}
  \end{overpic}
  \hspace{1.5cm}
  \begin{overpic}[width=0.15\textwidth]{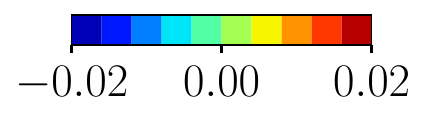}
         \put(-29.5,17.5){$\widetilde{u}_n/U_\infty$}
  \end{overpic}
\caption{Three-dimensional evolution of the coherent structures associated with the extreme positive event for the $9$ deg. angle of attack case at chordwise position $x_c=0.5$. The first column shows a three-dimensional perspective, while the second column presents a top view of the event dynamics. Red and blue isosurfaces correspond to regions of tangential velocity fluctuations of $\widetilde{u}_t=0.035$ and $\widetilde{u}_t=-0.035$, respectively, highlighting a high-speed streak flanked by low-speed streaks. A vortex structure is visualized using a $Q$-criterion isosurface colored in green. The third column shows a $z^+-y^+$ 
plane passing through the event center, where the background contours represent the tangential velocity fluctuations $\widetilde{u}_t$ and the ovelaid isolines denote the wall-normal velocity fluctuations $\widetilde{u}_n$. Movie 3 is provided as supplementary material showing the evolution of coherent structures and their interactions for this case.}
\label{fig:ep_3d_9}
\end{figure}

The turbulent structures associated with the extreme positive events are examined in figure  \ref{fig:ep_3d_9}. This figure presents the temporal evolution of the three-dimensional conditionally averaged structures corresponding to events identified predominantly on the left side of the $9$ deg. angle of attack case at chordwise position $x_c = 0.5$. The selected time instants capture the complete development of the event. The first and second columns show, respectively, a three-dimensional perspective view and a top view of the coherent structures. In these visualizations, the red and blue isosurfaces represent high- and low-speed streaks, denoted by tangential velocity fluctuations $\widetilde{u}_t = 0.035$ and $\widetilde{u}_t = -0.035$. The green isosurfaces highlight vortical structures identified using the $Q$-criterion. The third column in the figure displays a $z^+-y^+$ plane passing through the event location ($x_c=0.5$), where the background contours represent the tangential velocity fluctuations, while the isolines depict the wall-normal velocity fluctuations.
The arrows in the plots indicate the conditionally averaged velocity vectors composed of the spanwise $\widetilde{W}$ and wall-normal $\widetilde{U}_n$ velocity components, as defined in equation \ref{eq:conditional_ep}.

Figure \ref{fig:ep_3d_9}(\textit{a}) shows that the event initiates with the formation of a high-speed streak flanked by two low-speed streaks, the left one being longer than the right. Visualization of the $z^+-y^+$ plane in figure \ref{fig:ep_3d_9}(\textit{b}) reveals a region of negative wall-normal velocity fluctuation at the center of the high-speed streak, indicating a strong sweep motion. Owing to mass conservation, the mean flow is deflected in the spanwise direction, giving rise to a secondary motion. At $\Delta t_0 = 0.000$, when the extreme positive event reaches its maximum amplitude, figure \ref{fig:ep_3d_9}(\textit{c}) shows that the asymmetric low-speed streaks generate intense shear layers between themselves and the high-speed streak. Concurrently, a left-side vortex strengthens, exhibiting positive streamwise vorticity that drives fluid toward the wall within the high-speed streak region and away from the wall within the adjacent low-speed region, as evidenced by the $z^+-y^+$ plane visualization. On the right side, a secondary streamwise vortex develops through a similar mechanism, but it remains weaker and plays a lesser role in the overall dynamics. 
It is noteworthy that the event originates not only from a strong sweep motion toward the wall but also from a pronounced spanwise motion in the near-wall region, consistent with the observations of \citet{sheng2009}. In the subsequent stages, the vortices sustaining the event gradually decay in intensity, whereas the high-speed streak persists.

\begin{figure}
 \centering
 \begin{overpic}[width=0.32\textwidth]{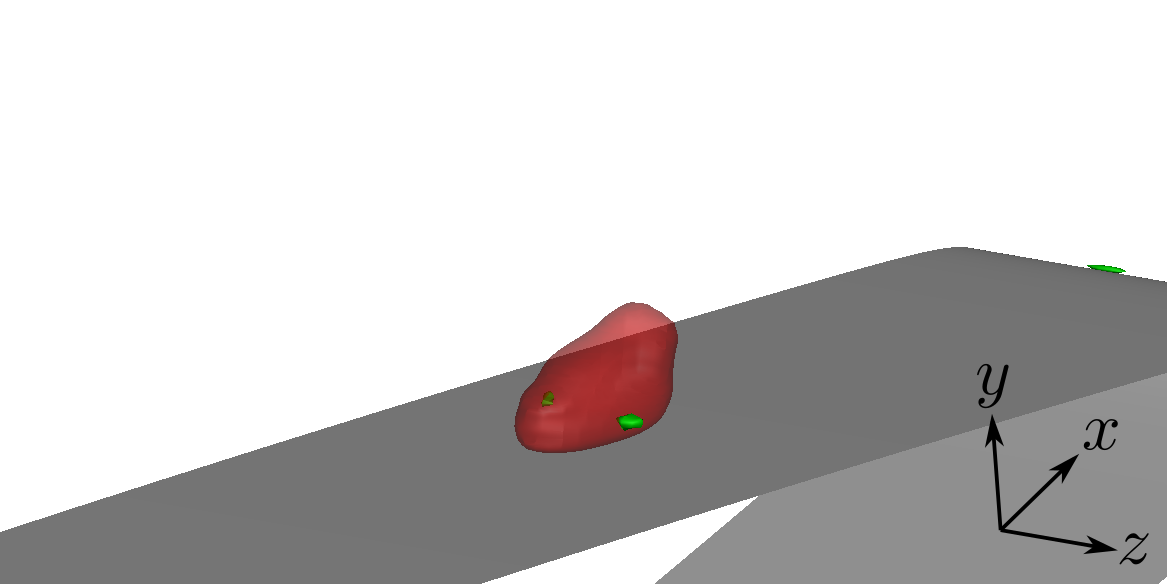}
	\put(1,43){(\textit{a})}
  \end{overpic}
  \begin{overpic}[width=0.32\textwidth]{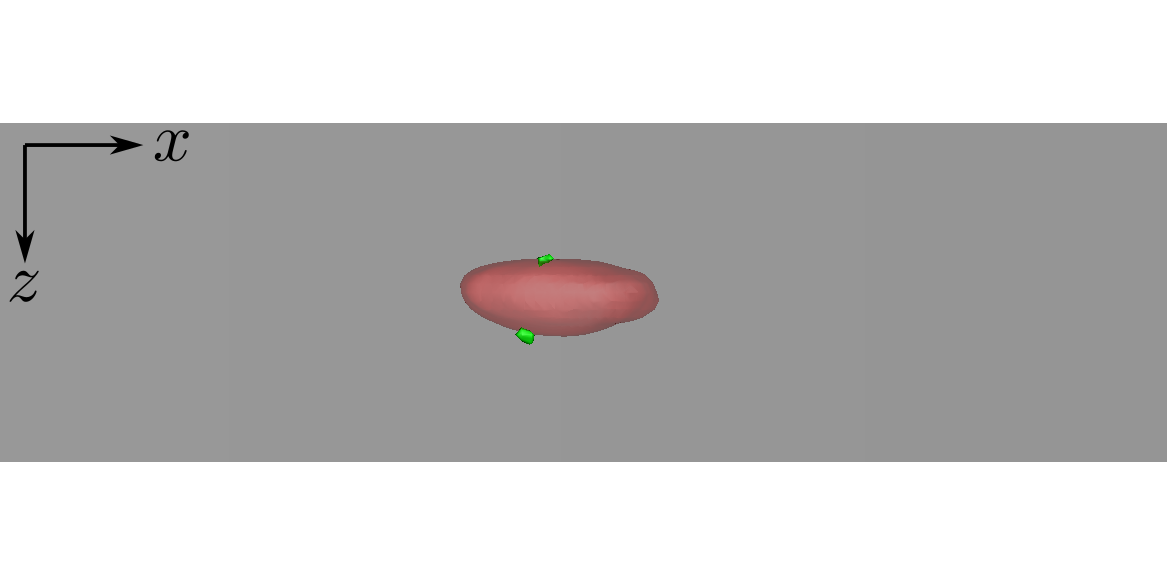}
          \put(27.5,42){$\Delta t_0 = -0.048$}
  \end{overpic}
    \begin{overpic}[width=0.32\textwidth]{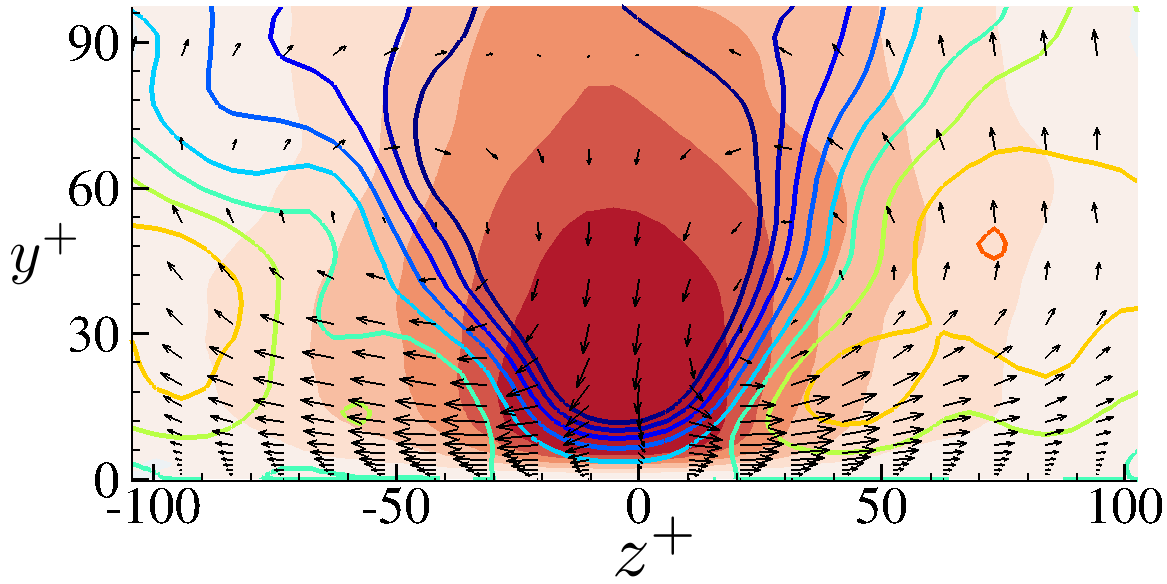}
  \end{overpic}
   \begin{overpic}[width=0.32\textwidth]{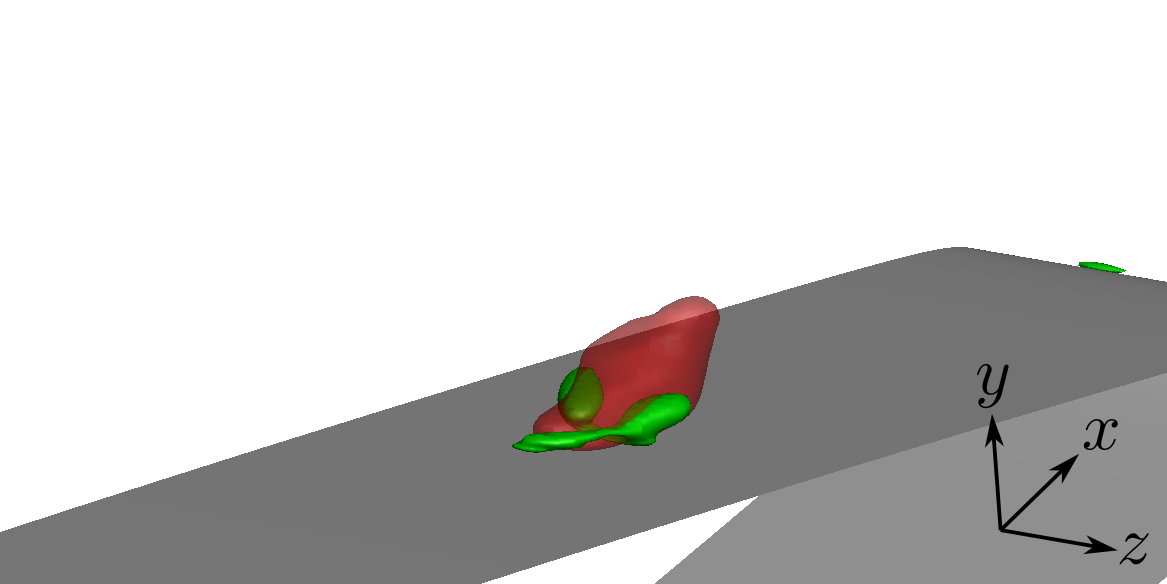}
	\put(1,43){(\textit{b})}
  \end{overpic}
  \begin{overpic}[width=0.32\textwidth]{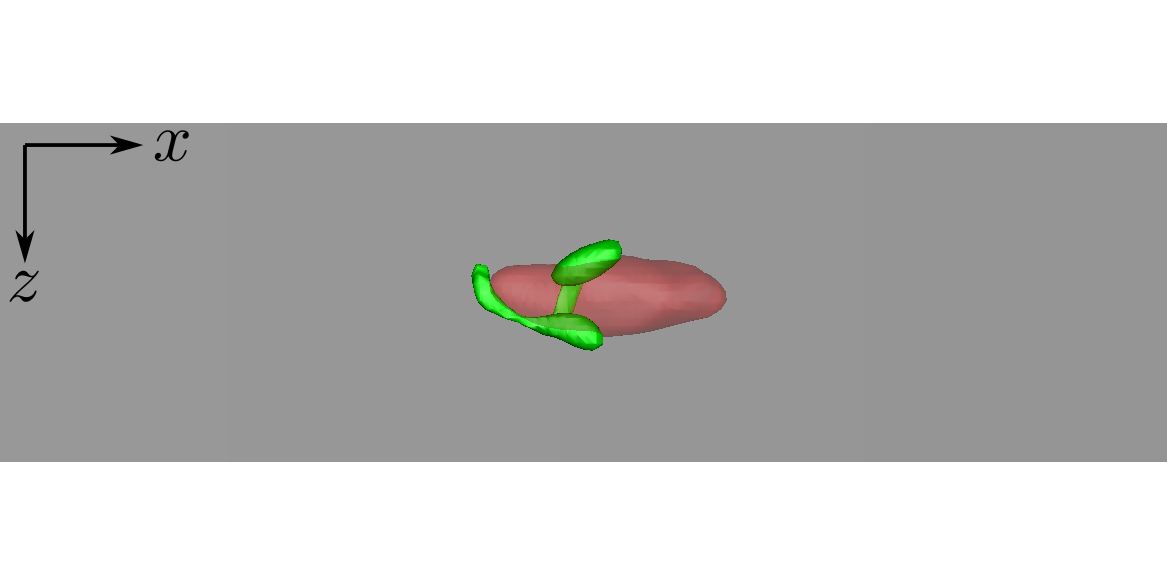}
          \put(27.5,42){$\Delta t_0 = -0.024$}
  \end{overpic}
    \begin{overpic}[width=0.32\textwidth]{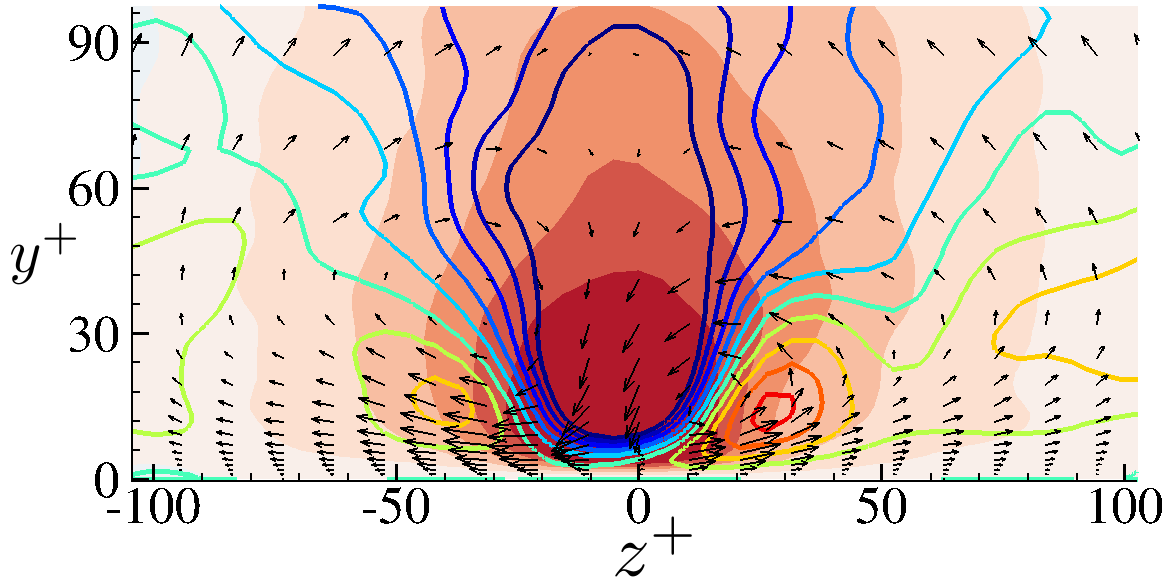}
  \end{overpic}
   \begin{overpic}[width=0.32\textwidth]{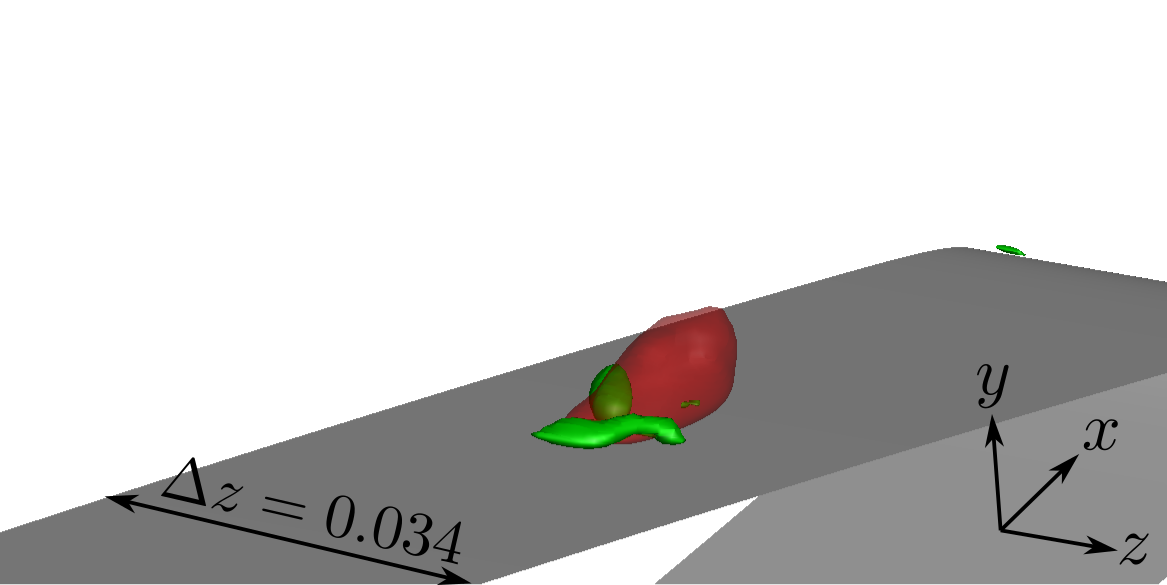}
	\put(1,43){(\textit{c})}
  \end{overpic}
  \begin{overpic}[width=0.32\textwidth]{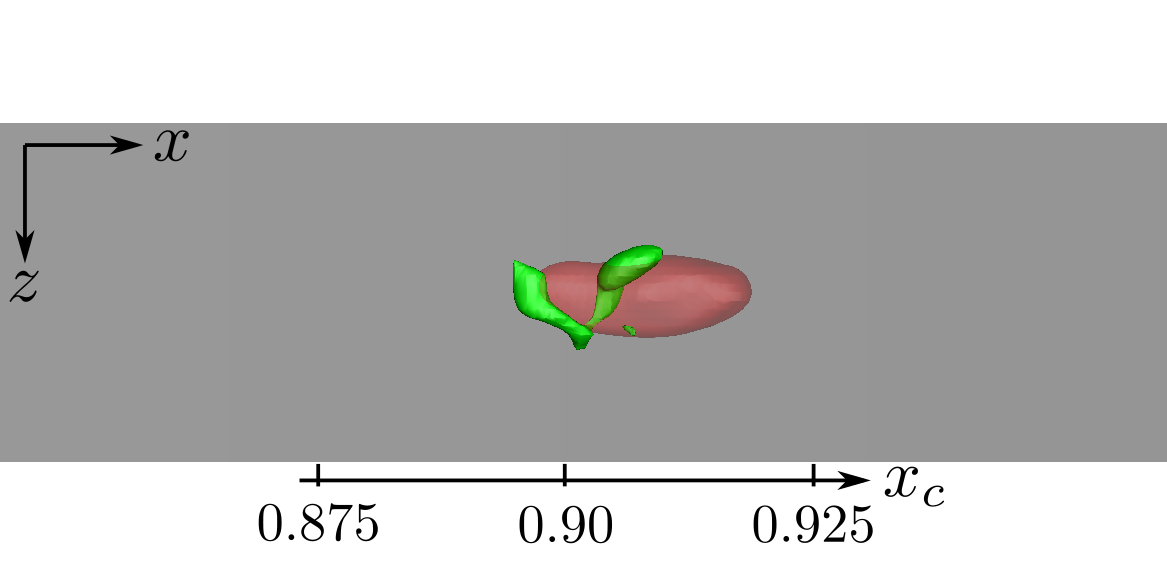}
          \put(27.5,42){$\Delta t_0 = 0.000$}
  \end{overpic}
    \begin{overpic}[width=0.32\textwidth]{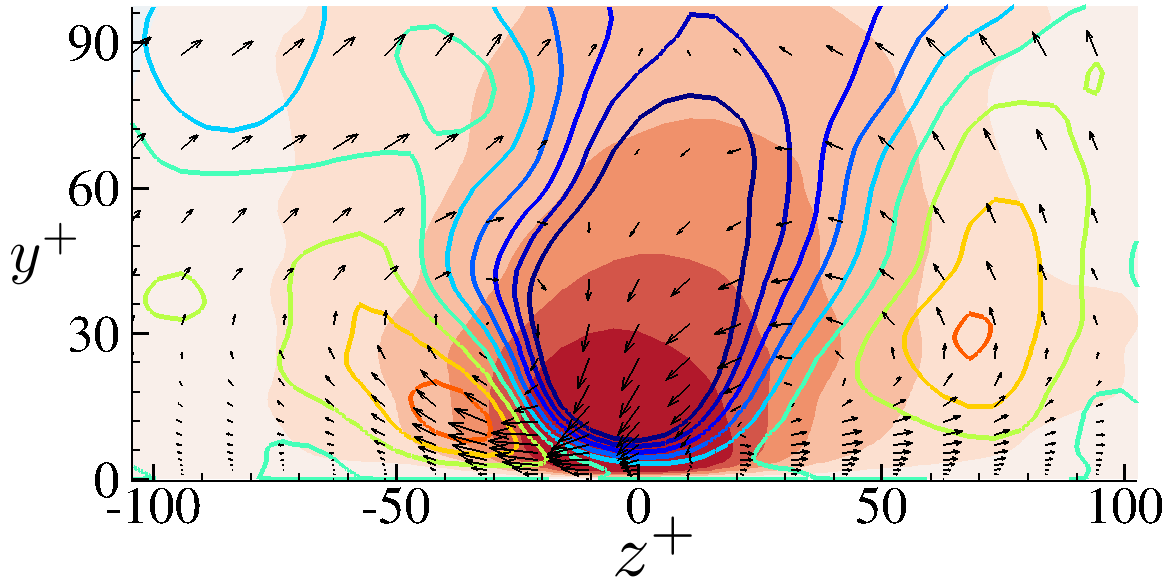}
  \end{overpic}
   \begin{overpic}[width=0.32\textwidth]{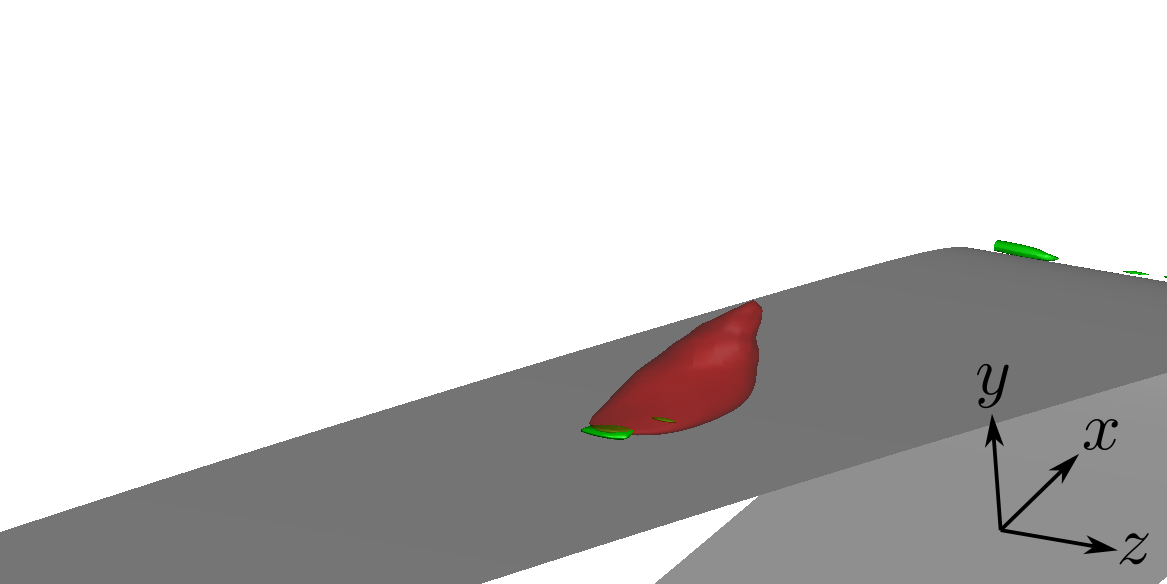}
	\put(1,43){(\textit{d})}
  \end{overpic}
  \begin{overpic}[width=0.32\textwidth]{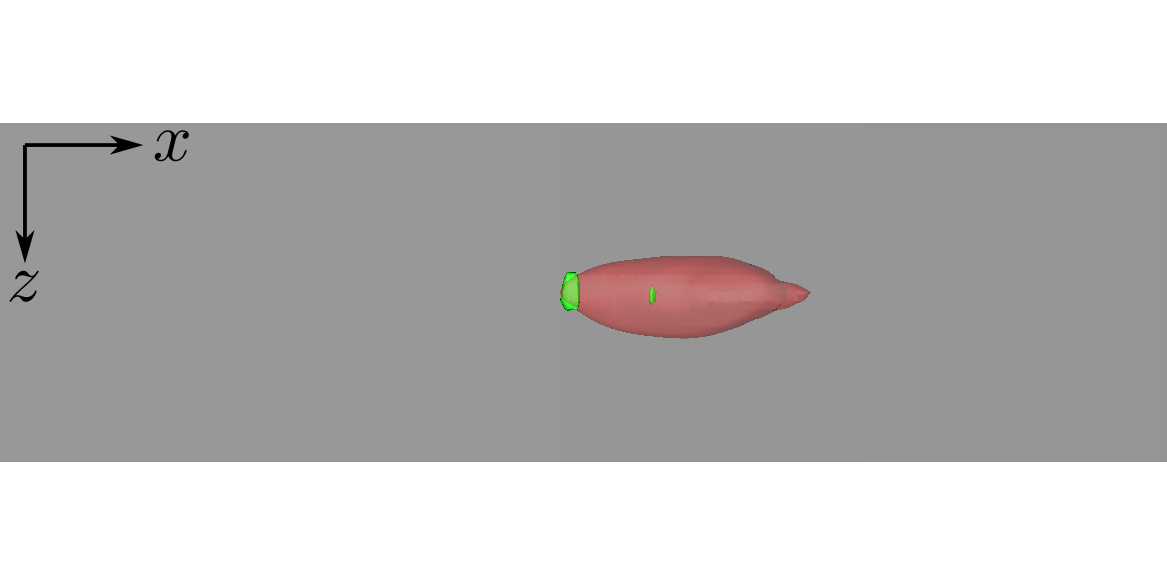}
          \put(27.5,42){$\Delta t_0 = 0.024$}
  \end{overpic}
    \begin{overpic}[width=0.32\textwidth]{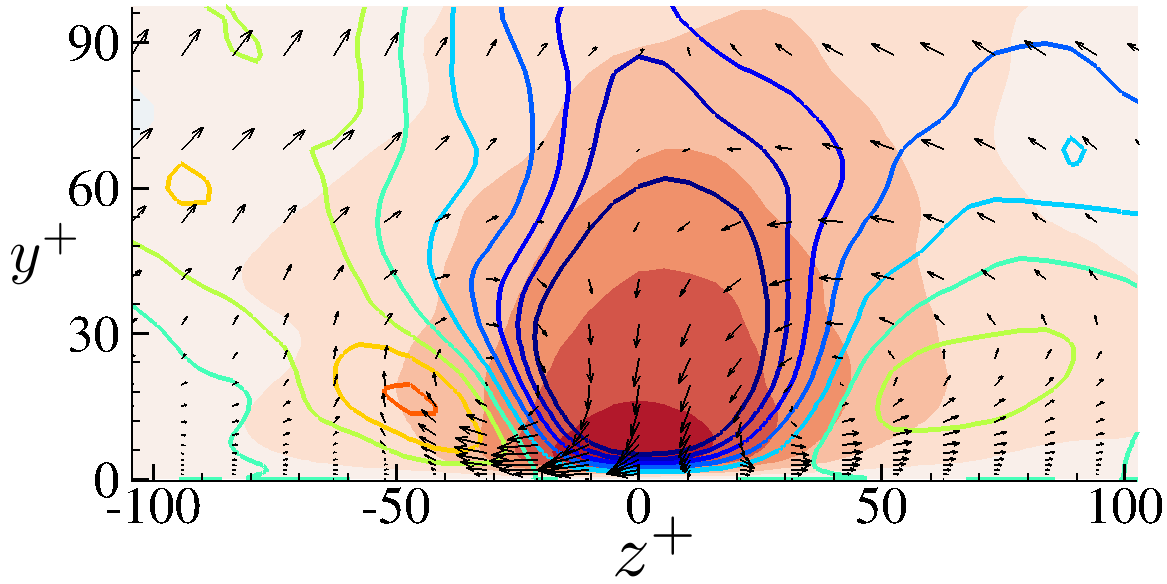}
  \end{overpic}
   \begin{overpic}[width=0.32\textwidth]{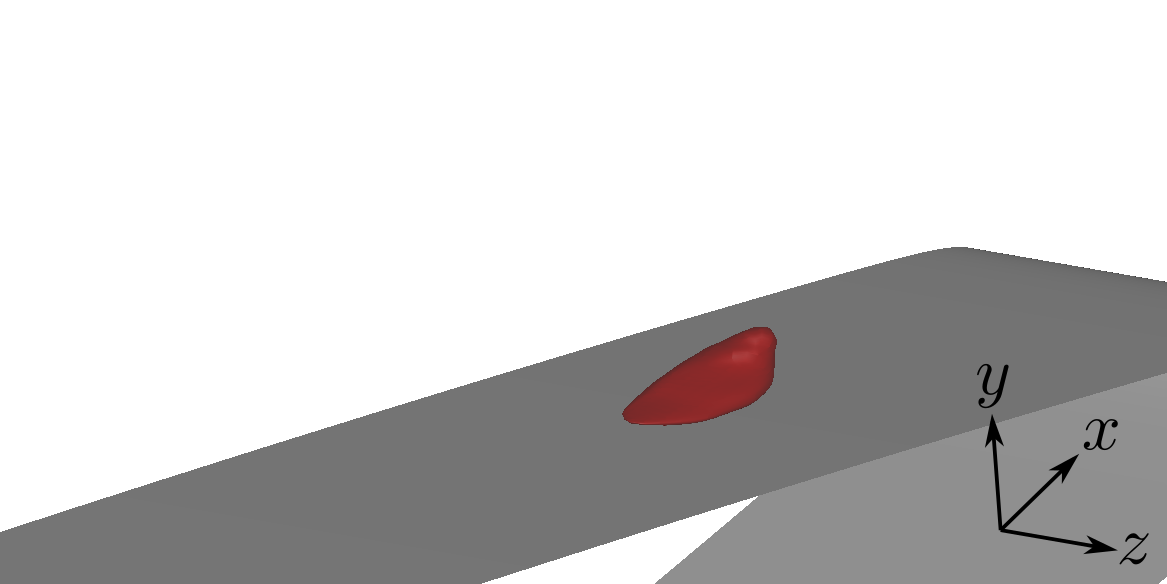}
	\put(1,43){(\textit{e})}
  \end{overpic}
  \begin{overpic}[width=0.32\textwidth]{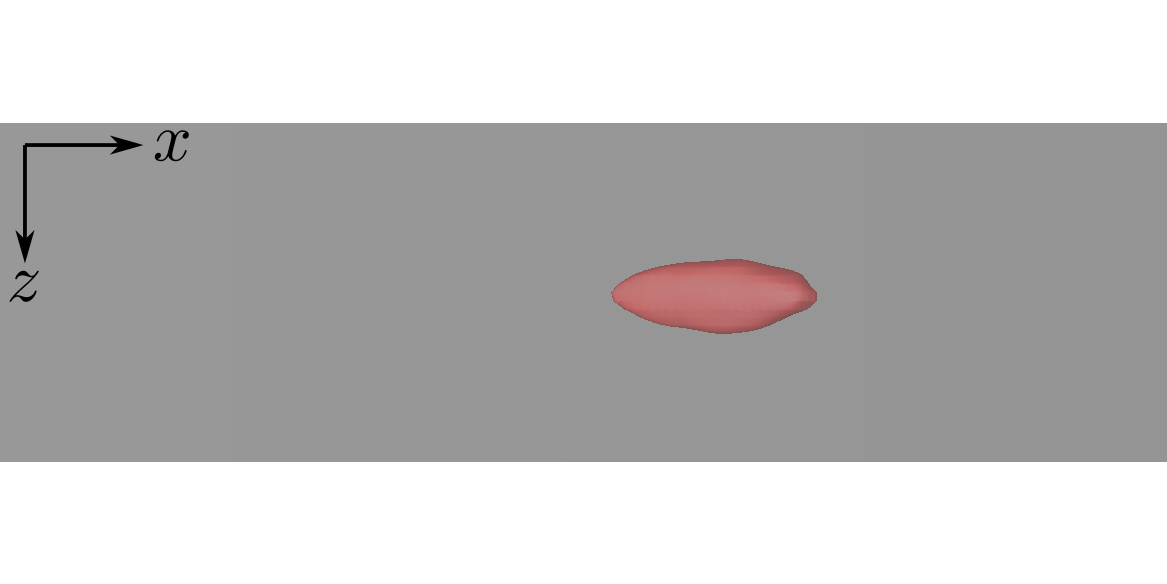}
        \put(27.5,42){$\Delta t_0 = 0.048$}
  \end{overpic}
    \begin{overpic}[width=0.32\textwidth]{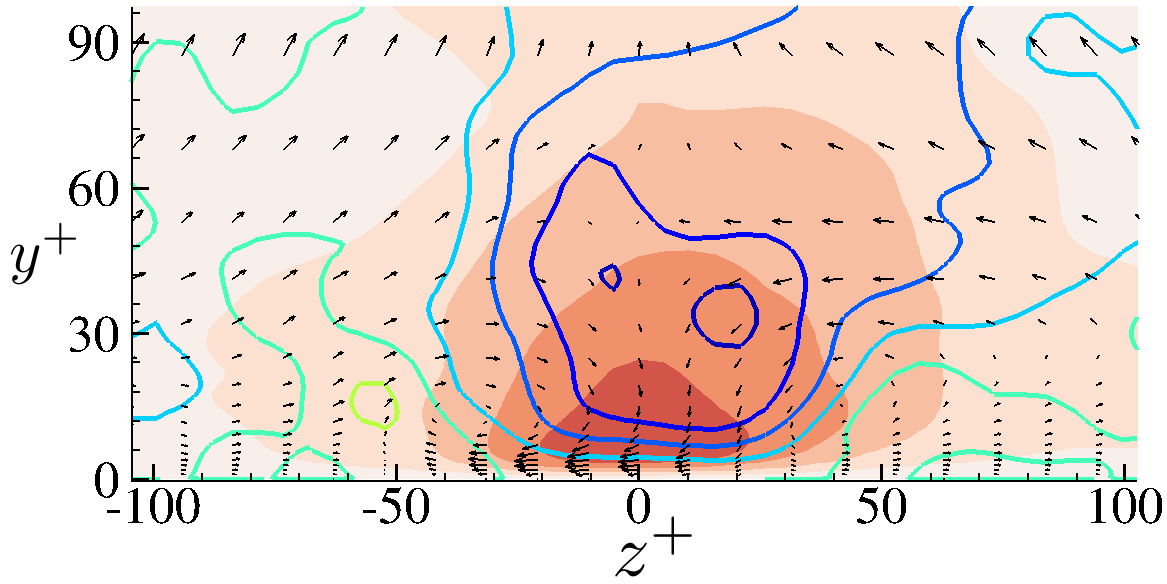}
  \end{overpic}
\begin{overpic}[width=0.15\textwidth]{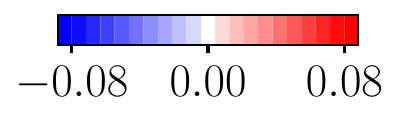}
         \put(-30,17.5){$\widetilde{u}_t/U_\infty$}
  \end{overpic}
  \hspace{1.5cm}
  \begin{overpic}[width=0.15\textwidth]{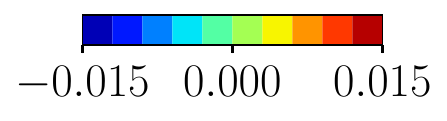}
         \put(-29.5,17.5){$\widetilde{u}_n/U_\infty$}
  \end{overpic}
\caption{Three-dimensional evolution of the coherent structures associated with the extreme positive event for the $12$ deg. angle of attack case at chordwise position $x_c=0.9$. The first column shows a three-dimensional perspective, while the second column presents a top view of the event dynamics. Red isosurfaces correspond to regions of tangential velocity fluctuations of $\widetilde{u}_t=0.08$, highlighting a short high-speed streak. Vortex structures are visualized using a $Q$-criterion isosurface colored in green. The third column shows a constant $x$ plane passing through the event center, where the background contours represent the tangential velocity fluctuations $u_t$ and the ovelaid isolines denote the wall-normal velocity fluctuations $\widetilde{u}_n$. Movie 4 is provided as supplementary material showing the evolution of coherent structures and their interactions for this case.}
\label{fig:ep_3d_12}
\end{figure}

The temporal evolution of the extreme positive event for the $12$ deg. angle of attack case at chordwise position $x_c=0.9$ is presented in figure \ref{fig:ep_3d_12}. The visualization follows the same format as the previous figure; however, the red isosurfaces represent regions of tangential velocity fluctuation $\widetilde{u}_t = 0.08$, while the green isosurfaces identify vortical regions extracted using the $Q$-criterion, generated by flow interactions during the event. A key difference from the $9$ deg. case is immediately evident: low-speed streaks are not observed. In figure \ref{fig:ep_3d_12}(\textit{a}), a high-speed streak is clearly visible, although it is significantly shorter than in the lower APG case. As shown in the $z^+-y^+$ plane, this structure is accompanied by a negative wall-normal velocity fluctuation at its center, indicating a strong sweeping motion. The plane centered on the event further reveals that this sweep motion induces a spanwise velocity component in the near-wall region, while positive wall-normal velocity fluctuations appear along the streak flanks. As shown in figure \ref{fig:ep_3d_12}(\textit{b}), the interaction between the spanwise and wall-normal motions generates small vortical structures on both sides of the high-speed streak. Subsequently, figure \ref{fig:ep_3d_12}(\textit{c}) shows that the right-hand vortex lifts away from the wall and rapidly dissipates, whereas the left-hand vortex remains attached and active throughout the event, sustained by the intensification of the sweep and strengthening of the induced spanwise motion. At the event peak, a quasi-spanwise vortex forms at the rear of the high-speed streak, leaving a distinct footprint in the space–time correlation shown in figure \ref{fig:corr_ep}(\textit{l}). This structure induces a positive wall-normal velocity that contributes to the local reduction of wall shear stress immediately after the event. Finally, figures \ref{fig:ep_3d_12}(\textit{d,e}) show that the vortical structures gradually weaken, while the high-speed streak persists, marking its dominant role in the overall event dynamics.

A comparison between figures \ref{fig:ep_3d_9} and  \ref{fig:ep_3d_12} reveals that the APG strongly influences the dynamics of the extreme positive events. A key difference is the absence of low-speed streaks flanking the high-speed streak in the $12$ deg. case. \citet{guerrero2020} reported that, in turbulent pipe flows, extreme positive events typically arise from the side-by-side interaction of low- and high-speed streaks. This behavior is consistent with the $9$ deg. case shown in figure \ref{fig:ep_3d_9}, where the flow experiences only a mild APG. Analysis of the other chordwise positions for both angles of attack, shown in the Appendix \ref{appB}, confirms that the contribution of the low-speed streaks to the event dynamics progressively diminishes as the APG increases. This trend is associated with changes in the coupling between wall-normal and spanwise velocities \citep{sheng2009}, as evidenced by the $z^+-y^+$ 
plane visualizations for both cases. In the strongest APG condition investigated, the spanwise velocity component becomes increasingly relevant in the overall dynamics. This feature likely explains the more spread out probability density function of the wall shear stress orientation observed by \citet{silva2024_aiaa} for the present flow configurations, also consistent with the findings of \citet{vinuesa2017}.

\section{Conclusions} \label{sec:conclusions}

Adverse pressure gradient turbulent boundary layers (APG-TBLs) are investigated with emphasis on the dynamics of extreme wall shear stress events. The flow configurations analyzed correspond to a NACA0012 airfoil at angles of attack of 9 and 12 deg, for which varying streamwise adverse pressure gradients develop along the suction side. The datasets are obtained from wall-resolved large eddy simulations performed at a chord-based Reynolds number of $Re_c = 4\times10^5$ and Mach number $M = 0.2$. We begin the analysis with a characterization of the mean flow, including the streamwise pressure-gradient parameter and skin-friction coefficient, followed by detailed balances of the streamwise and wall-normal momentum equations and a quadrant decomposition of the Reynolds shear stress. Subsequently, the occurrence of extreme wall shear stress events is examined using space-time correlations and conditional statistics, which are employed to analyze the temporal evolution of the conditional mean velocity fields and their corresponding Reynolds shear stress profiles. The turbulent coherent structures responsible for the generation of the extreme events are then identified and characterized through conditional flow field analysis, and the effects of mild and strong APGs on their mutual interaction and dynamics are investigated.

Due to the high angles of attack, the TBLs develop over progressively stronger APGs. Different chordwise locations are examined corresponding to mild, moderate and strong APG conditions. The strongest APG occurs near the trailing edge of the 12 deg. case, for which the pressure gradient rises sharply. Despite the high angles of attack, the mean flow remains attached along the airfoil suction side. Analysis of the streamwise mean momentum balance equation shows that the Reynolds shear stress term becomes progressively more intense in the buffer region as the APG increases. In this region, the turbulence fluctuations remove energy from the mean flow field. In a similar fashion, both the streamwise and wall-normal mean flow convection terms become more prominent with the APG in the outer region of the TBL. While the former term reflects a streamwise flow deceleration, the latter contributes to an outward momentum gain, this characteristic being described by other studies of APG-TBLs as a wall-normal convection effect. 

A quadrant decomposition of the Reynolds shear stress shows that sweeps and ejections are the dominant mechanisms responsible for momentum exchange between the mean flow and the turbulent fluctuations. The contributions of these motions increase with the APG in both the inner and outer layers. In the inner layer, momentum is transferred from the mean flow to the fluctuation field, whereas the opposite is observed in the outer layer, near the edge of the boundary layer. We also observe that sweeps are more prominent near the wall, and intensify more rapidly with the APG than ejections. The adverse pressure gradient also affects the inward and outward interactions of the Reynolds shear stress, which become more pronounced for stronger APGs. These terms are responsible for transferring energy from the fluctuation field to the mean flow, being relevant only in the inner layer.

Probability density functions of the wall shear stress indicate that the frequency of backflow events increases with the strength of the APGs. The distributions are skewed toward positive wall shear events, but become more symmetric as the APG intensifies. A threshold of $|\tau_w'/\tau_{w_{\text{rms}}}| > 2$ is adopted to identify extreme positive and negative (backflow) wall shear events. Space-time correlations between the instantaneous wall shear stress and velocity components in the viscous sublayer are used to characterize the spatiotemporal scales of the events. For both BF and EP events, the characteristic timescales increase with APG strength, consistent with the overall flow deceleration induced by stronger adverse pressure gradients.

Conditional averages of velocity components and Reynolds stress profiles are computed to examine the temporal evolution of the coherent motions associated with the extreme events. Combined with the space-time correlations, these results show that BF events are initiated by a sweep motion originating in the inner layer, which transports high-momentum fluid toward the wall. This motion is accompanied by inward interactions near the wall that precede the onset of backflow, leading to a local deceleration within the viscous sublayer. As the event develops, the wall-directed motions are replaced by ejections and outward interactions. Although the overall dynamics of BF events are similar across conditions, the strength of the APG affects the dynamics of Reynolds shear stress components. Under weaker APGs, strong ejections emerge at the onset of the BF event and briefly exceed the sweep contribution. However, near the peak intensity of the event, sweeps dominate the inner-layer dynamics before being overtaken again by ejections during the decay phase. In contrast, for stronger APGs, sweeps remain dominant throughout most of the BF event, with ejections becoming slightly more pronounced near the wall only after the event termination. A similar analysis of extreme positive events shows that sweeps are the dominant mechanism, although their interaction with outward interactions modulates the intensity of these events. 

The interplay between the coherent turbulent structures associated with EP and BF events is further investigated using conditional flow field analysis. Backflow events are initiated by high-speed coherent structures from the inner layer that interact with near-wall low-speed streaks through a sweep motion that intercepts the tail of the streaks. In the rear portion of the streak, a positive wall-normal velocity fluctuation develops, inducing an ejection motion. These interactions generate vortical structures that stretch and tilt as the extreme event evolves. Following this stage, the low-speed streak and the sweep motion of the high-speed structure weaken, while the vortex structure tilts in the streamwise direction before dissipating. The analyses for mild and strong APGs indicate that the fundamental mechanisms governing the backflow dynamics remain essentially unchanged with the APG intensity. However, stronger APGs increase the probability of BF events and modify the orientation of the associated vortical structures, which become more spanwise aligned. Under stronger APGs, the high-speed structures involved in the sweep motion grow in size, whereas the low-speed streaks become shorter. On the other hand, under weaker APGs, the low-speed streaks are flanked by high-speed streaks, and the resulting vortex structures are more inclined in the streamwise direction and more elongated. In this regime, the interacting high-speed structures that approach the low-speed streak from above are considerably smaller.

Extreme positive events are more strongly influenced by APGs. Under weak pressure gradients, these events initiate with the formation of a high-speed streak flanked by low-speed streaks. A sweep motion drives the high-speed streak toward the wall, and mass conservation deflects the flow spanwise, generating a secondary motion. The sideline low-speed streaks are typically of unequal size, producing an asymmetric interaction with the high-speed streak. This interaction generates asymmetric, streamwise-oriented vortices that gradually weaken, while the high-speed streak persists. Under strong APG conditions, however, EP events no longer exhibit flanking low-speed streaks. Moreover, the streamwise vortices are smaller and tend to merge forming a spanwise vortical structure which induces a positive wall-normal velocity, locally reducing the wall shear stress immediately after the event. In this regime, the high-speed streaks are considerably shorter, and the associated sweep motions generate spanwise velocities that play an increasingly important role in the near-wall dynamics, influencing the orientation of the wall shear stress.


\backsection[Supplementary data]{\label{SupMat} Movies and figures are submitted as supplementary material together with the manuscript.}


\backsection[Funding]{The authors acknowledge the financial support received from Fundação de Amparo à Pesquisa do Estado de São Paulo, FAPESP, under grant 2021/06448-0. FAPESP is also acknowledged for the scholarship provided to the first author under grant No. 2024/09969-9. The authors also thank the Center for Computing in Engineering and Sciences at Unicamp (FAPESP grant 2013/08293-7) for providing the computer time in the Coaraci Supercomputer (FAPESP grant 2019/17874-0).}

\backsection[Declaration of interests]{The authors report no conflict of interest.}

\backsection[Data availability statement]{The data that support the findings of this study can be made available upon request.} 

\backsection[Author ORCIDs]{L. Silva, https://orcid.org/0000-0001-5896-3553; W. Wolf, https://orcid.org/0000-0001-8207-8466}

\backsection[Author contributions]{L.S.: conceptualization, methodology, software, validation, investigation, data curation, writing, visualization; W.W.: conceptualization, methodology, writing, supervision, project administration, funding acquisition.}

\appendix

\section{Tensor calculus and generalized curvilinear coordinates}\label{appA}

In the present work, the flow governing equations are written in generalized curvilinear coordinates ($\xi^j, j = 1,2,3$). In this form, the variation of the basis vectors is accounted for in the covariant derivative \citep{aris1989}, which defines a derivative along the tangent vectors of a manifold. This is obtained by applying the product rule to a vector represented by contravariant components, as follows:
\begin{equation}\label{eq:derivative_vector}
\begin{aligned}
    \partial_j\mathbf{A} &= \partial_j(A^i\mathbf{E}_i)\\
    &= \mathbf{E}_i\partial_jA^i + A^i\partial_j\mathbf{E}_i \quad &\text{(product rule)}\\
    &= \mathbf{E}_i\partial_jA^i + A^i\Gamma_{ij}^k\mathbf{E}_k\quad &\text{ (where } \partial_j\mathbf{E}_i = \Gamma_{ij}^k\mathbf{E}_k\text{)} \\
    &= \mathbf{E}_i\partial_jA^i + A^k\Gamma_{kj}^i\mathbf{E}_i\quad &\text{(relabeling indices $i$ and $k$)} \\
    &= (\partial_jA^i + A^k\Gamma_{kj}^i)\mathbf{E}_i \\
    &= (A^i)_{,j}\mathbf{E}_i \mbox{ .}
\end{aligned}
\end{equation}
Here, $A^i$ is the contravariant component of vector $\mathbf{A}$, expressed as $A^i = A_j\pdv{\xi^i}{x^j}$. The covariant basis vector is $\mathbf{E}_i = \pdv{\mathbf{x}}{\xi^i}$, 
and $\Gamma_{kj}^i$ is the Christoffel symbol of the second kind, defined as:
\begin{equation}
    \Gamma_{jk}^i = \dfrac{g^{il}}{2}(\partial_kg_{jl} + \partial_jg_{kl} -\partial_lg_{jk}) \mbox{ .}
\end{equation}
The covariant and contravariant metric tensors are given by $g_{ij} = \pdv{\xi^i}{x^k}\pdv{\xi^j}{x^k}$ and $g^{ij} = \pdv{x^k}{\xi^i}\pdv{x^k}{\xi^j}$, respectively. 
As shown by \citet{aris1989}, the covariant derivative of a contravariant tensor of arbitrary order can be written as  
\begin{equation}
A^{ij\dots k}_{,q} = \partial_qA^{ij\dots k} +\Gamma_{aq}^iA^{aj\dots k} + \Gamma_{aq}^jA^{ia\dots k} +\cdots +\Gamma_{aq}^kA^{ij\dots a} \mbox{ .}
\end{equation}
The flow governing equations are obtained by writing their components in contravariant form along with their respective basis vectors. This is followed by the evaluation of the covariant derivatives, as shown in Eq. \ref{eq:derivative_vector}. 
Once this procedure is completed, the vector components can be rewritten in terms of their physical representation as $\hat{A}^i = \sqrt{g_{ii}}A^i$. It is worth noting that, for mutually orthogonal covariant basis vectors, the metric tensor diagonal satisfies $g^{ii} = 1/g_{ii}$.
Below, the covariant derivatives of first- and second-order tensors are presented for the mass and momentum fluxes, as they appear in Eqs. \ref{eq:continuity_equation} and \ref{eq:momentum_equation}
\begin{equation}
    (\rho u^j)_{,j} = \partial_j(\rho u^j) +\Gamma_{kj}^j\rho u^k \mbox{ ,}
\end{equation}
\begin{equation}
    (\rho u^iu^j)_{,j} = \partial_j(\rho u^iu^j) +\Gamma_{kj}^i\rho u^ku^j +\Gamma_{kj}^j\rho u^iu^k \mbox{ .}
\end{equation}

\section{Side-orientation dependence of conditionally averaged coherent structures}\label{appB}

\begin{figure}
 \centering
    \begin{overpic}[width=0.32\textwidth]{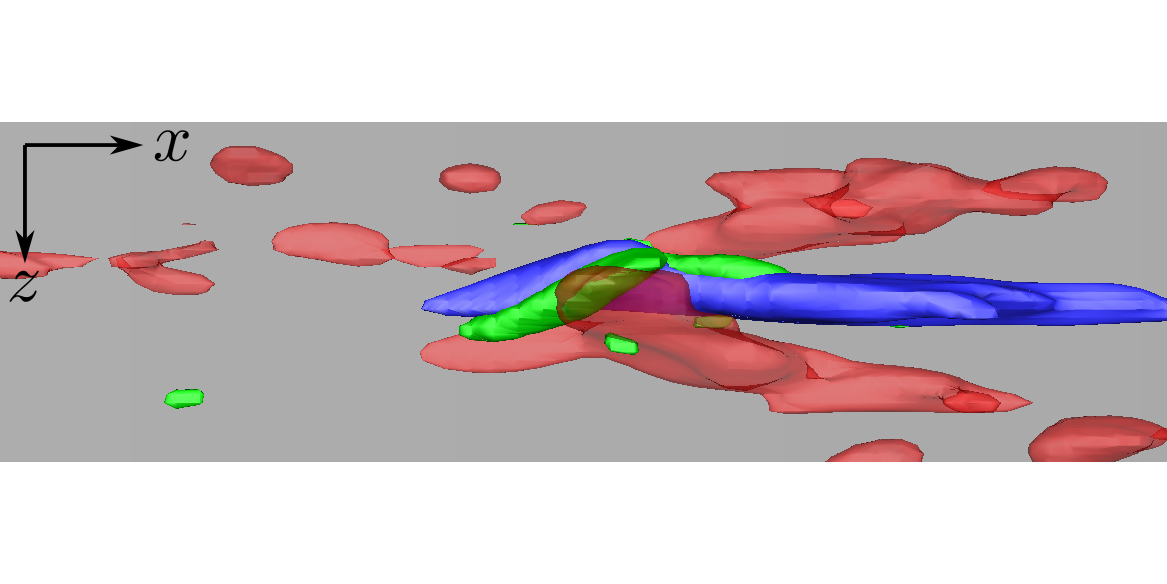}
	\put(1,43){(\textit{a})}
  \end{overpic}
  \begin{overpic}[width=0.32\textwidth]{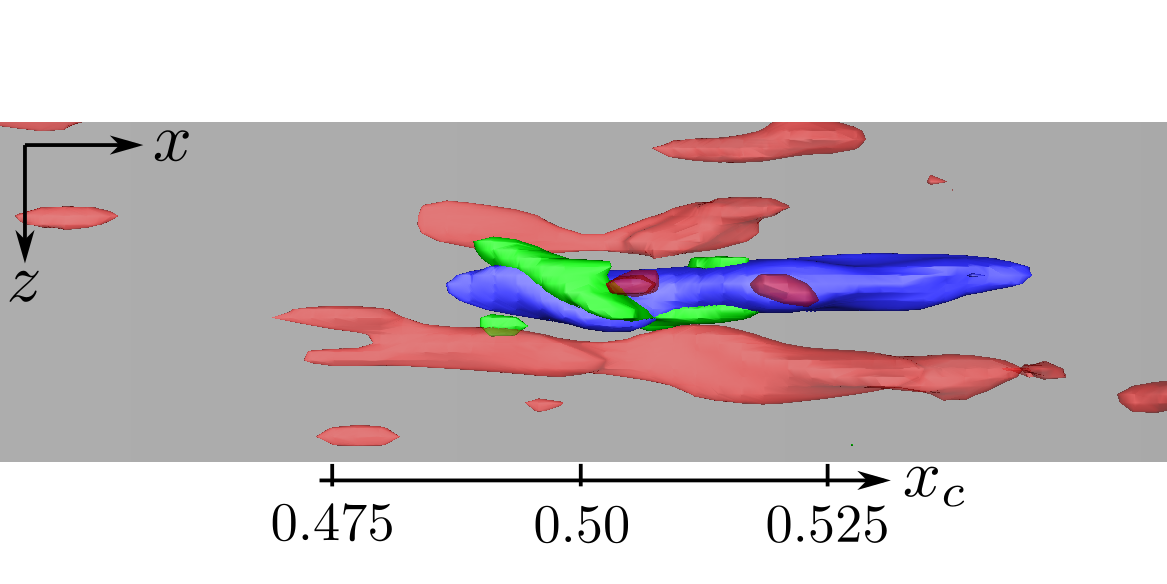}
   \end{overpic}
   \begin{overpic}[width=0.32\textwidth]{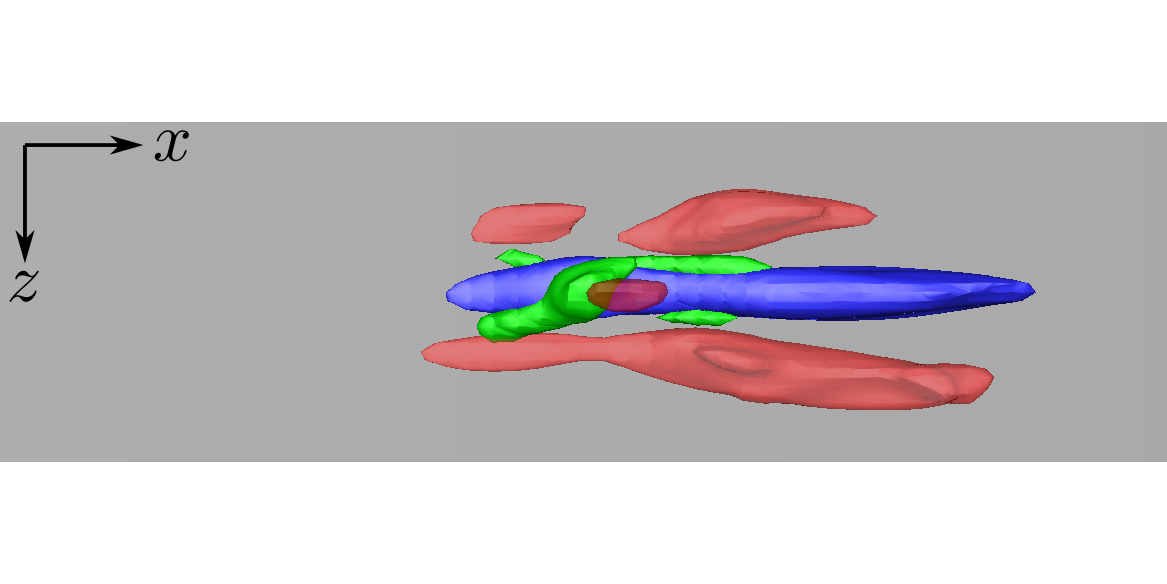}
   \end{overpic}
   \begin{overpic}[width=0.32\textwidth]{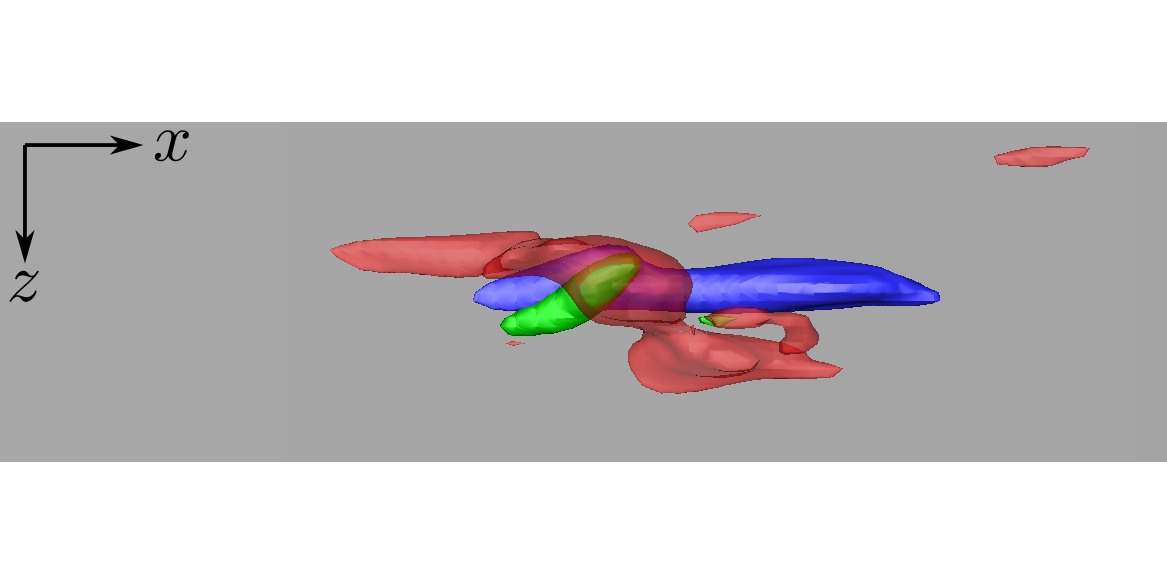}
	\put(1,43){(\textit{b})}
  \end{overpic}
  \begin{overpic}[width=0.32\textwidth]{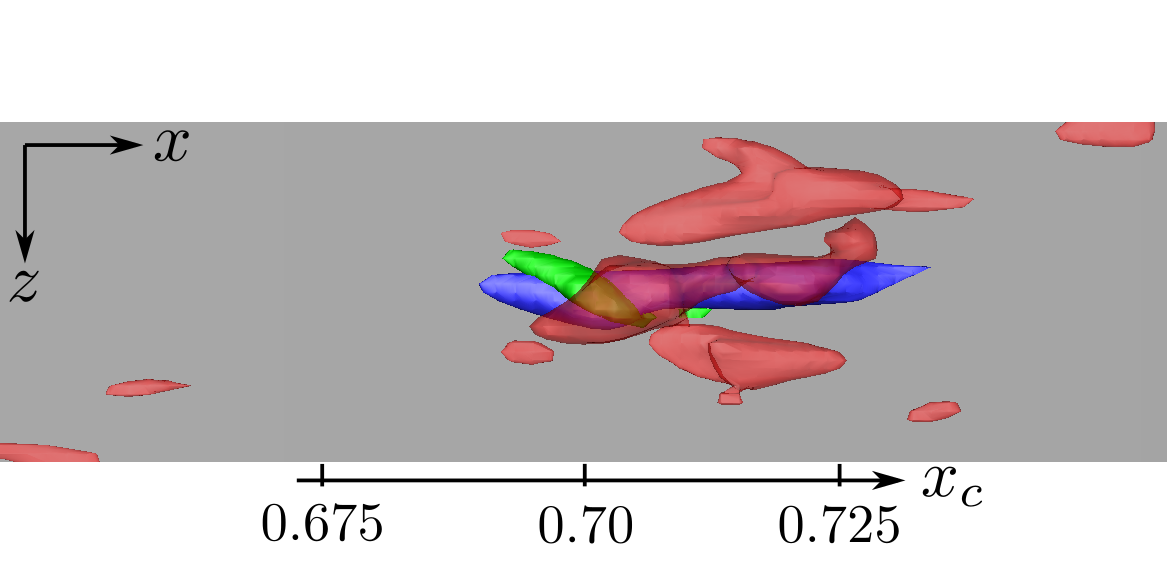}
   \end{overpic}
   \begin{overpic}[width=0.32\textwidth]{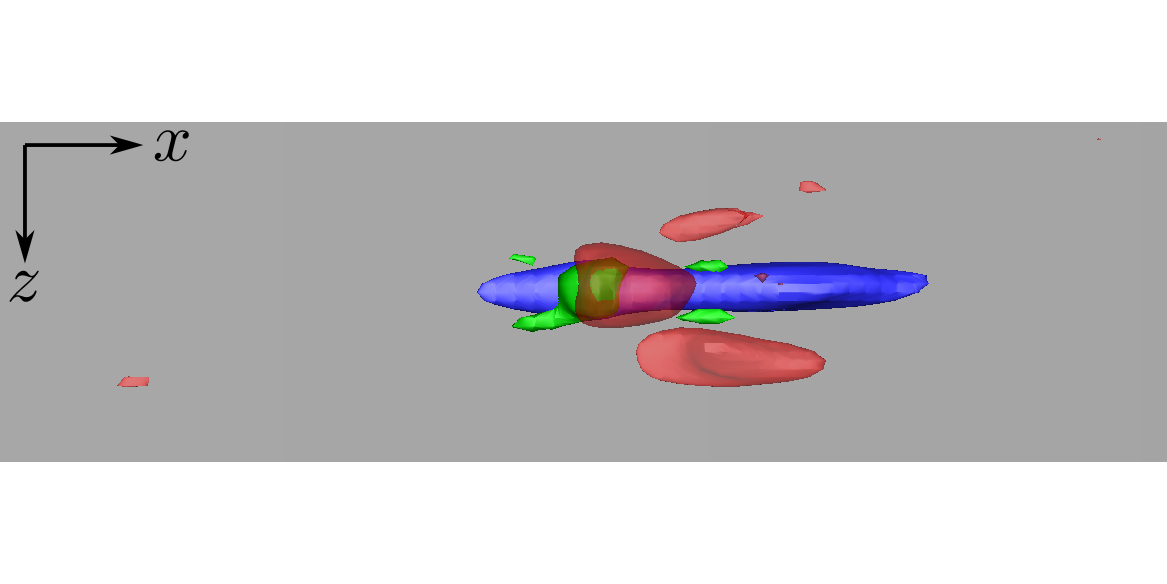}
   \end{overpic}
  \begin{overpic}[width=0.32\textwidth]{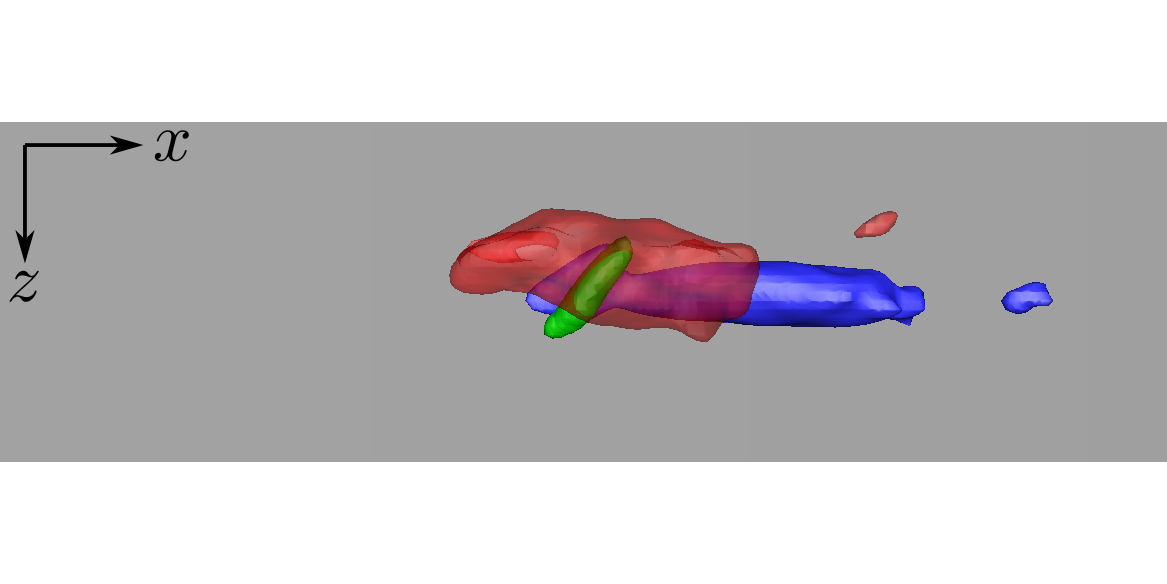}
	\put(1,43){(\textit{c})}
  \end{overpic}
  \begin{overpic}[width=0.32\textwidth]{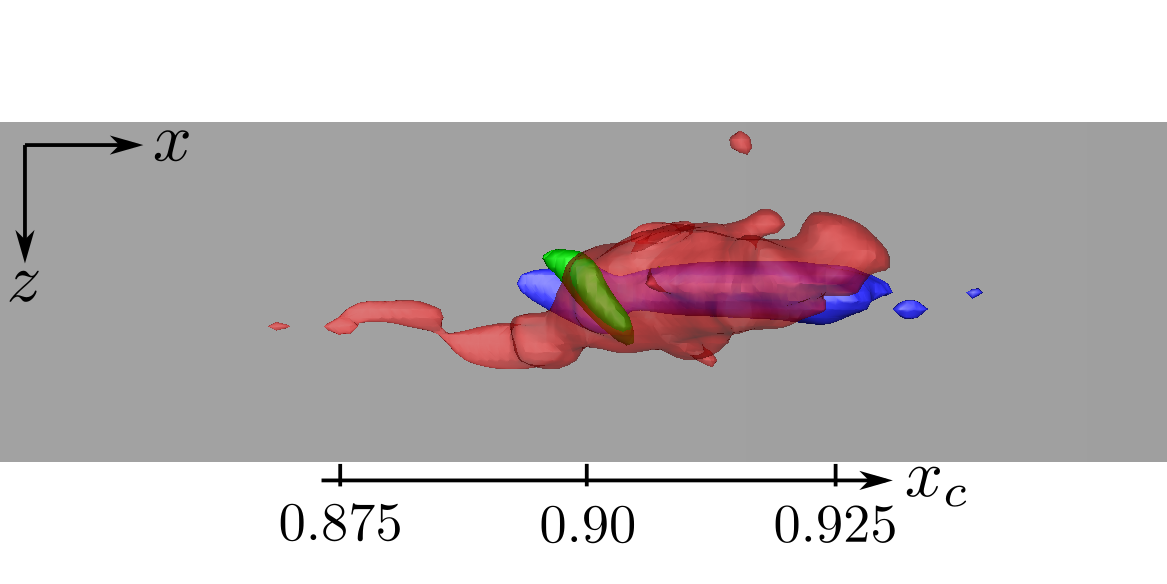}
   \end{overpic}
   \begin{overpic}[width=0.32\textwidth]{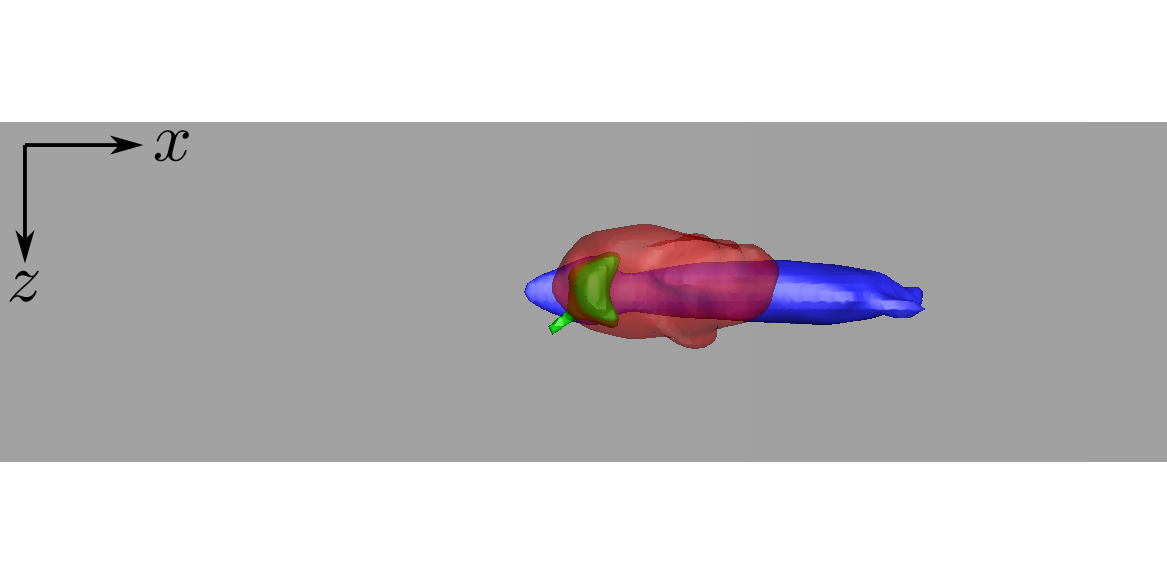}
   \end{overpic}
   \begin{overpic}[width=0.32\textwidth]{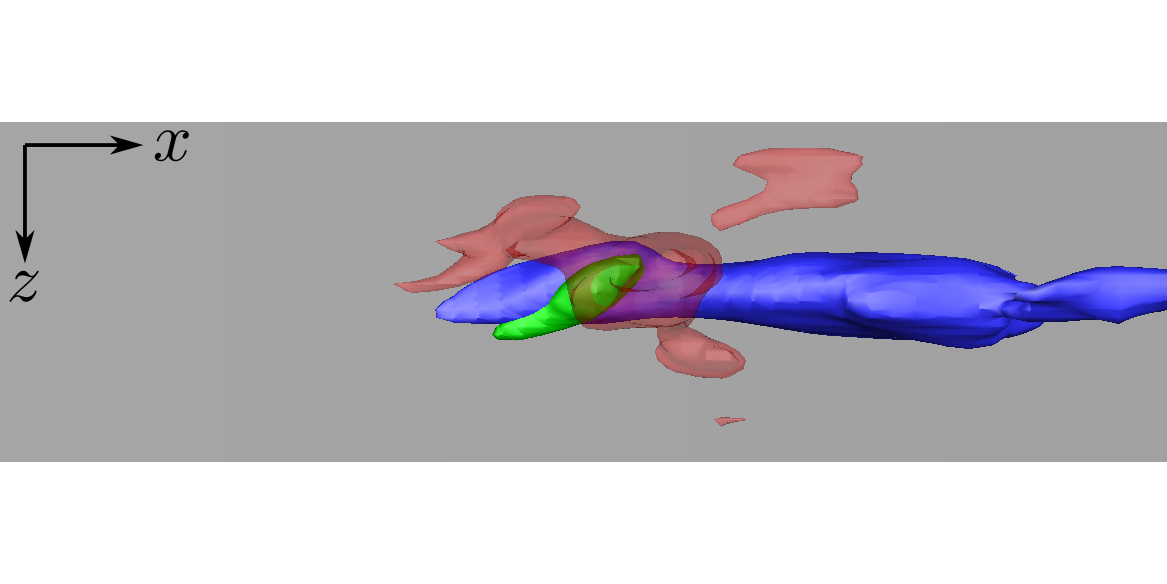}
	\put(1,43){(\textit{d})}
  \end{overpic}
  \begin{overpic}[width=0.32\textwidth]{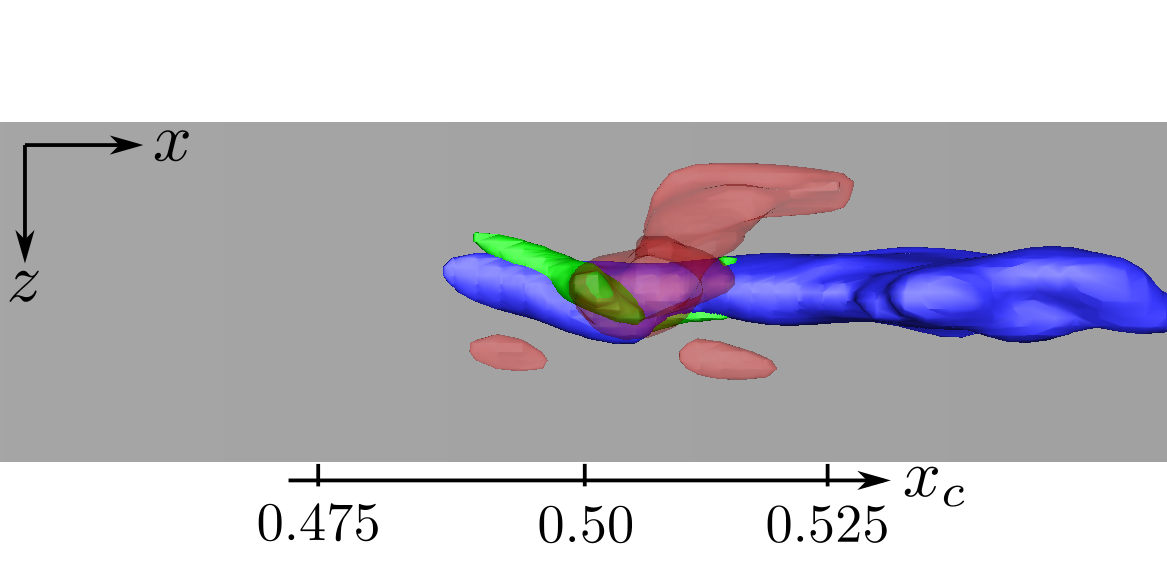}
  \end{overpic}
    \begin{overpic}[width=0.32\textwidth]{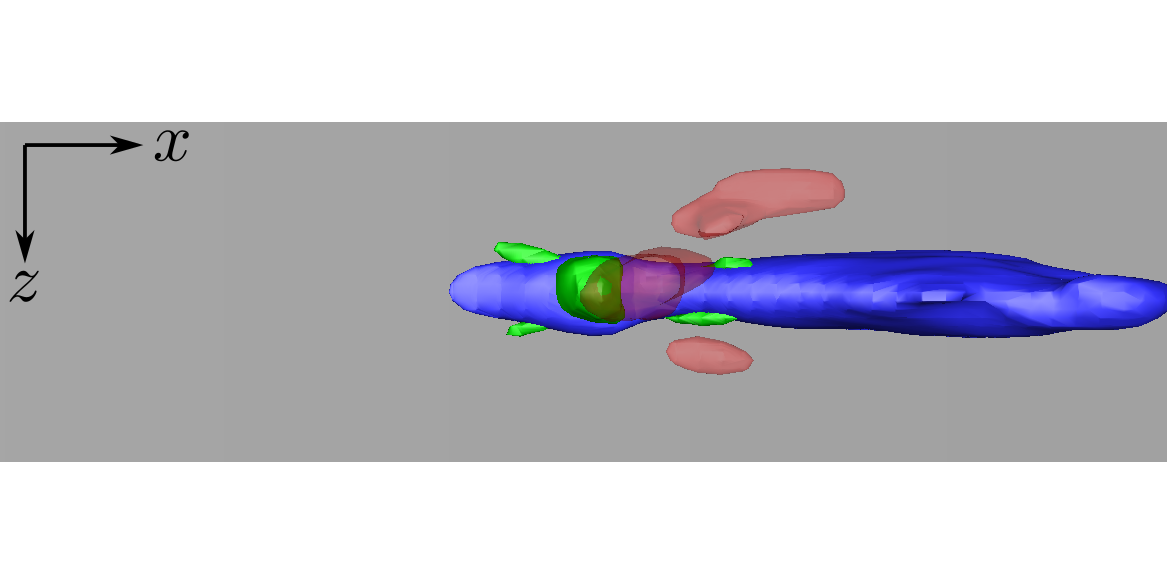}
  \end{overpic}
  \begin{overpic}[width=0.32\textwidth]{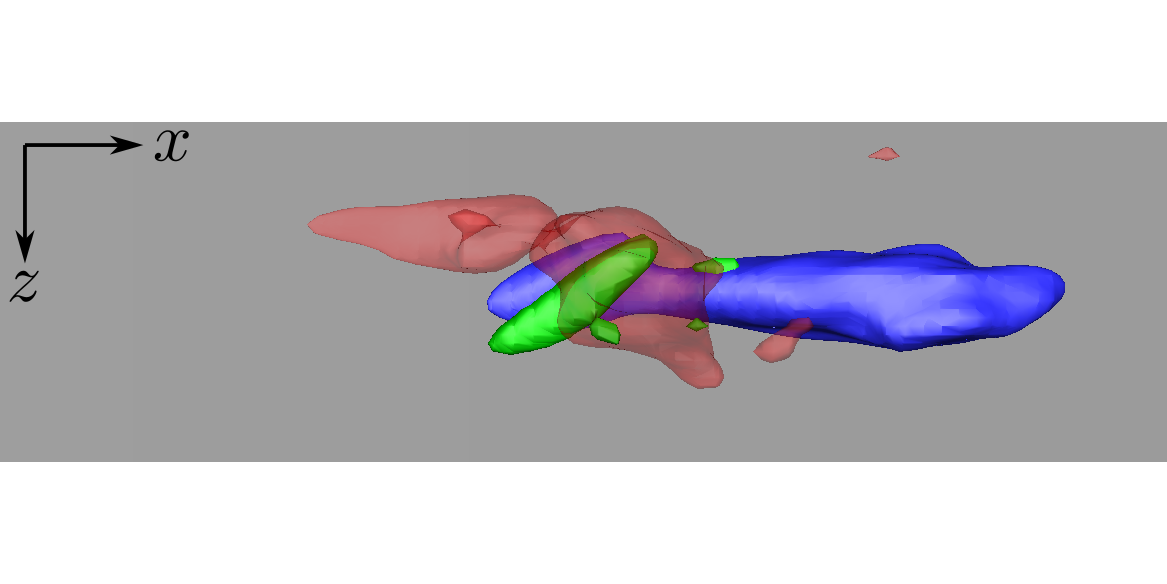}
	\put(1,43){(\textit{e})}
  \end{overpic}
  \begin{overpic}[width=0.32\textwidth]{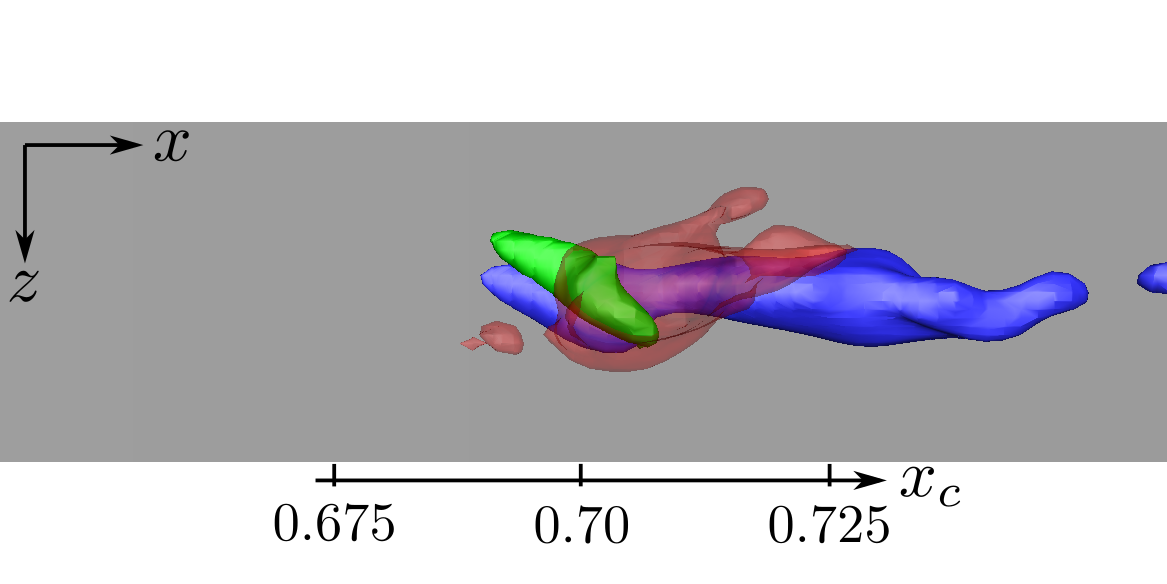}
  \end{overpic}
    \begin{overpic}[width=0.32\textwidth]{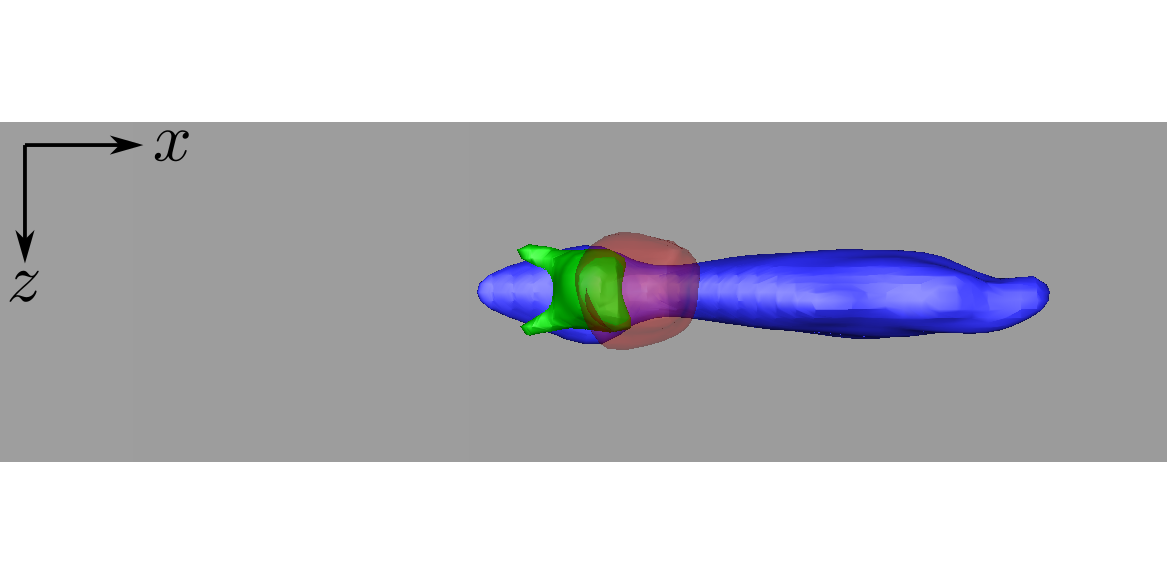}
  \end{overpic}
  \begin{overpic}[width=0.32\textwidth]{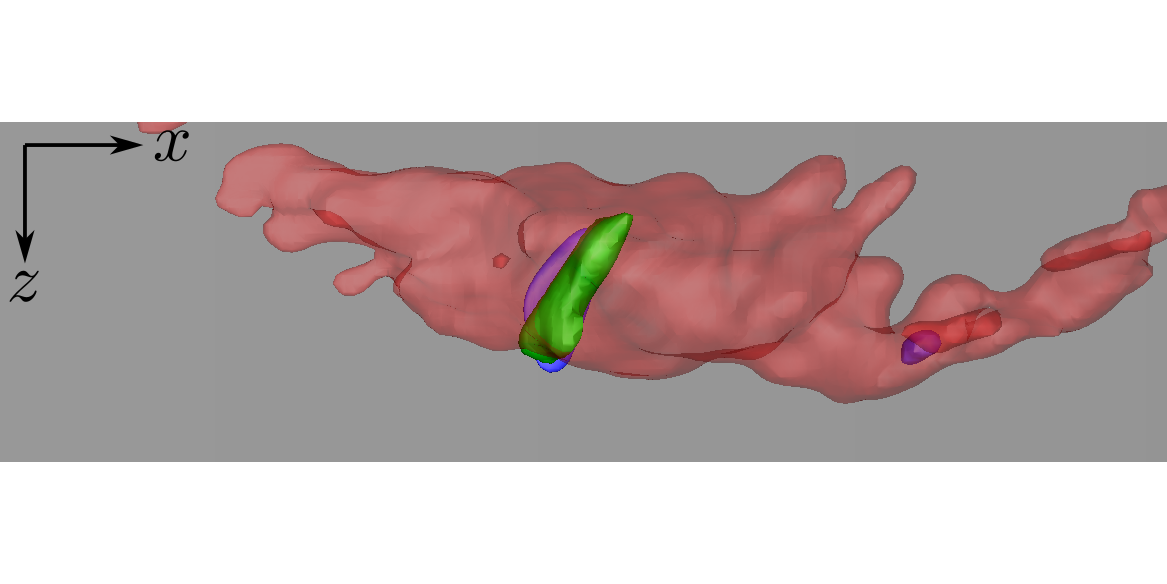}
	\put(1,43){(\textit{f})}
  \end{overpic}
  \begin{overpic}[width=0.32\textwidth]{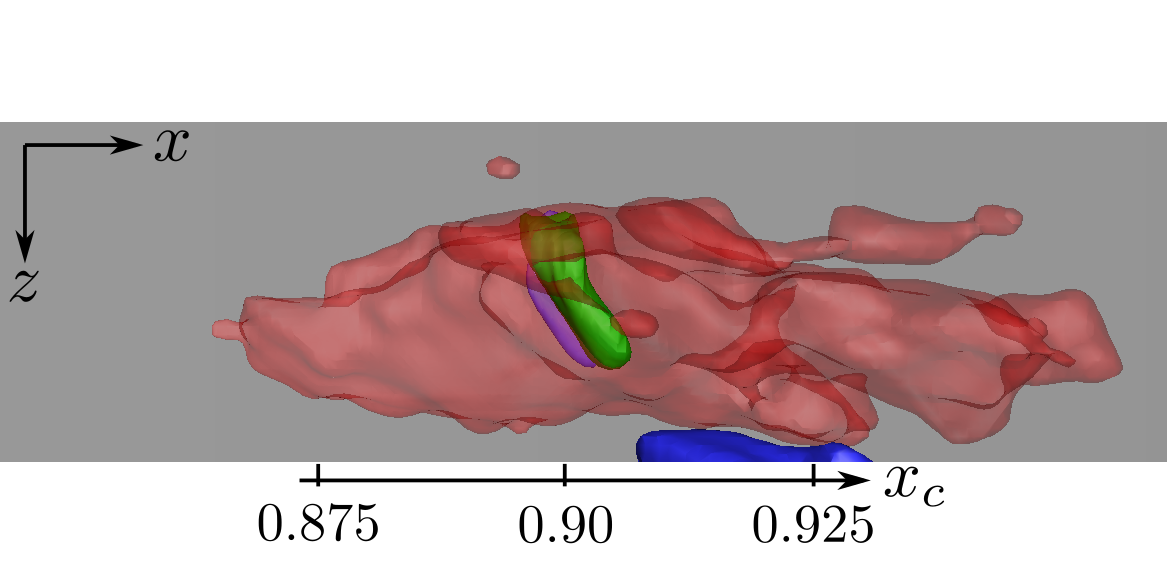}
  \end{overpic}
    \begin{overpic}[width=0.32\textwidth]{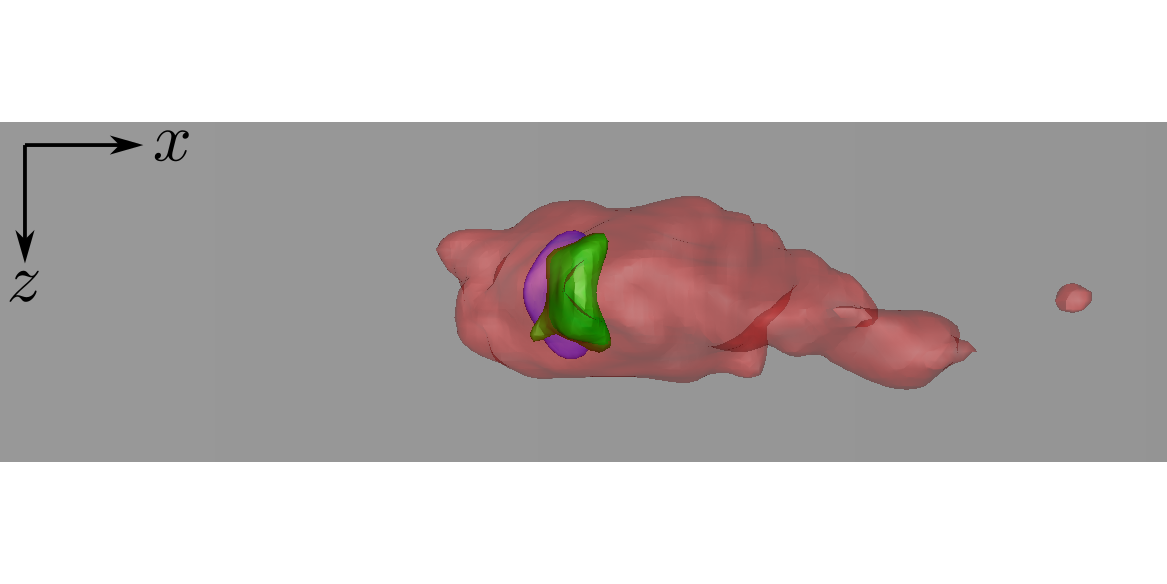}
  \end{overpic}
\caption{Top view of coherent structures associated with backflow events at $\Delta t_0 = 0.000$ for chordwise positions (\textit{a}, \textit{d}) $x_c = 0.5$, (\textit{b}, \textit{e}) $0.7$, and (\textit{c}, \textit{f}) $0.9$. The first three rows present the results for the $9$ deg. angle of attack case, while the last three rows show the results for $12$ deg. Red and blue isosurfaces correspond to regions of positive and negative tangential velocity fluctuations $\widetilde{u}_t$, respectively. Vortex structures are visualized by green isosurfaces using the $Q$-criterion.}
\label{fig:bk_top_view}
\end{figure}

As discussed in section \ref{sec:coherent_structures} and also reported by \citet{sheng2009}, extreme wall shear stress events exhibit a preferential side orientation. An inspection of the instantaneous fields confirms the same tendency. If this preferential orientation is not taken into account, the results obtained from conditional averaging may lead to misleading interpretations. In this section, we examine the impact of the conditional averaging procedure and highlight the tendencies described in section \ref{sec:coherent_structures}. Figure \ref{fig:bk_top_view} presents the results for the coherent structures associated with backflows at chordwise positions $x_c = 0.5$, $0.7$, and $0.9$ for angles of attack of $9$ deg. (panels \textit{a}-\textit{c}) and $12$ deg. (panels \textit{d}-\textit{f}). The first and second columns correspond to conditional averages of events classified on the left and right sides, respectively, while the third column shows the average of all events without any distinction. The red and blue isosurfaces represent regions of positive and negative tangential velocity fluctuations $\widetilde{u}_t$, respectively, and the green isosurfaces display vortical structures identified by the $Q$-criterion. The tendencies discussed in section \ref{sec:bf_struc} are clearly observed: as the APG intensity increases downstream, the low-speed streak reduces in size, while the high-speed sweeping structure over the streak grows. The vortex orientation also becomes increasingly aligned with the spanwise direction with the APG. If all events are averaged without separating their orientation, the resulting flow topology resembles a $\Lambda$-vortex structure, which does not accurately reflect the actual coherent structure responsible for the backflow and its flow dynamics.

Figure \ref{fig:ep_top_view} presents the results for the extreme positive events, using the same visualization approach as in the previous figure. The results reveal a clear tendency for the flanking low-speed streaks to weaken and for the high-speed streak and vortex structures responsible for the event to decrease in size as the APG magnitude increases. As discussed in section \ref{sec:ep_struc}, these events are predominantly associated with a dominant quasi-streamwise vortex located either on the left or right side of the high-speed streak. However, when all occurrences are conditionally averaged without distinguishing their orientation (third column of figure \ref{fig:ep_top_view}), the resulting flow topology suggests the presence of two interacting counter-rotating vortices that sweep high-speed fluid toward the wall.
\textbf{\begin{figure}
 \centering
    \begin{overpic}[width=0.32\textwidth]{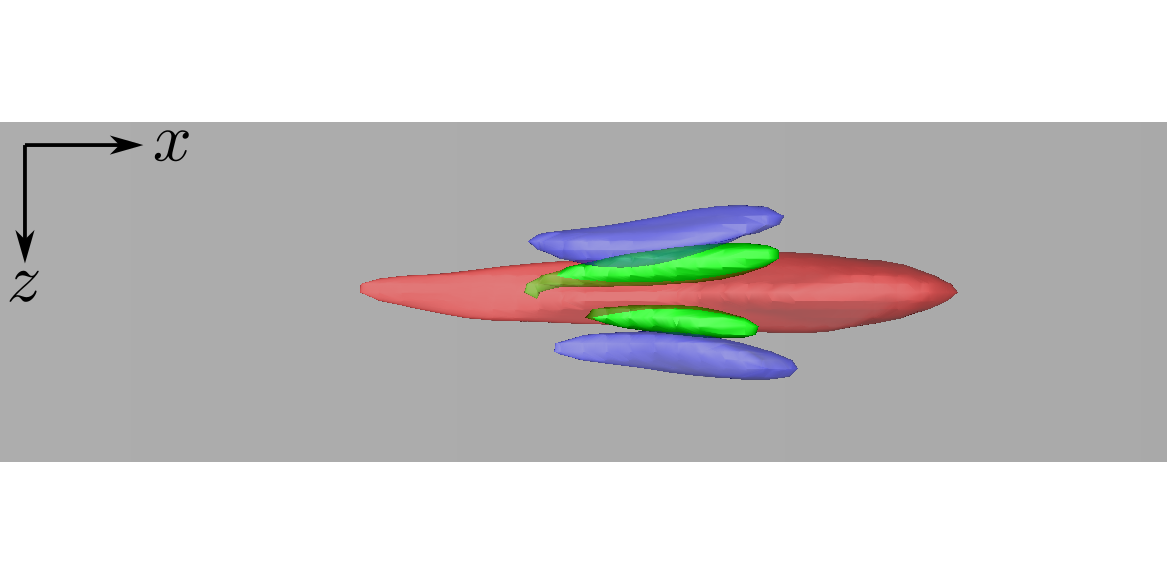}
	\put(1,43){(\textit{a})}
  \end{overpic}
  \begin{overpic}[width=0.32\textwidth]{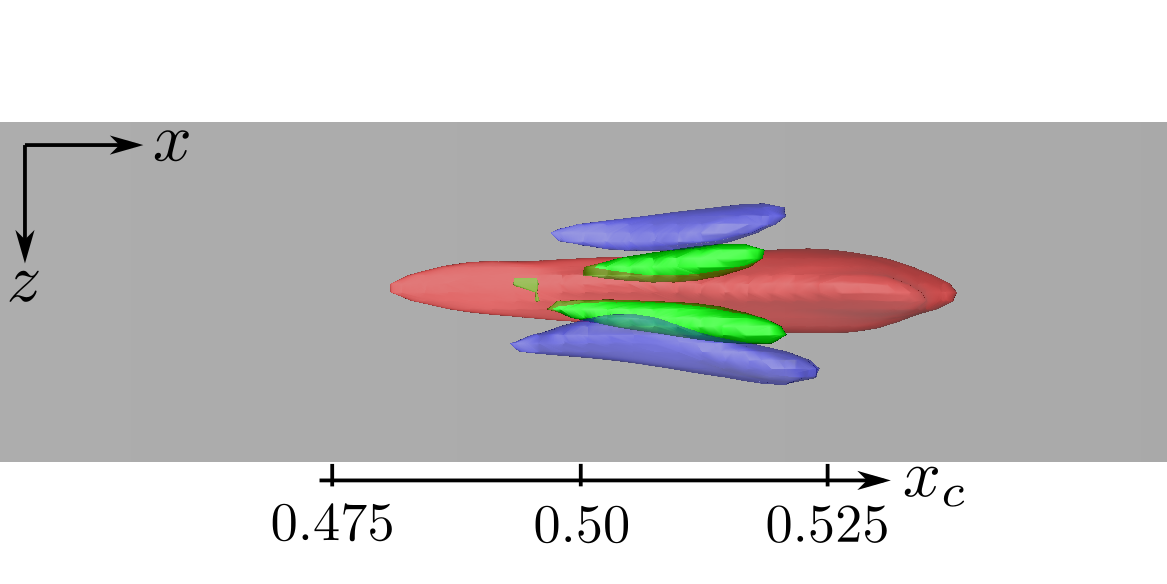}
   \end{overpic}
   \begin{overpic}[width=0.32\textwidth]{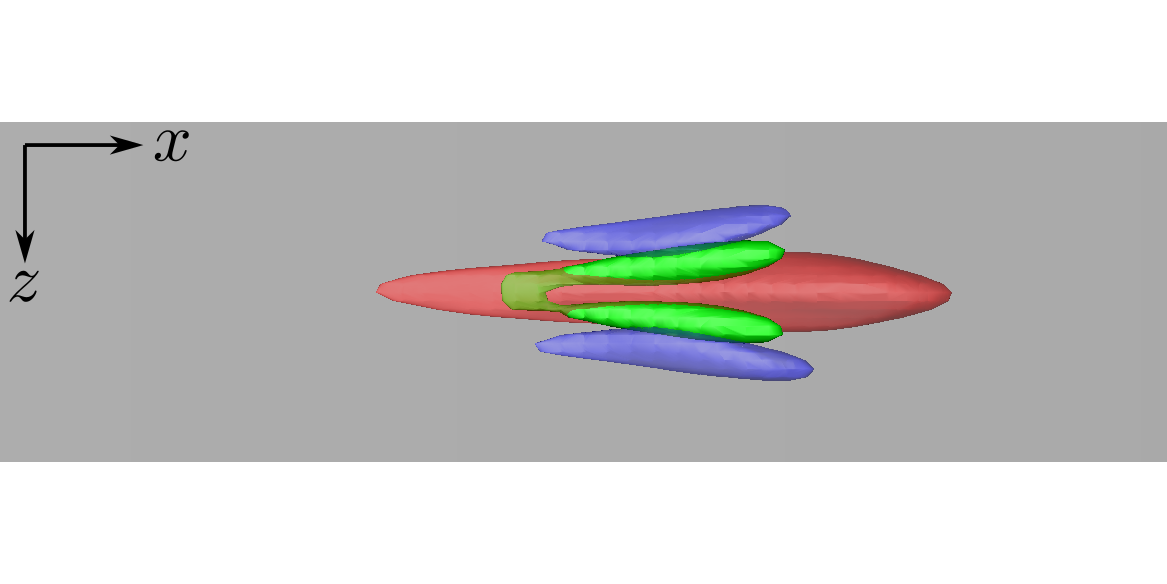}
   \end{overpic}
   \begin{overpic}[width=0.32\textwidth]{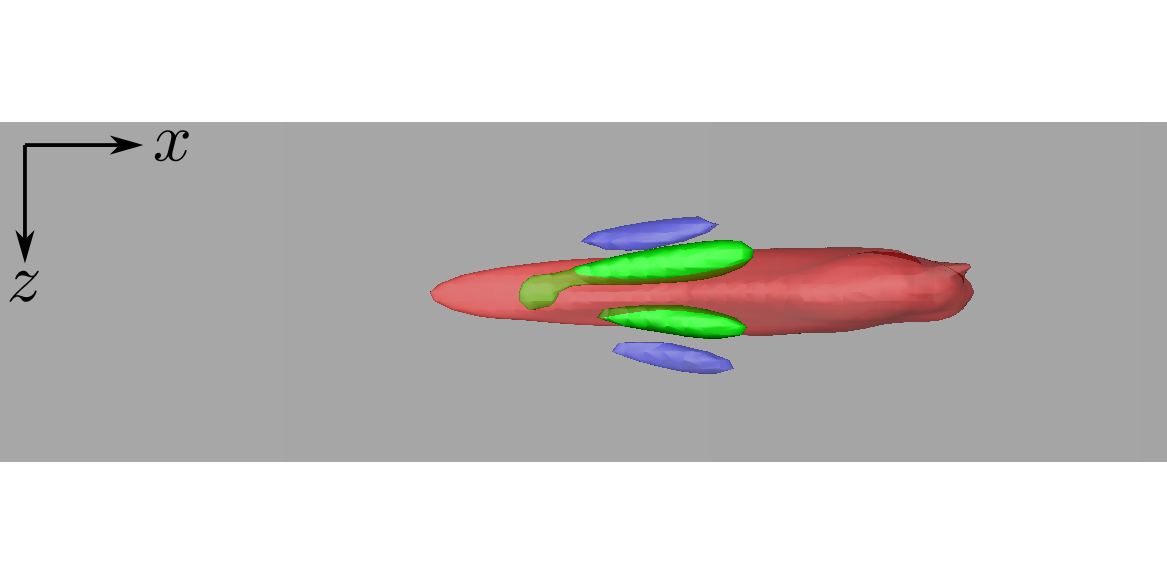}
	\put(1,43){(\textit{b})}
  \end{overpic}
  \begin{overpic}[width=0.32\textwidth]{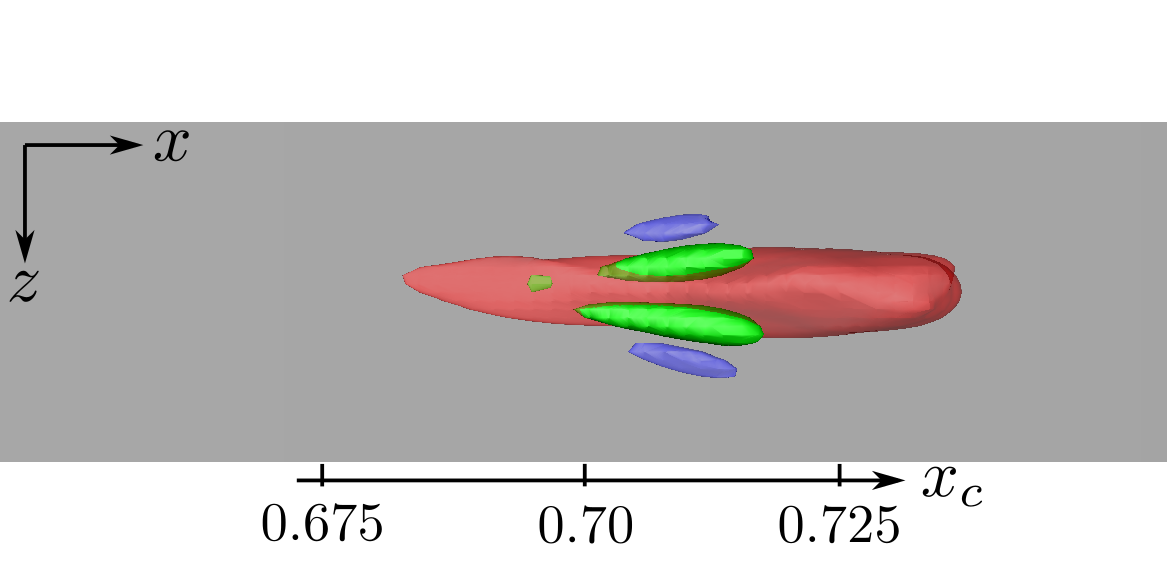}
   \end{overpic}
   \begin{overpic}[width=0.32\textwidth]{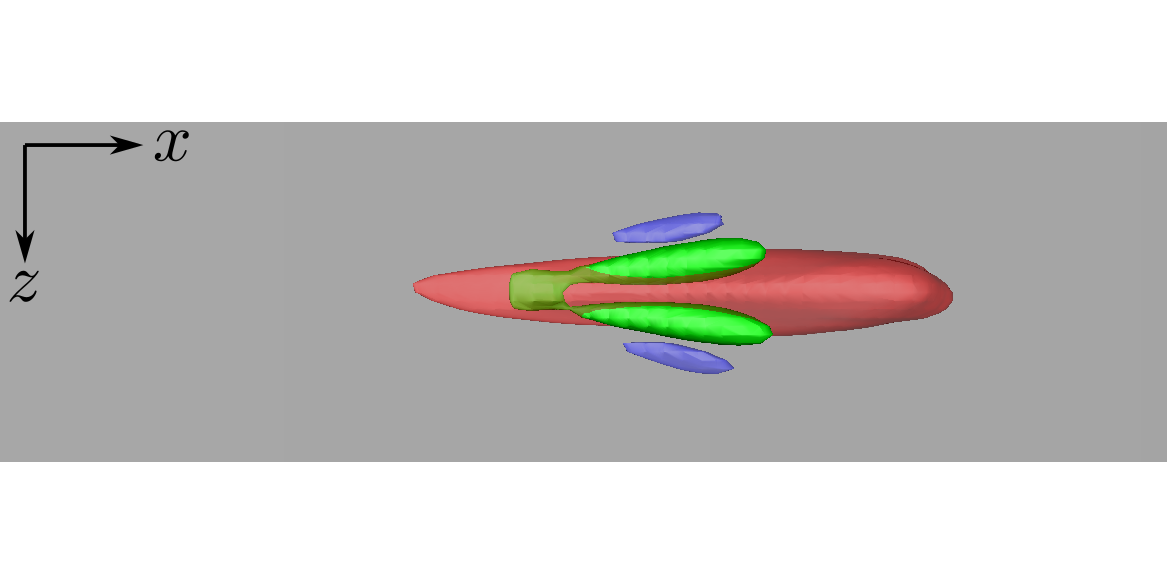}
   \end{overpic}
  \begin{overpic}[width=0.32\textwidth]{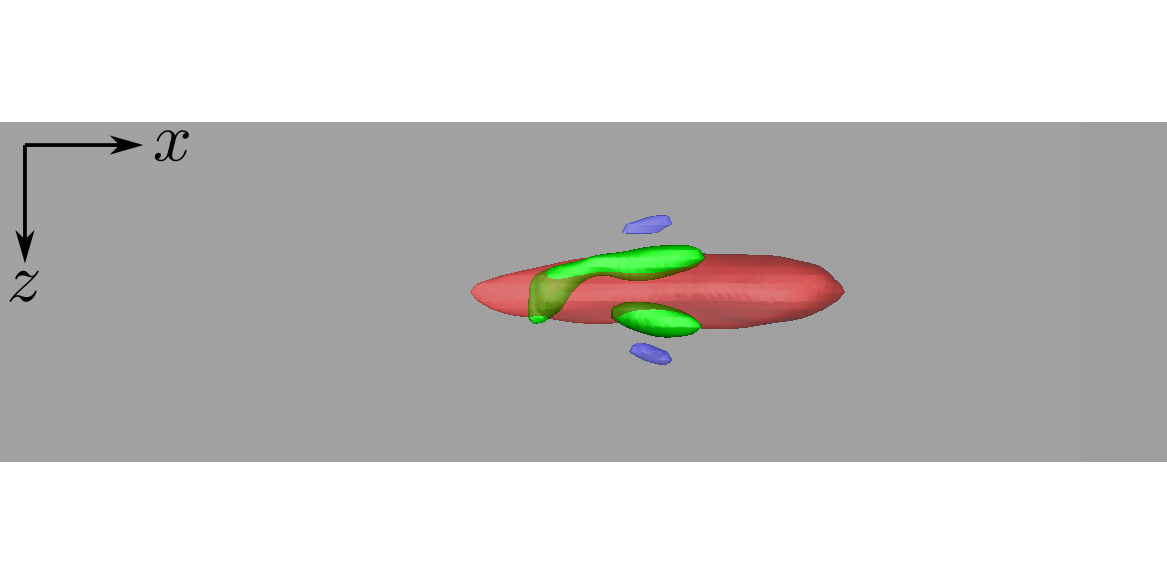}
	\put(1,43){(\textit{c})}
  \end{overpic}
  \begin{overpic}[width=0.32\textwidth]{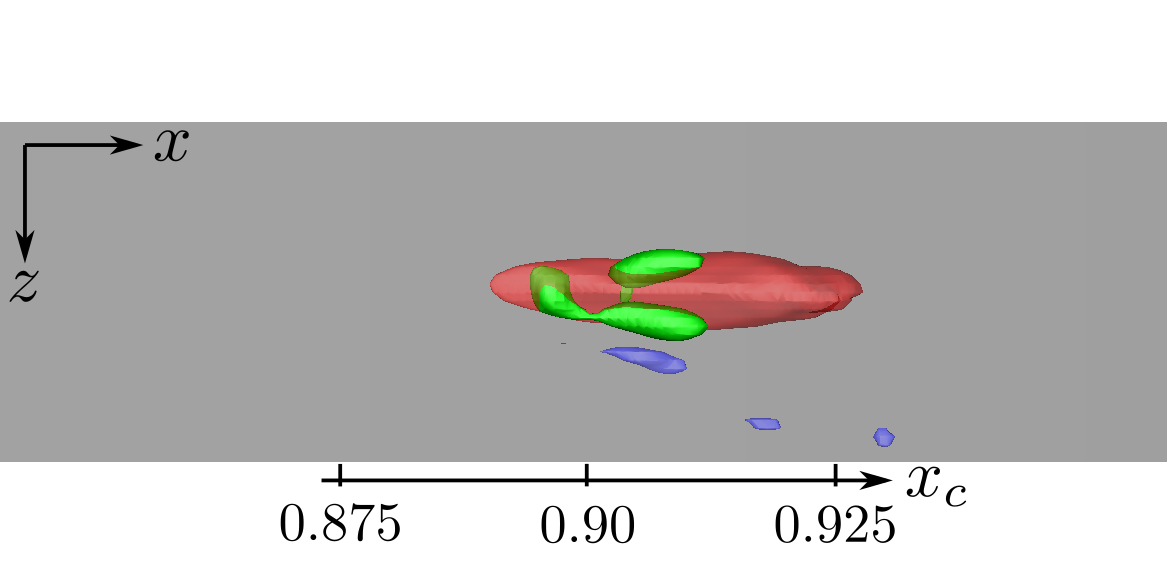}
   \end{overpic}
   \begin{overpic}[width=0.32\textwidth]{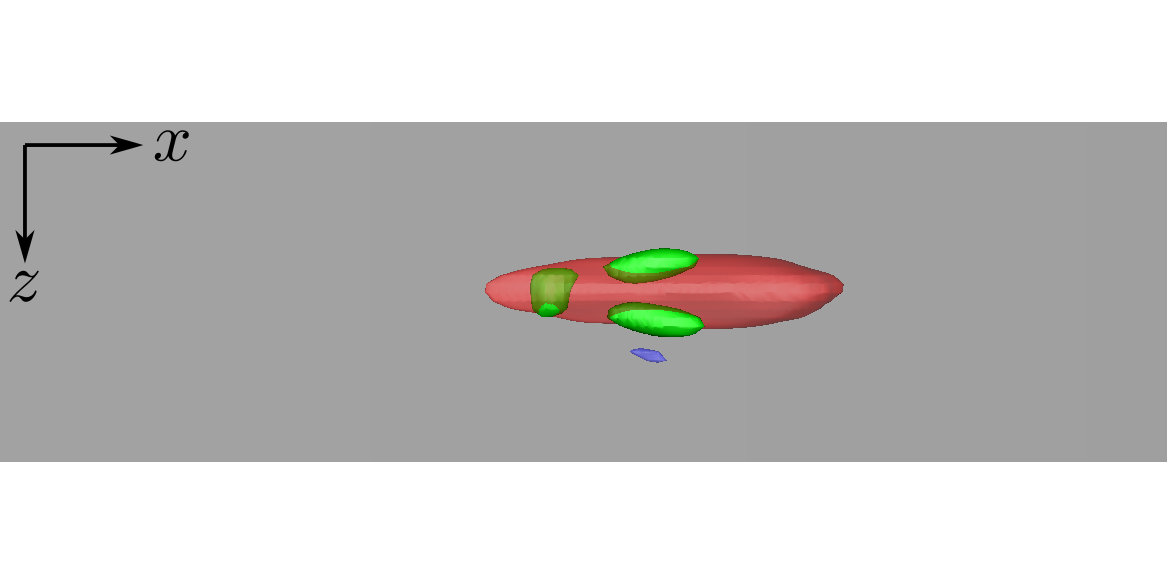}
   \end{overpic}
   \begin{overpic}[width=0.32\textwidth]{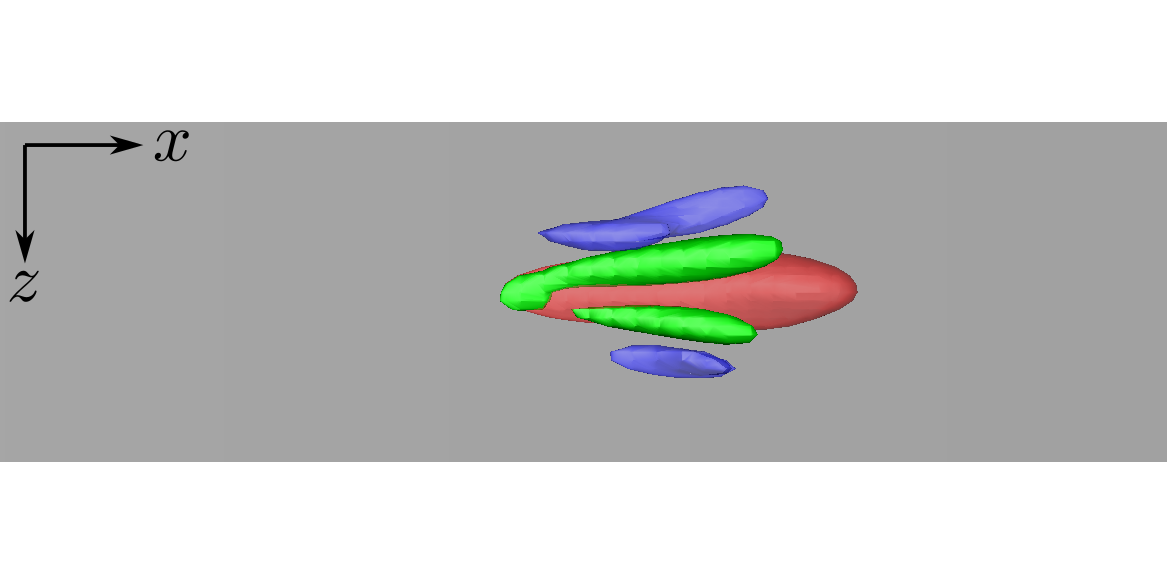}
	\put(1,43){(\textit{d})}
  \end{overpic}
  \begin{overpic}[width=0.32\textwidth]{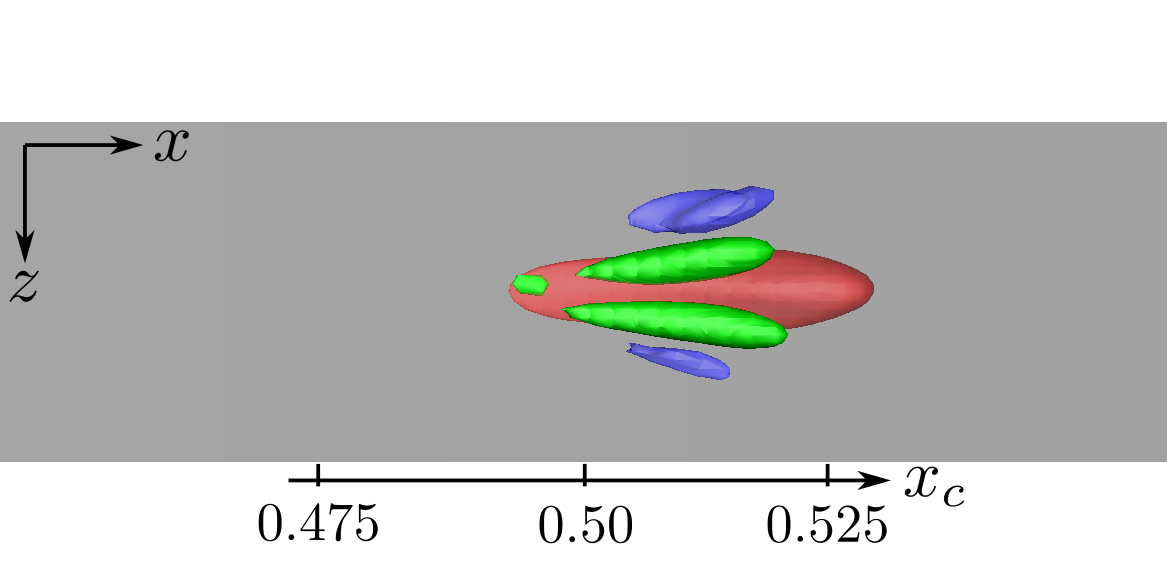}
  \end{overpic}
    \begin{overpic}[width=0.32\textwidth]{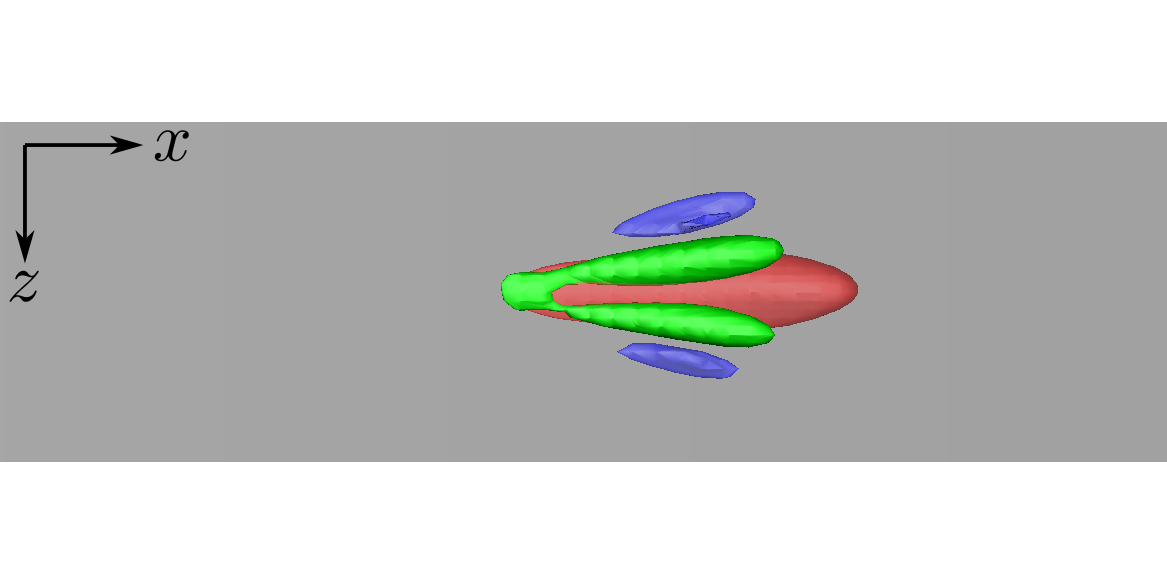}
  \end{overpic}
  \begin{overpic}[width=0.32\textwidth]{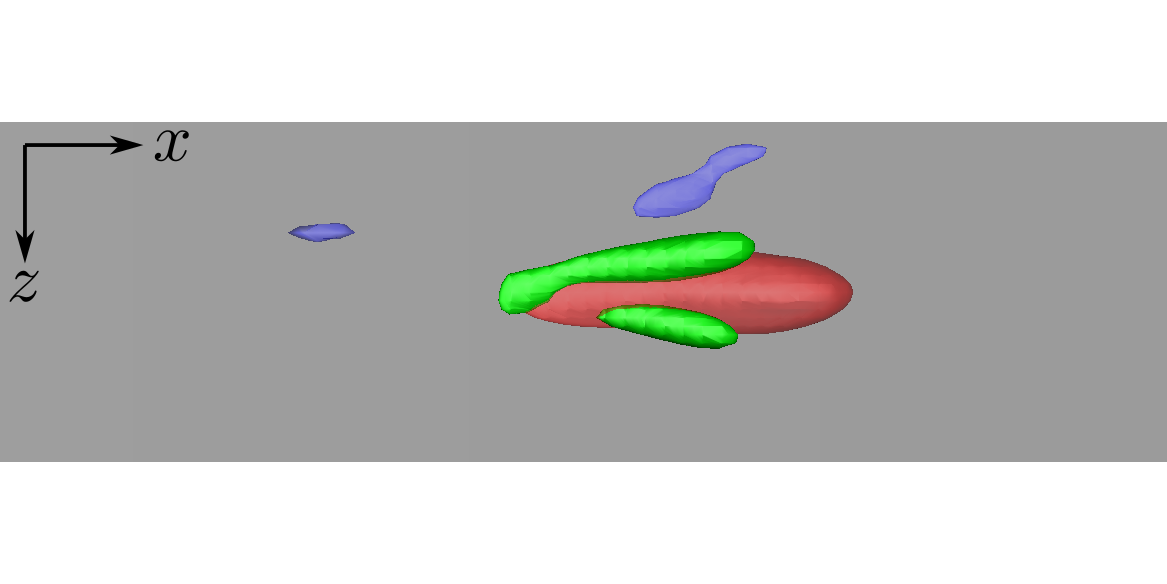}
	\put(1,43){(\textit{e})}
  \end{overpic}
  \begin{overpic}[width=0.32\textwidth]{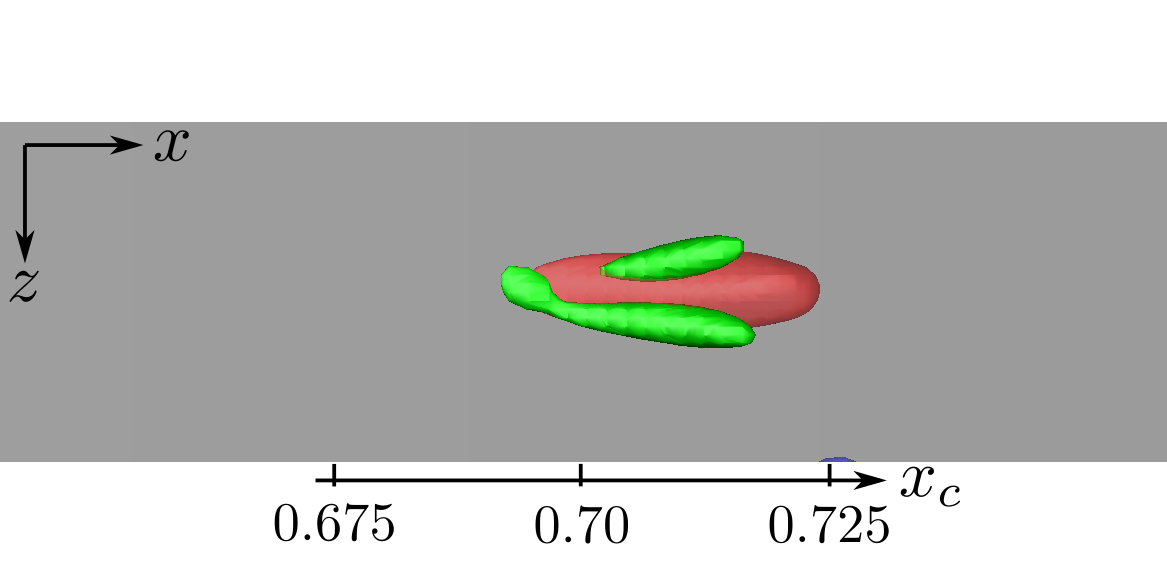}
  \end{overpic}
    \begin{overpic}[width=0.32\textwidth]{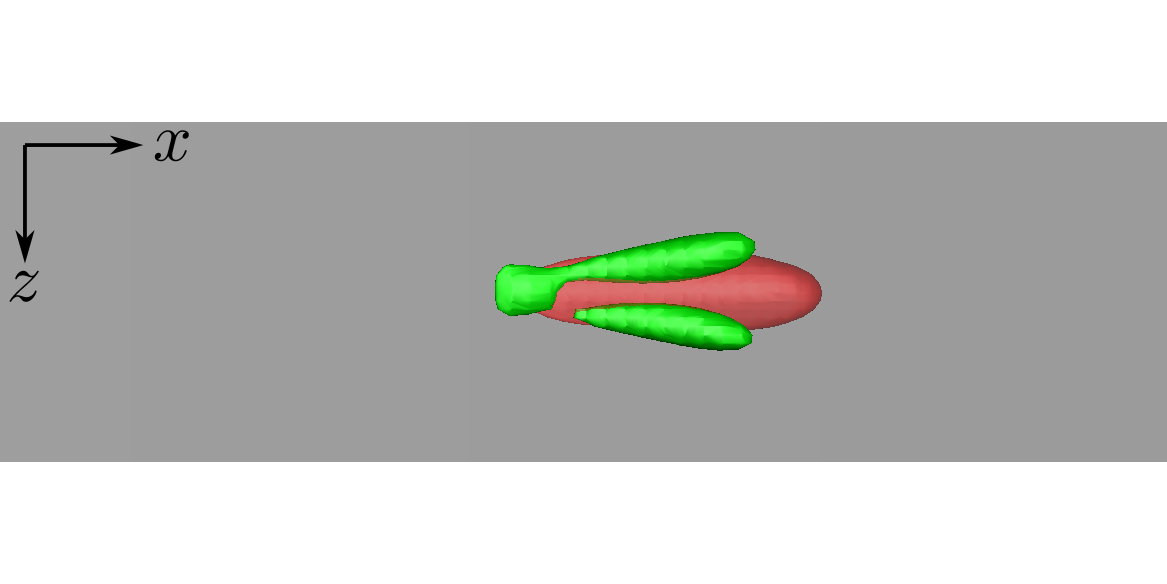}
  \end{overpic}
  \begin{overpic}[width=0.32\textwidth]{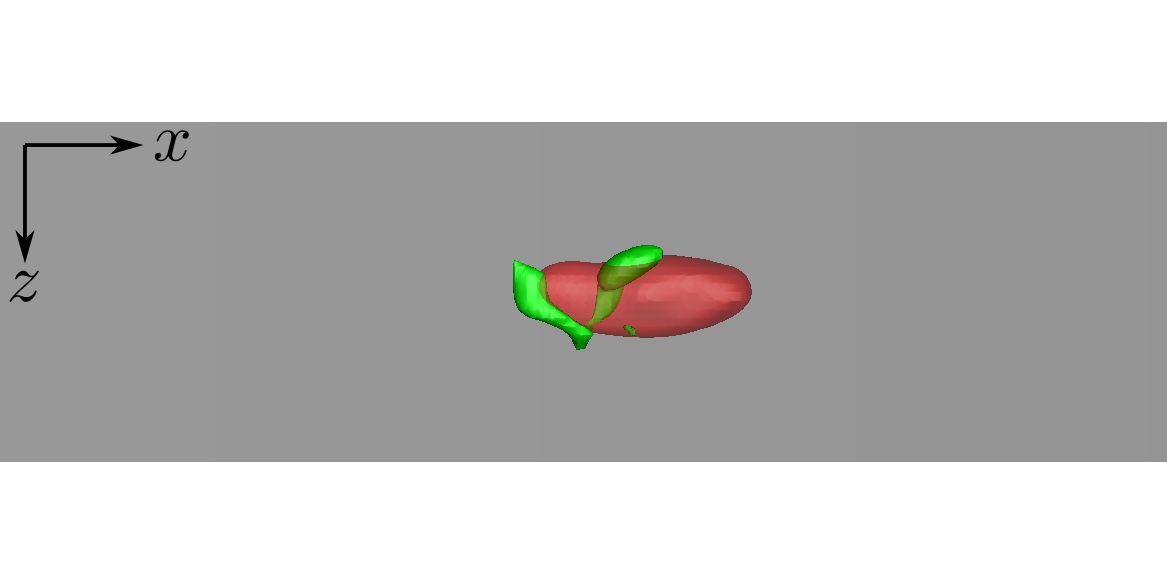}
	\put(1,43){(\textit{f})}
  \end{overpic}
  \begin{overpic}[width=0.32\textwidth]{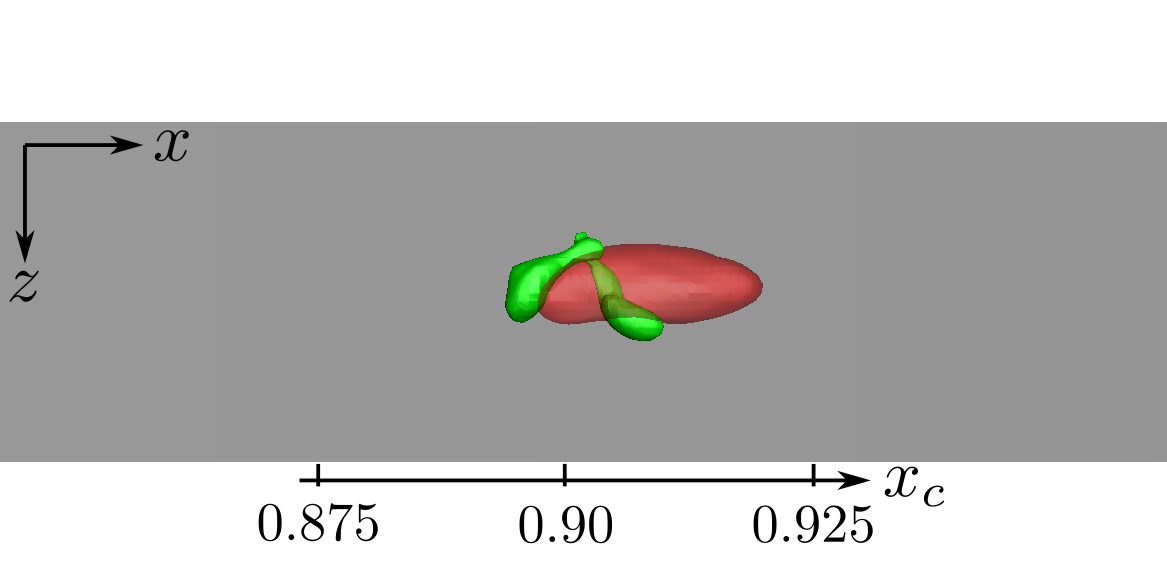}
  \end{overpic}
    \begin{overpic}[width=0.32\textwidth]{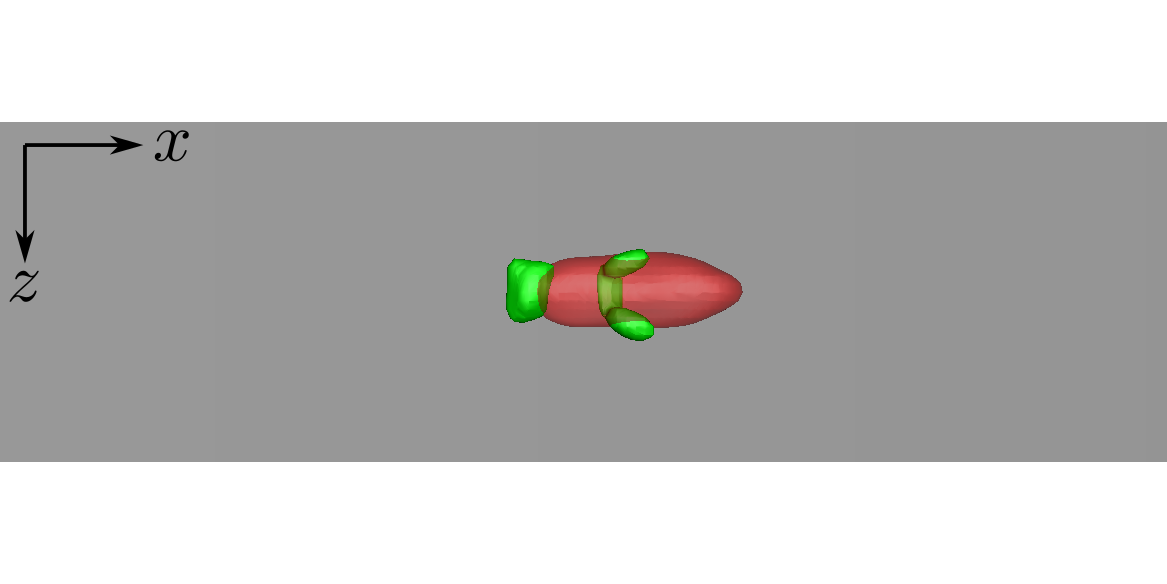}
  \end{overpic}
\caption{Top view of coherent structures associated with extreme positive events at $\Delta t_0 = 0.000$ for chordwise positions (\textit{a}, \textit{d}) $x_c = 0.5$, (\textit{b}, \textit{e}) $0.7$, and (\textit{c}, \textit{f}) $0.9$. The first three rows present the results for the $9$ deg. angle of attack case, while the last three rows show the results for $12$ deg. Red and blue isosurfaces correspond to regions of positive and negative tangential velocity fluctuations $\widetilde{u}_t$, respectively. Vortex structures are visualized by green isosurfaces using the $Q$-criterion.}
\label{fig:ep_top_view}
\end{figure} }

\bibliographystyle{jfm}
\bibliography{jfm}


\end{document}